\documentclass[11pt,a4paper]{article}
\usepackage{longtable}
\usepackage{amsmath}
\usepackage{amssymb}
\usepackage{graphicx}
\usepackage{bm}
\usepackage{footmisc}
\usepackage{afterpage}
\usepackage{float}
\usepackage{lscape}
\usepackage{longtable}
\usepackage{float}
\usepackage{slashed}

\newcommand*{\no}{\noindent}
\newcommand*{\bea}{\begin{eqnarray}}
\newcommand*{\eea}{\end{eqnarray}}
\newcommand*{\be}{\begin{equation}}
\newcommand*{\ee}{\end{equation}}
\newcommand*{\pd}{\partial}
\newcommand*{\pdm}{\pd_{\mu}}

\newcommand*{\pref}[1]{(\ref{#1})}

\newcommand*{\nn}{\nonumber}

\newcommand{\bma}{\begin{pmatrix}}
\newcommand{\ema}{\end{pmatrix}}

\usepackage[square,comma,sort&compress,numbers]{natbib}

\title{The quenched SU(2) fundamental scalar propagator in minimal Landau gauge}

\author{Axel Maas\\
Institute of Physics, NAWI Graz, University of Graz,\\
Universit\"atsplatz 5, A-8010 Graz, Austria}

\begin{document}

\maketitle

\begin{abstract}

It is a long-standing question whether the confinement of matter fields in QCD has an imprint in the (gauge-dependent) correlation functions, especially the propagators. As the analytic structure plays an important role in this question, high-precision data is necessary for lattice investigations. Also, it is interesting how this depends on the dimensionality of the theory. To make a study over a wide range of parameters possible this suggests to use scalar particles. This is done here: The propagator of a fundamental scalar is studied in two, three, and four dimensions in quenched SU(2) Yang-Mills theory in minimal Landau gauge, both in momentum space and position space. Particular emphasis is put on the effects of renormalization. The results suggest a quite intricate volume dependence and the presence of an intrinsic mass scale, but no obvious connection to confinement.

\end{abstract}

\section{Introduction}

The confinement of matter particles in QCD is a very long-standing problem \cite{Alkofer:2006fu}. In particular, the question is, whether the properties of the propagator describing the elementary particles, both gluon and matter, show signs of confinement\footnote{Confinement is here understood, if not noted otherwise, in the sense that a particle cannot be observed as an asymptotic, physical state. In this sense also QCD is confining. A definition of confinement based on the Wilson string tension is in no obvious way related to this. In fact, with this view on confinement QCD is not a confining theory. See \cite{Alkofer:2006fu} for a more detailed discussion on this difference.}  \cite{Alkofer:2000wg,Maas:2011se,Fischer:2006ub,Binosi:2009qm,Boucaud:2011ug,Vandersickel:2012tg,Roberts:2015lja}. Of course, such investigations require to fix a gauge to discuss the corresponding propagator, which will be here chosen to be the best-studied case so far, the Landau gauge, especially the so-called minimal Landau gauge \cite{Maas:2011se}.

One of the major tools used for this purpose is the spectral density. This spectral density is found to be positivity violating for gluons \cite{Alkofer:2000wg,Maas:2011se,Fischer:2006ub,Binosi:2009qm,Boucaud:2011ug,Vandersickel:2012tg,Roberts:2015lja,Strauss:2012as,Cucchieri:2016jwg}, though the precise form this violation takes, e.\ g.\ by a non-trivial cut structure, complex poles, or otherwise, is still under debate. At any rate, any violation of positivity immediately implies that the particle cannot be part of the physical state space, and thus not observable. Sufficient, but not necessary, conditions for violation of positivity can be either a non-positive definite position-space correlation function or a non-monotonous behavior of the derivatives of the momentum-space correlation functions \cite{Maas:2011se}. The form of such violations can then be used to constrain the type of analytic structure. E.\ g.\ an oscillatory behavior in position space points to a complex pole structure \cite{Maas:2011se}.

The quark case is more intricate, see \cite{Roberts:1994dr,Roberts:2015lja,Alkofer:2000wg,Alkofer:2003jj,August:2013jia}, but there are also strong indications for a violation of positivity.

For scalar matter, so far little is known, though their much simpler Lorentz structure suggests them as a testbed. There have been investigations using functional methods which provide two possible scenarios for the behavior of the propagator, either being conventionally massive or possibly showing some quasi-conformal behavior \cite{Fister:2010yw,Macher:2011ad,Capri:2013oja,Hopfer:2013via,Maas:2011yx,Maas:2013aia}.

There is also another intriguing question. In two (Euclidean) dimensions, i.\ e.\ one space and one time dimension, where gluons are not dynamical degrees of freedom, the violation of positivity for gluons appears to occur in much the same way as in higher dimensions \cite{Maas:2007uv,Cucchieri:2007rg,Maas:2014xma,Cucchieri:2016jwg}. However, in two dimensions gluons are not physical particles, but merely gauge degrees of freedom. At the same time, a confinement according to the Wilson potential occurs already for purely geometrical reasons \cite{Dosch:1978jt}. Scalar particles are, however, also in two dimensions physical particles. Whether their analytical structure changes may therefore give a hint how geometrical and dynamical confinement differ. Of course, this requires to study the propagator in the quenched case, where the Wilson potential is indefinitely rising also in higher dimensions. Fermions, on the other hand, are quite differently affected by changing the dimensionality, thus intertwining many different aspects.

All of this suggests to study the propagator of scalar matter in the quenched case to obtain further insights on the analytical structure, including the dependence on dimensionality. Thus, the aim of this work is to determine the quenched scalar fundamental propagator in a wide range of parameters and for two, three, and four dimensions, using lattice gauge theory. Having obtained the renormalized propagator, usage of the Schwinger function \cite{Alkofer:2003jj,Maas:2011se} should then provide information on the analytic structure.

Besides these fundamental issues, the quenched calculation provides also an excellent testbed to study the lattice artifacts and renormalization properties of the scalar propagator beyond perturbation theory, which is helpful in studies of the dynamical case \cite{Maas:2010nc,Maas:2011yx}.

The technical setup of the calculations are presented in section \ref{s:tech}. The issue of renormalization is studied in detail in section \ref{s:ren}. The results in momentum space and for the analytical structure of the renormalized propagator are then presented in section \ref{s:ana}. These are the main results of this work. The findings are summarized in section \ref{s:sum}. Some preliminary results can be found in \cite{Maas:2010nc,Maas:2011yx}.

\section{Technical setup}\label{s:tech}

In the following the propagator of a scalar particle in the fundamental representation in SU(2) in the quenched theory will be determined in two, three, and four dimensions. The technical setup is based on \cite{Cucchieri:2006tf,Maas:2007uv,Maas:2010nc}. Hence, the Wilson action for SU(2) Yang-Mills theory is simulated using a cycle of heatbath and overrelaxation updates. The lattice setups are listed in table \ref{tcgf} in appendix \ref{a:ls}. The determination of the lattice spacing has been performed as in \cite{Maas:2014xma}.

Each configuration selected for measurement has been fixed to minimal Landau gauge \cite{Maas:2011se} using adaptive stochastic overrelaxation \cite{Cucchieri:2006tf}. The quenched fundamental propagator has been obtained as in \cite{Maas:2010nc}. In the continuum, it is given by the inverse of the covariant fundamental Laplacian including the mass term
\be
-D^2=-\left(\pdm-i\frac{g\tau^a}{2}A_\mu^a\right)^2\nn+m_0^2,\nn
\ee
\no where the generators of the gauge algebra $\tau^a$ are the usual Pauli matrices, the $A_\mu^a$ are the gauge fields, $g=\sqrt{4/\beta}$ the (bare) coupling constant, and $m_0$ the bare mass of the scalar. As the lattice version of this operator its naive discretization \cite{Greensite:2006ns}
\be
-D^2_L=-\sum_\mu\left(U_\mu(x)\delta_{y(x+e_\mu)}+U_\mu^\dagger(x-\mu)\delta_{y(x-e_\mu)}-2\delta_{xy}\right)+m_0^2\delta_{xy}\label{cov},
\ee
\no has been used, where $U_\mu$ are the link variables and $e_\mu$ are lattice unit vectors in the corresponding directions. Since this operator is positive semi-definite, it can be inverted. This has been done using the same method as for the Faddeev-Popov operator in \cite{Cucchieri:2006tf}. It should be noted that even a zero mass is not a problem for this method\footnote{In contrast to the Faddeev-Popov operator, this operator has no trivial zero modes, and thus an inversion even at zero momentum is possible. However, since constant modes affect the result on a finite lattice, this is not done here.}. The final result has been averaged over color. The momenta have been evaluated along the $x$-axis as edge momenta and along the $xy$, $xyz$, and $xyzt$ diagonal directions, when available in a given number of dimensions.

This leaves to fix the bare mass $m_0$ in \pref{cov}. Since the lattice spacings are known in advance, it can be set to the desired tree-level value $m=am_0$ at the ultraviolet cutoff $1/a$. Four different values will be used, zero, 100 MeV, 1 GeV, and 10 GeV. The corresponding bare values $m_0$ for the case of 1 GeV are listed in table \ref{tcgf}. While the values of 100 MeV and 1 GeV are suitable for the wide range of lattice parameters used here to expect no serious lattice artifacts, both zero mass and 10 GeV are naively expected to be strongly affected by lattice artifacts, and therefore are intended to serve as benchmarks for these artifacts. However, it will turn out that these artifacts are often surprisingly small.

\section{Renormalization}\label{s:ren}

\subsection{Definition of the scheme}\label{ss:scheme}

Assuming that the renormalization of the propagator can be performed as in the perturbative case \cite{Bohm:2001yx}, which will be supported by the results, there are two necessary renormalization constants. One is a multiplicative wave-function renormalization $Z_H$, and one an additive mass renormalization $\delta m^2$, leading to the renormalized propagator
\be
D^{ij}(p^2)=\frac{\delta^{ij}}{Z(p^2+m_r^2)+\Pi(p^2)+\delta m^2}\nn,
\ee
\no where $m_r^2$ is the renormalized mass, $p^2$ is the momentum and $\Pi(p^2)$ is the self-energy obtained from the unrenormalized color-averaged propagator $D_u=D_u^{ii}/N_c$,
\be
\Pi(p^2)=\frac{1-p^2D_u(p^2)}{D_u(p^2)}\nn
\ee
\no and therefore encodes the deviation from the tree-level propagator as
\be
D_u=\frac{1}{p^2+\Pi(p^2)}\nn.
\ee
\no The inclusion of the tree-level mass $m^2$ in the self-energy is for technical convenience only, as it avoids to involve explicitly the scale $a$. This only shifts the renormalization constant $\delta m^2$ by the tree-level mass $m^2$, and has no other implications.

The renormalization scheme to fix both renormalization constants is
\bea
D^{ij}(\mu^2)&=&\frac{\delta^{ij}}{\mu^2+m_r^2}\label{prc}\\
\frac{\pd D^{ij}}{\pd p}(\mu^2)&=&-\frac{2\mu\delta^{ij}}{(\mu^2+m_r^2)^2}\label{dprc},
\eea
\no with the renormalization scale $\mu$. This requires the propagator and its derivative to have the tree-level values at $p^2=\mu^2$. In most of the paper the choice $\mu=1.5$ GeV and $m_r=m$ will be made. The effect of different choices will be investigated in sections \ref{ss:ssd} and \ref{ss:sren}.

Solving for the renormalization constants, this yields
\bea
Z&=&\frac{2\mu-\frac{d\Pi(p^2)}{dp}(\mu^2)}{2\mu}\nn\\
\delta m^2&=&\frac{(\mu^2+m_r^2)\frac{d\Pi(p^2)}{dp}(\mu^2)-2\mu\Pi(\mu^2)}{2\mu}\nn.
\eea
\no Numerically, these constants are determined by linear interpolation between the two momenta values along the $x$-axis between which the actual value of $\mu$ is. The derivative of $\Pi$ is obtained by deriving the linear interpolation of $\Pi$ between both points analytically. Errors are then determined by error propagation from the original propagator, whose statistical error in turn has been determined using bootstrap \cite{Cucchieri:2006tf}. Note that throughout only values for $\mu$ have been chosen such that the lower of the two momentum values has been non-zero, and the higher of the two momentum values has not been the maximum momentum along the $x$-axis of $2/a$.

Using the derivative with respect to $p$ rather than $p^2$ for the discretization is numerically convenient, as the lattice momentum in the relevant region are approximately linearly spaced, at least for moderately small discretizations. Of course, this is completely equivalent to the more conventional derivative with respect to $p^2$.

\subsection{Numerical results and discretization dependence}\label{ss:numr}

\begin{figure}[!htbp]
\includegraphics[width=0.95\linewidth]{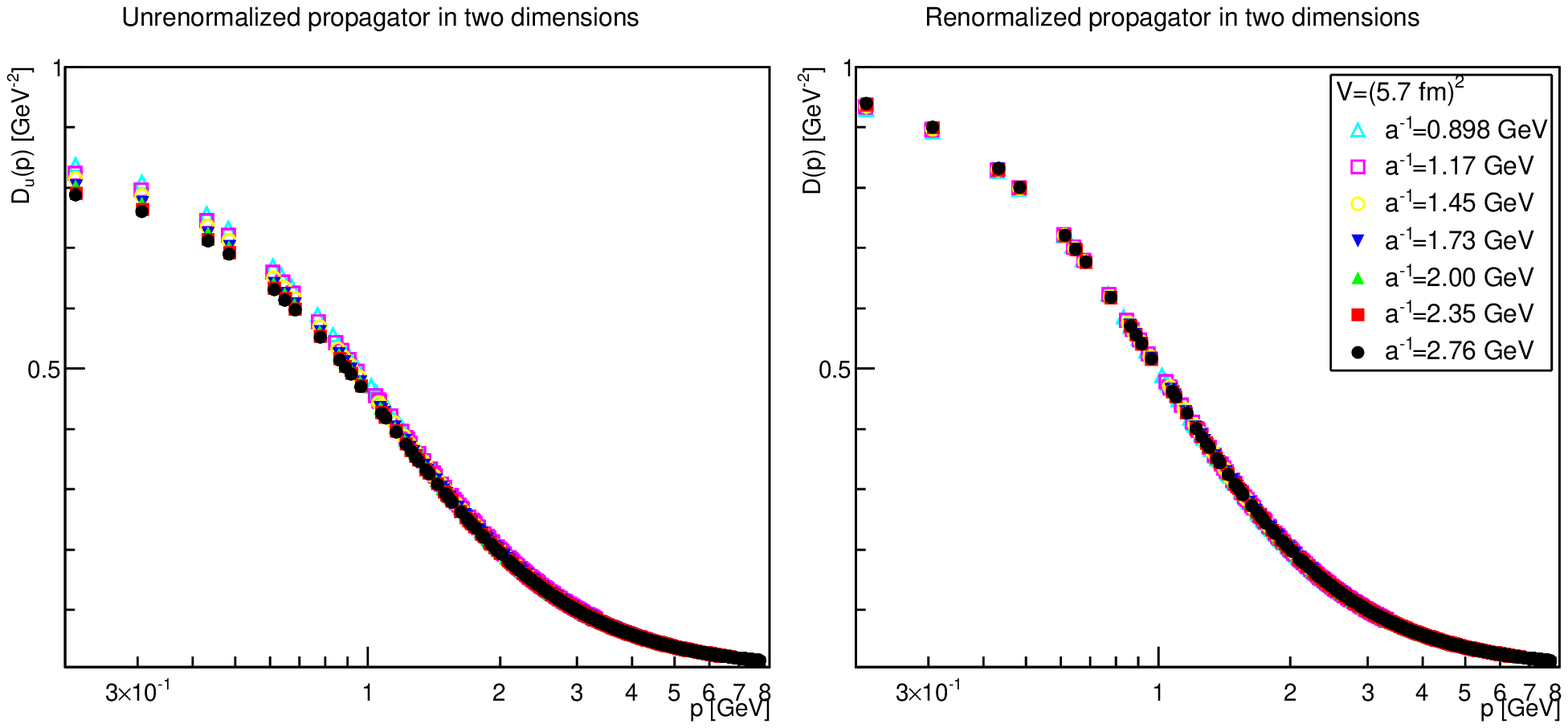}\\
\includegraphics[width=0.95\linewidth]{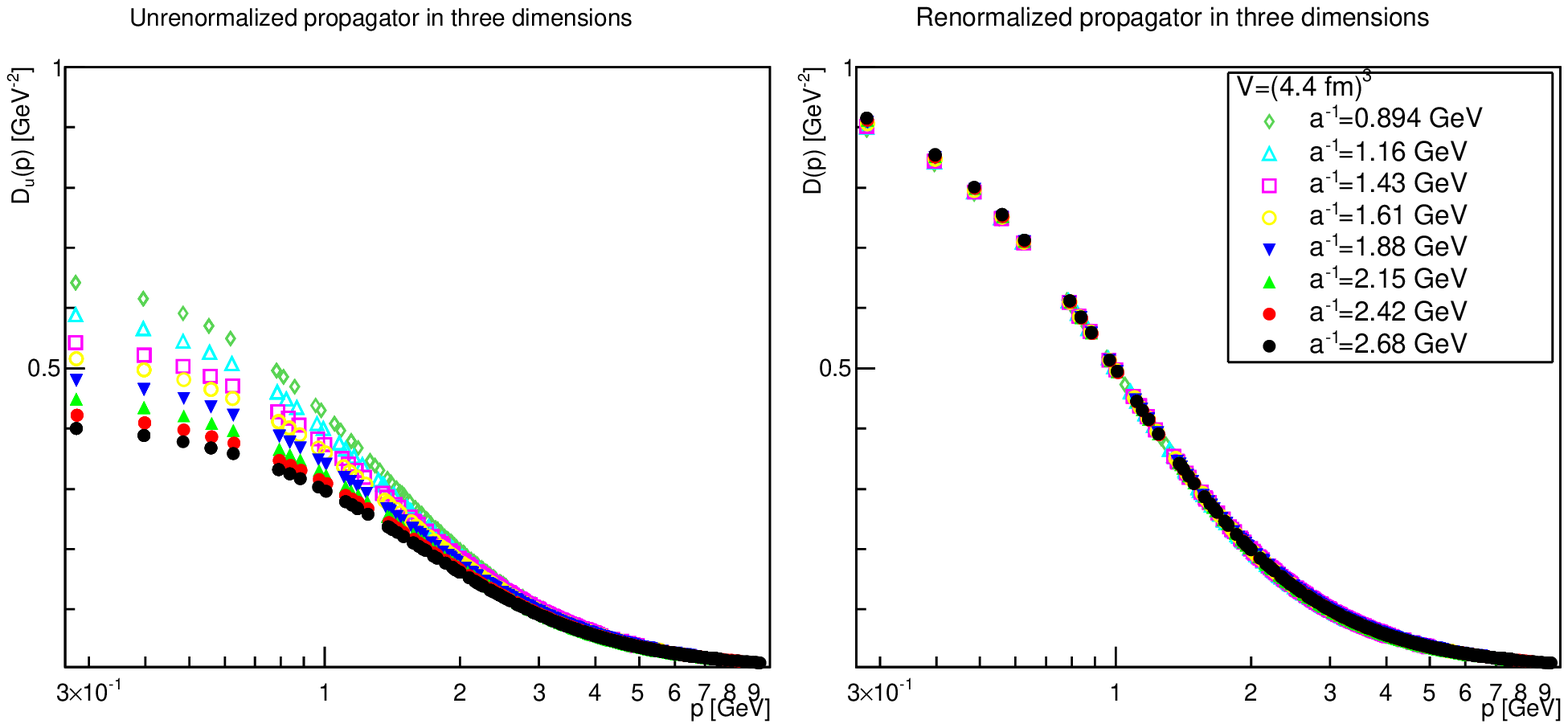}\\
\includegraphics[width=0.95\linewidth]{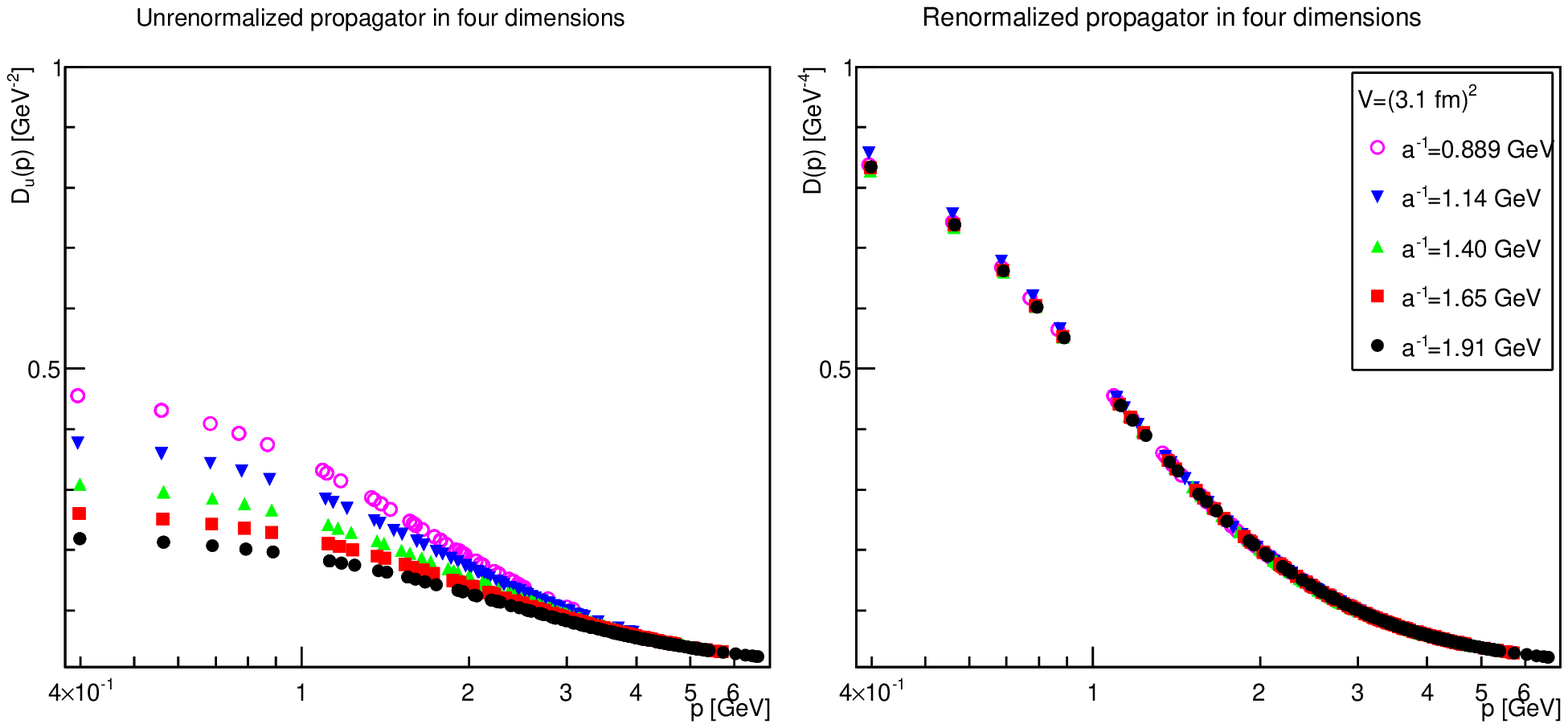}
\caption{\label{fig:ur}Unrenormalized propagator (left panels) and renormalized propagator (right panels). The top panels are for two dimensions, the middle panels for three dimensions, and the bottom panels for four dimensions. The values are $m_r=m$ and $\mu=1.5$ GeV. If the (statistical 1$\sigma$) error bars here and hereafter are not visible then they are smaller than the symbol size.}
\end{figure}

The effect of renormalization is shown in figure \ref{fig:ur}. It is visible that for all dimensions, though only very slightly in two dimensions, the propagators deviate from each other for different values of $a$. This deviation is not only a multiplicative factor, as they still coincide in the ultraviolet, but also a change of mass as the difference in the infrared shows. In fact, the results already suggest, and this will be confirmed below, that the wave-function renormalization is close to one, while the mass renormalization is sizable. After renormalization, the propagators show essentially no difference anymore, and thus only a very weak dependence on the lattice spacing. The same pattern is also observed for the other masses in principle, but altered by two effects. For the smaller masses, as will be seen, finite-volume effects play a role in the infrared. For the larger mass, the propagator is very close to tree-level for almost all momenta accessible and the effects of renormalization are therefore substantially suppressed. Nonetheless, in all cases the renormalized propagators essentially coincide.

This also shows that there are only rather small discretization artifacts present after renormalization. This will be discussed more in detail in section \ref{ss:renvc}, where the volume and cutoff-dependence of the renormalization constants will be analyzed.

However, there is some systematic uncertainties due to the linear interpolation, as is, e.\ g., visible for $a^{-1}=1.14$ GeV in four dimensions in figure \ref{fig:ur}. It has been attempted to improve the situation using a four-point spline interpolation. However, the statistical error then accumulated, while the systematic error was not strongly improved. Thus, this did not lead to a substantial reduction of the error. However, the problem is reduced more and more with finer and finer lattices at the relevant momentum regime of $\mu$, and this type of systematic error remains hence as a discretization artifact. However, it does not influence the qualitative conclusions, nor, as visible in figure \ref{fig:ur}, is it a substantial quantitative error.

\subsection{Scale and scheme dependence}\label{ss:ssd}

In section \ref{ss:scheme} the scheme was defined by setting the mass on the right-hand side of \pref{prc} and \pref{dprc} equal to the tree-level mass. In section \ref{ss:numr}, the renormalization scale was furthermore fixed to $\mu=1.5$ GeV. Both conditions are, of course, not necessary. Here, the dependence on both is studied for the case of $m=1$ GeV. Of course, strong deviations are expected if either $\mu$ or the masses $m$ and $m_r$ are close to the lattice scales $1/a$ and $N/a$. This is the reason why only the case $m=1$ GeV will be considered in more detail, and only the largest volume for which a lattice spacing $a\gtrsim(2$ GeV)$^{-1}$ is available.

\begin{figure}[htbp]
\includegraphics[width=0.475\linewidth]{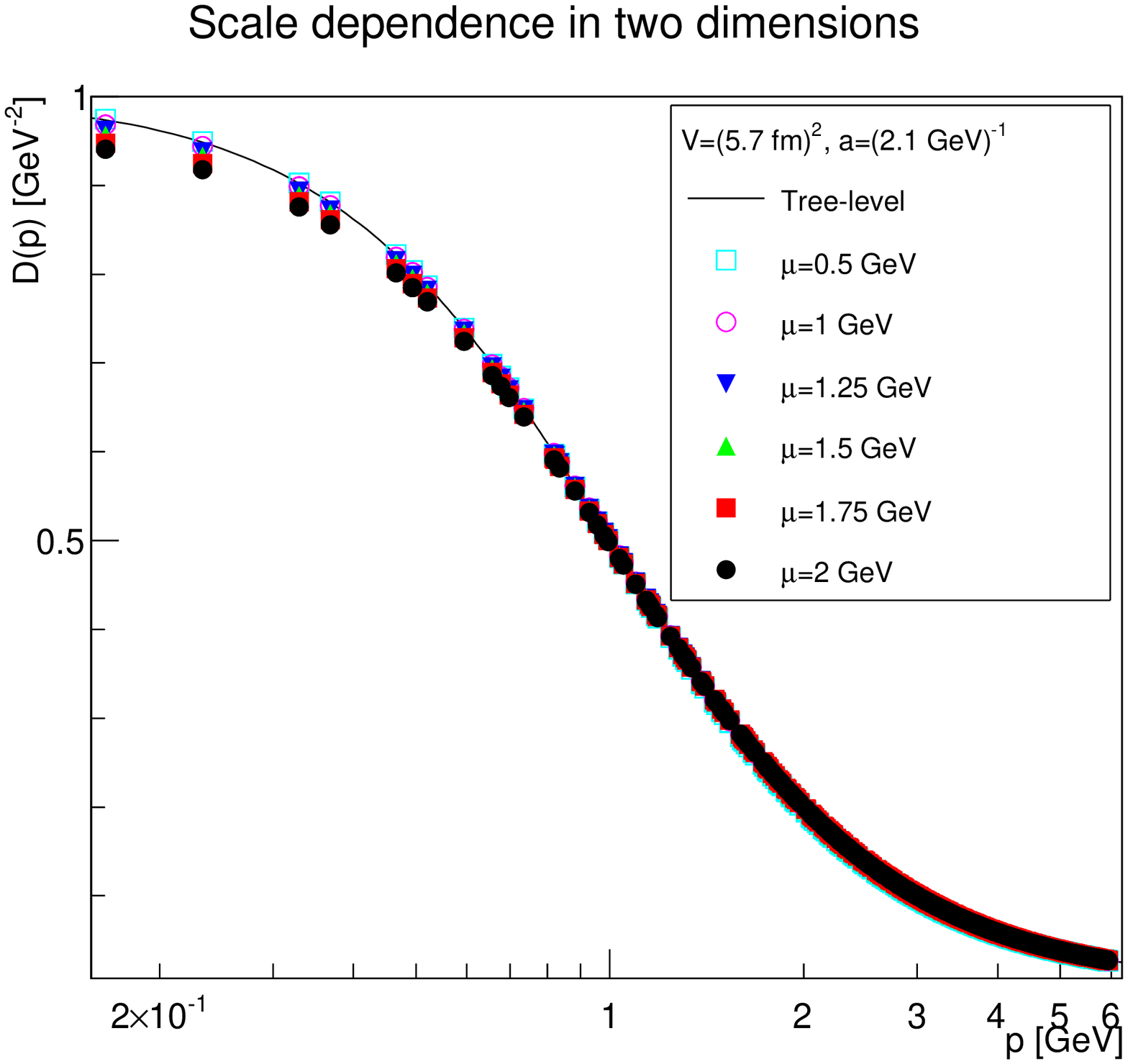}\includegraphics[width=0.475\linewidth]{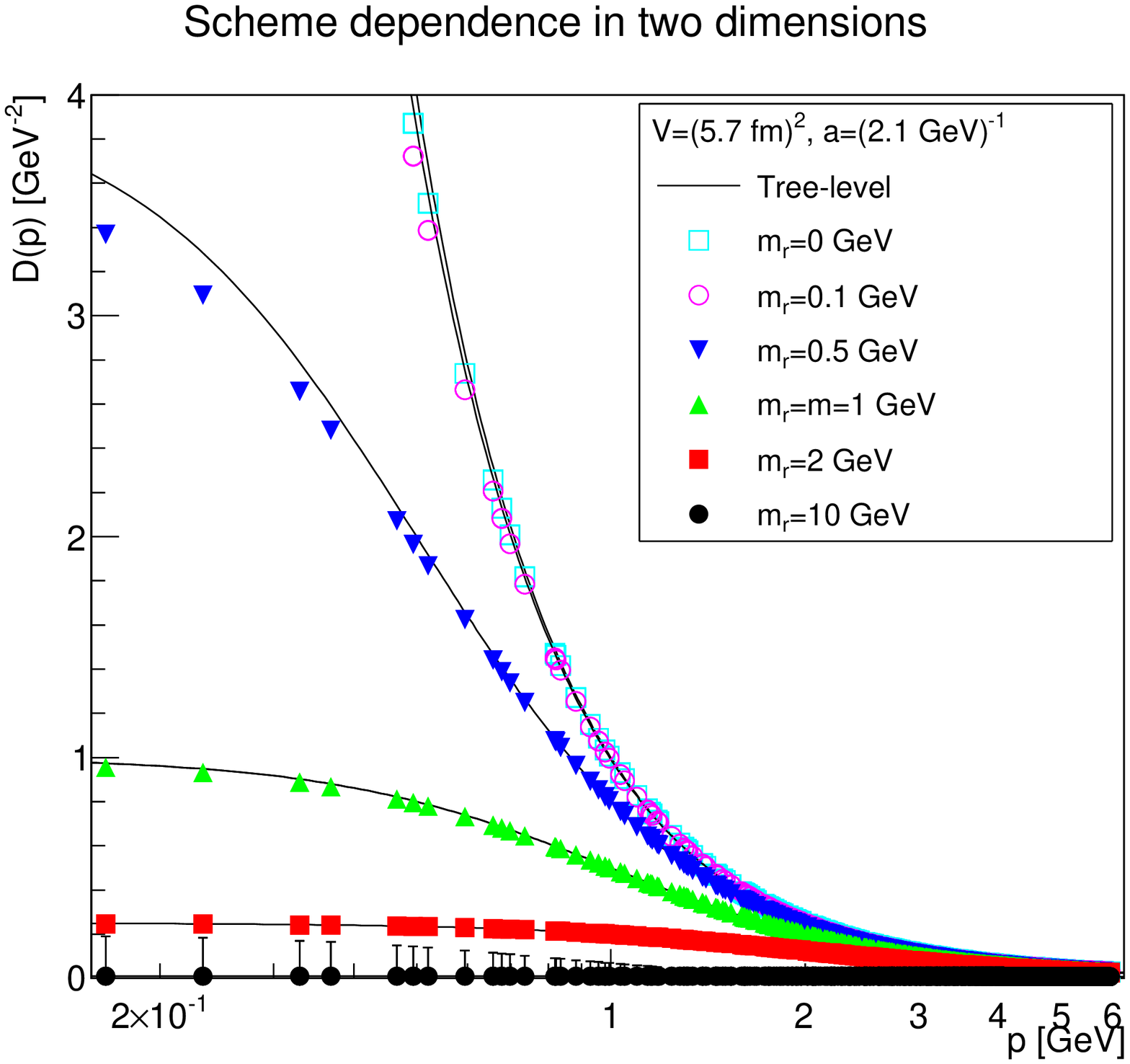}\\
\includegraphics[width=0.475\linewidth]{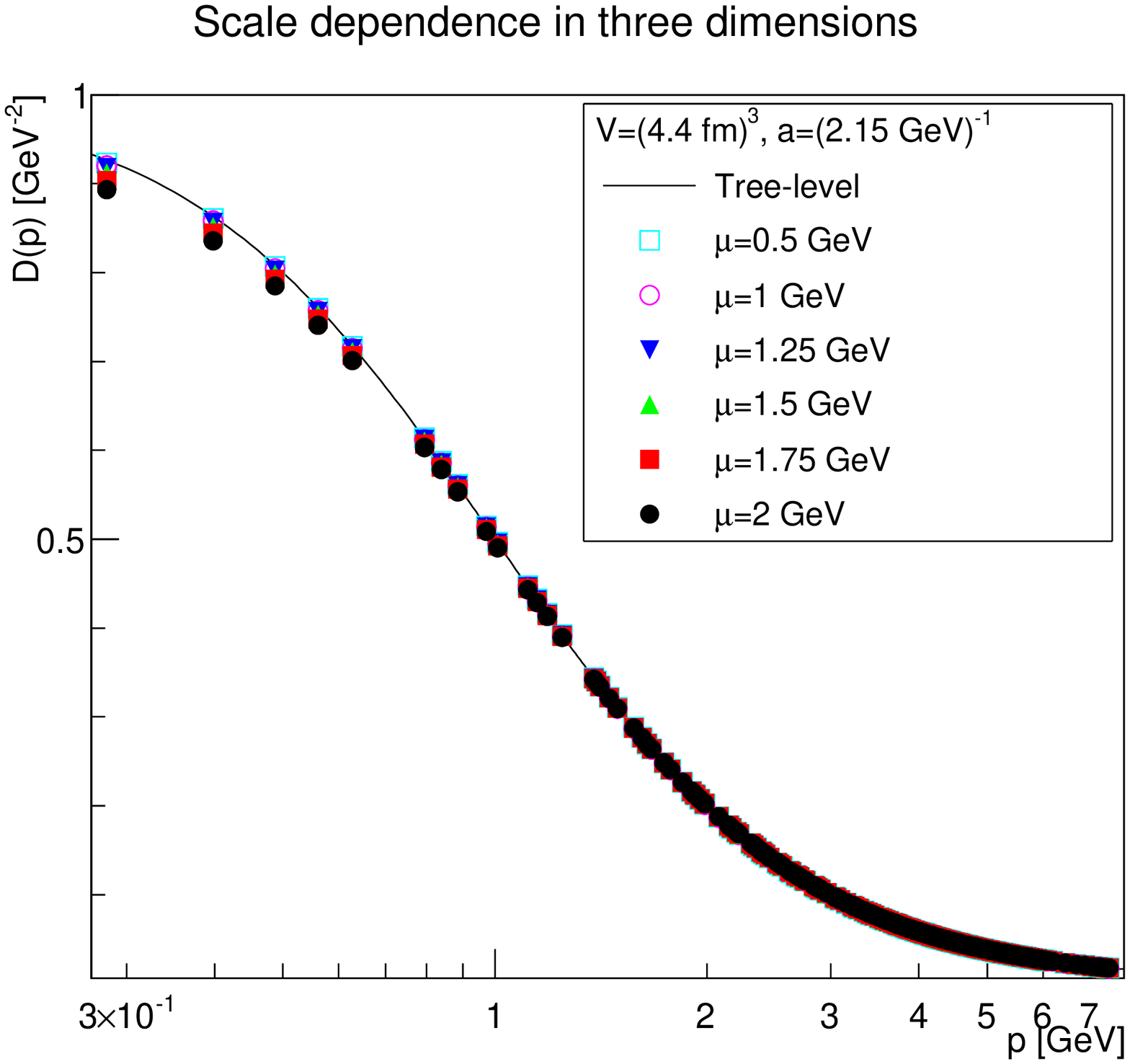}\includegraphics[width=0.475\linewidth]{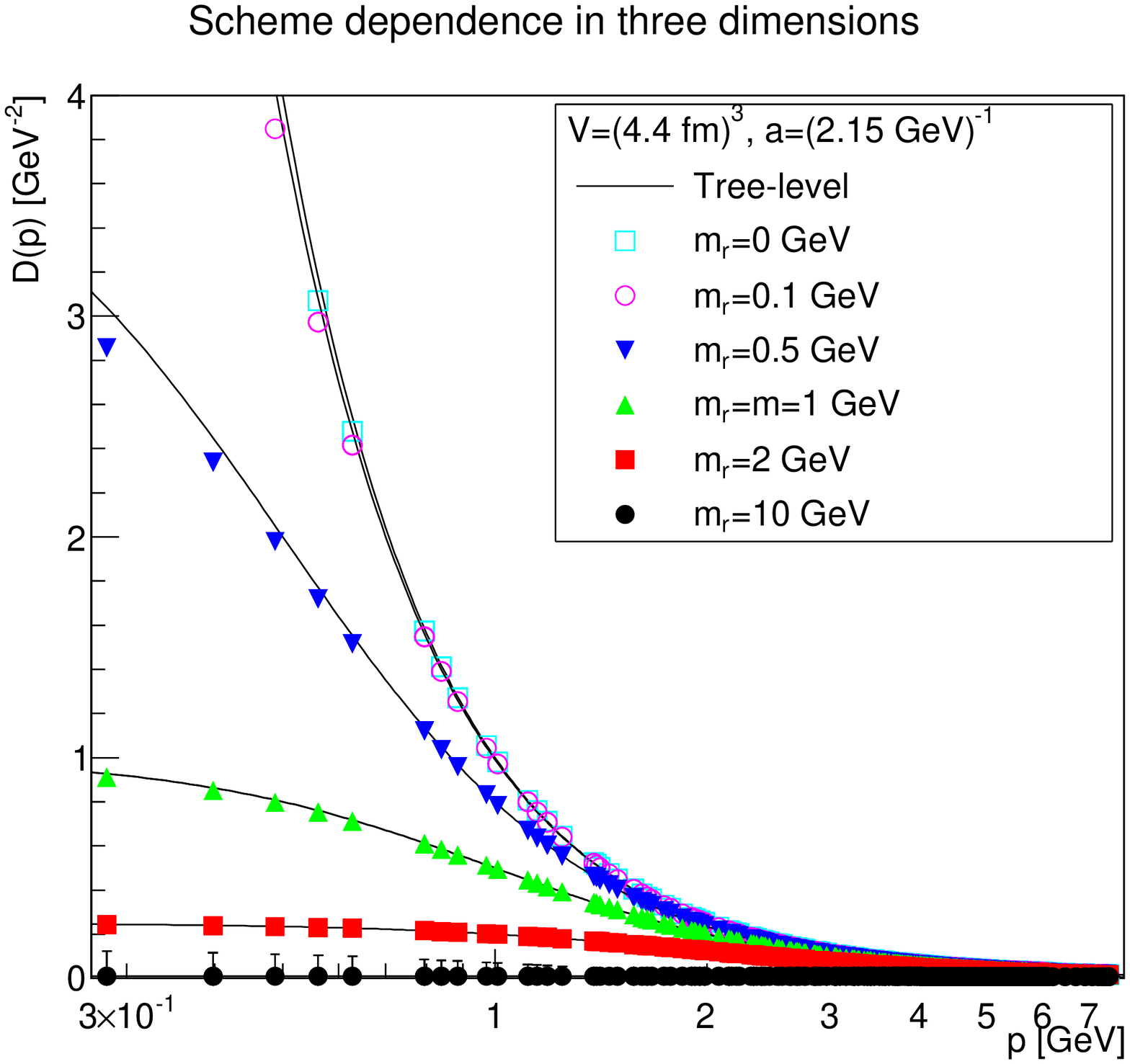}\\
\includegraphics[width=0.475\linewidth]{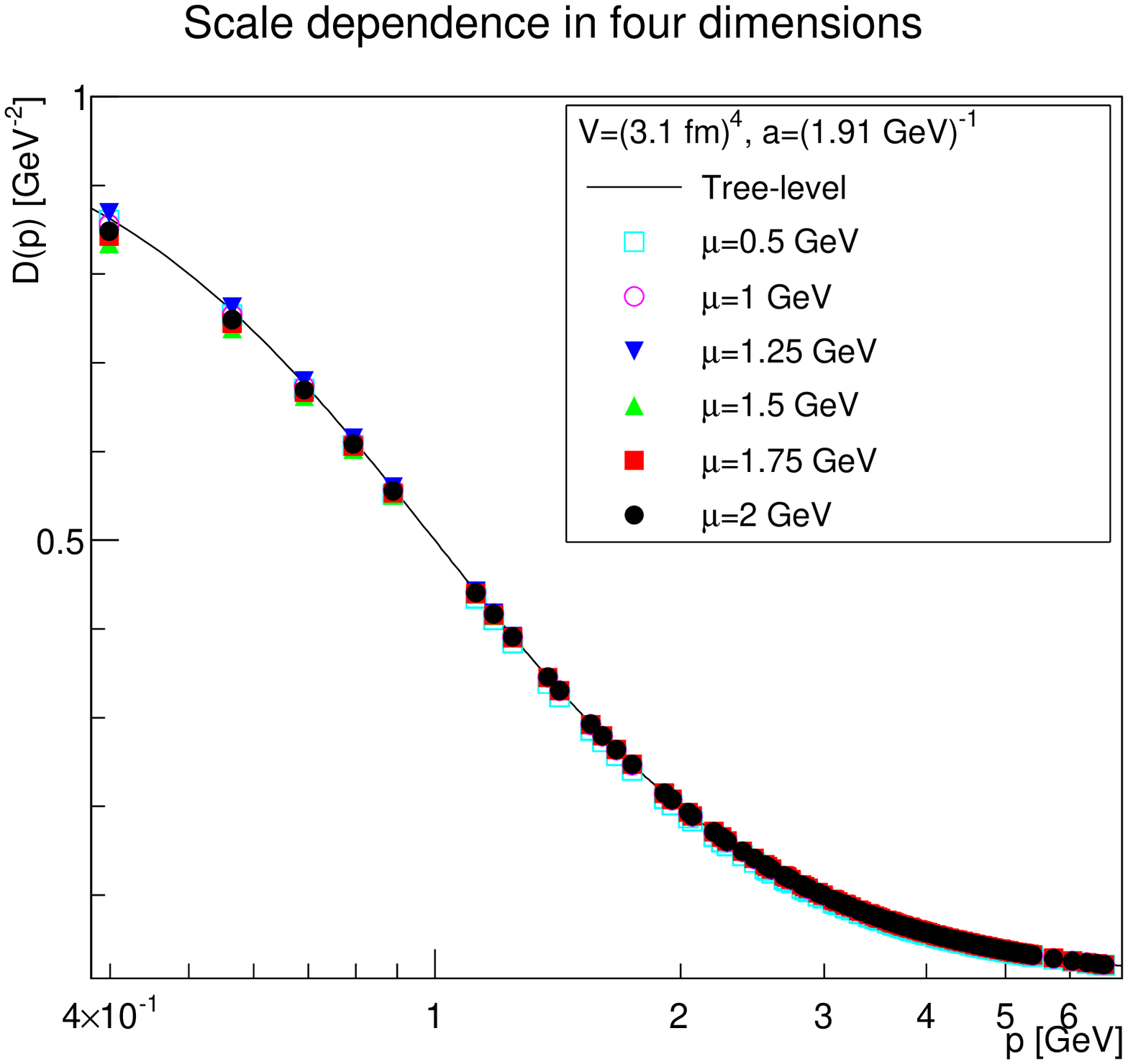}\includegraphics[width=0.475\linewidth]{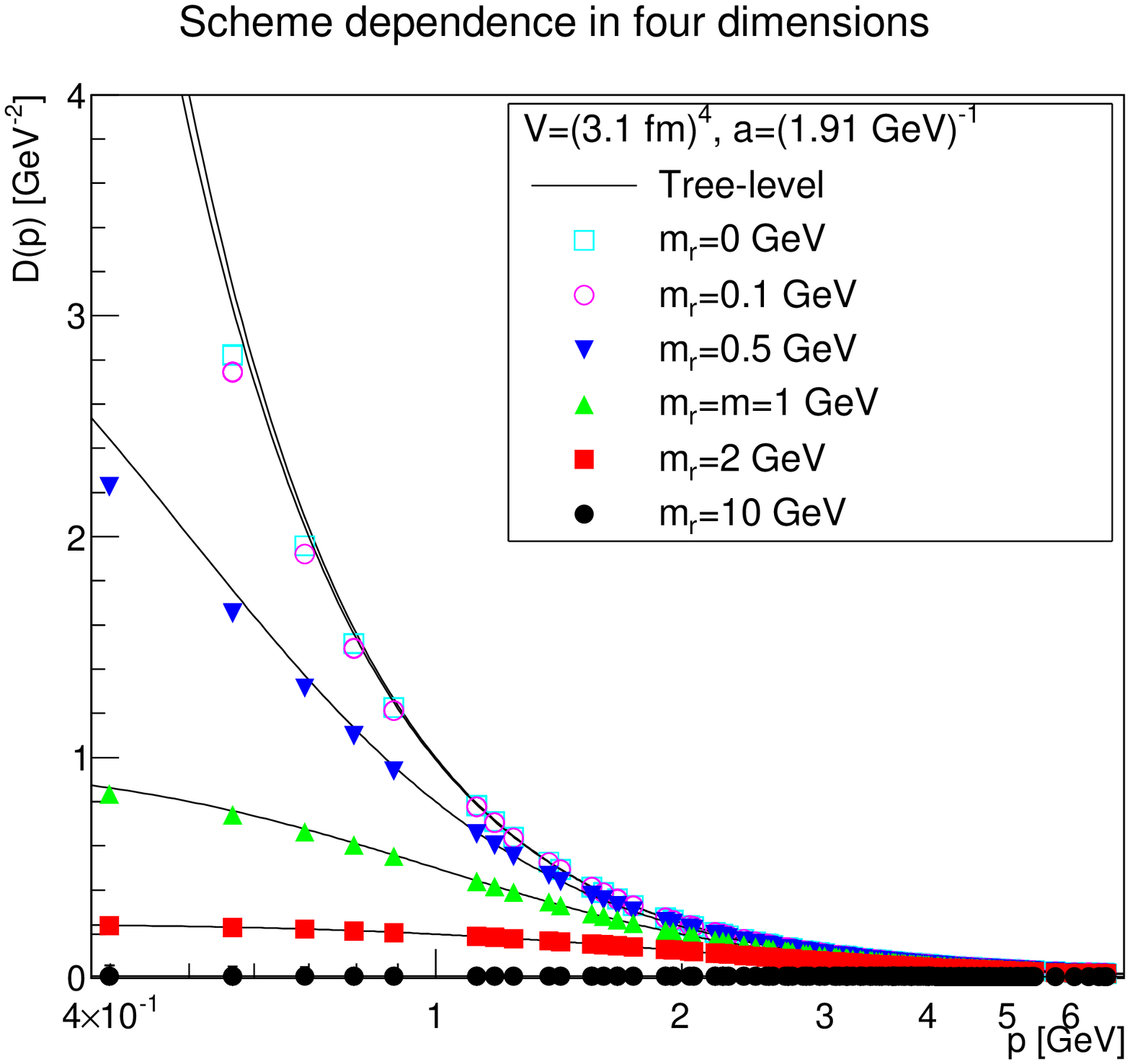}
\caption{\label{fig:ren}Scale-dependence at $m_r=m=1$ GeV (left panels) and scheme dependence at $\mu=1.5$ GeV (right panels) of the renormalized propagator. The top panels are two dimensions, the middle panels three dimensions, and the bottom panels four dimensions. The tree-level propagator is shown for comparison as a full line.}
\end{figure}

The results are shown for both scale and scheme dependence in figure \ref{fig:ren}. The dependence of the results on the renormalization scale $\mu$ is extremely minor, and not larger than any systematic error. The situation is distinctively different for the scheme dependence. With decreasing renormalized mass $m_r$ the propagator starts to deviate increasingly from the tree-level propagator, and the stronger the higher the dimension. Interestingly, the renormalized propagator is below the tree-level propagator, a behavior indicative of an effective mass larger than the renormalized mass. This will be confirmed in section \ref{s:ana}.

\subsection{Dependence of the renormalization constants on the volume and the cutoff}\label{ss:renvc}

One relevant question is the dependence of the renormalization constants on the lattice parameters. The dependence on volume is quite interesting, since if it is weak, it would allow to determine the renormalization constants on small lattices with large statistics, and thus very precisely. This would reduce the effect of error propagation considerably. The dependence on the cutoff is especially interesting not only as a systematic error source, but also to see whether the naive perturbative expectation \cite{Bohm:2001yx} coincides with the actual behavior. In the following the standard scheme $m_r=m$ with a renormalization scale $\mu=1.5$ GeV will be used.

\begin{figure}[!htb]
\includegraphics[width=0.475\linewidth]{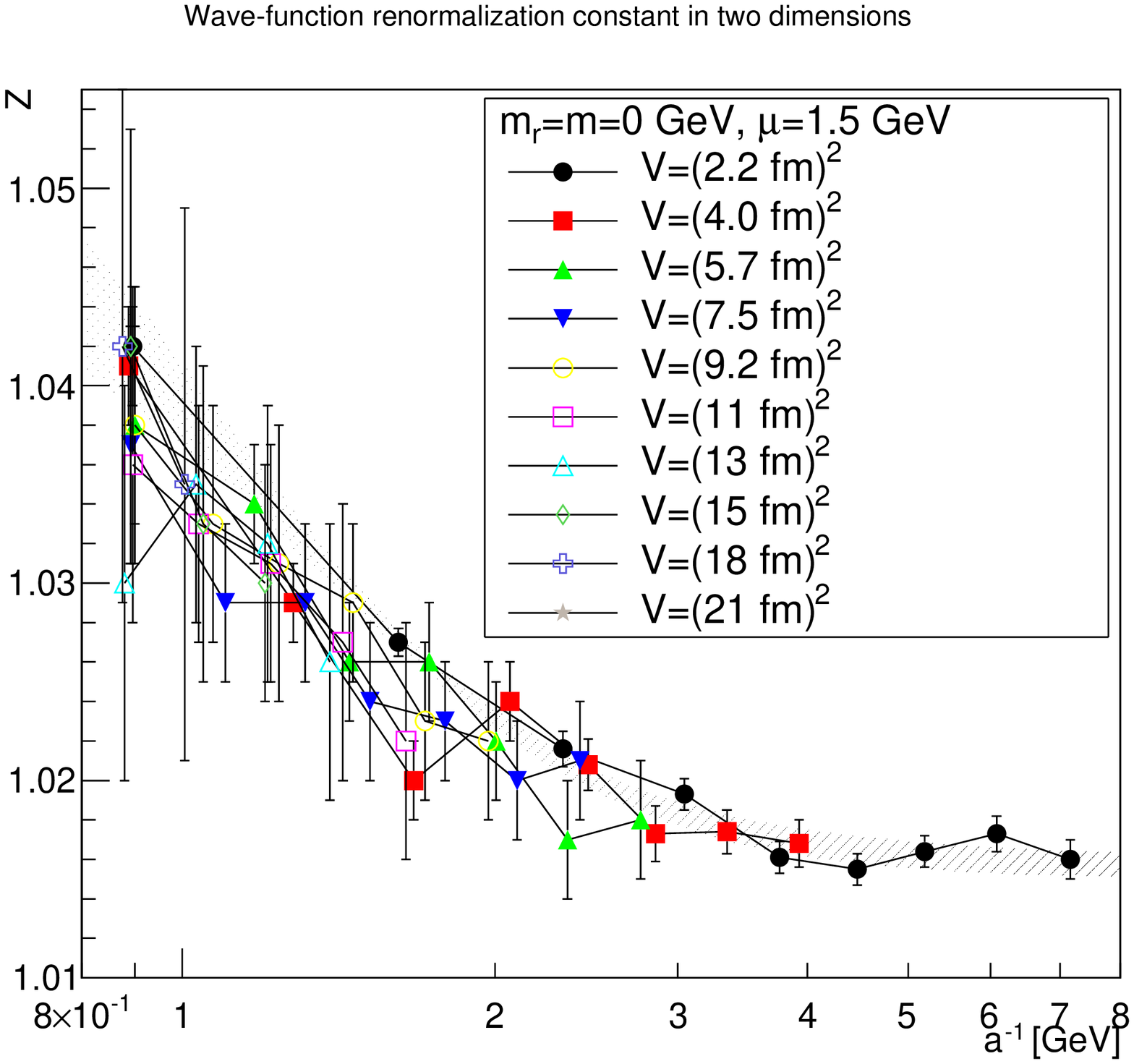}\includegraphics[width=0.475\linewidth]{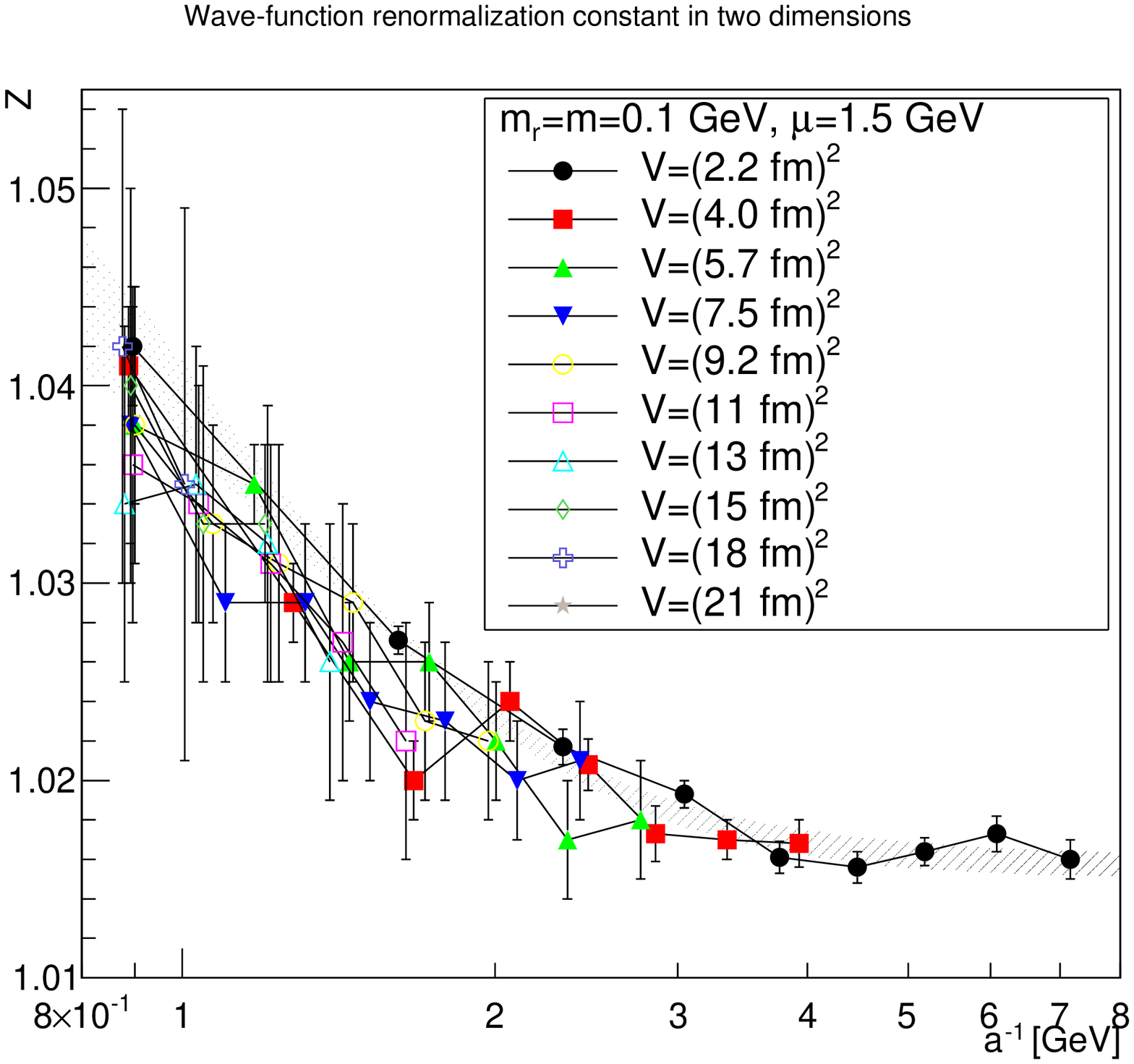}\\
\includegraphics[width=0.475\linewidth]{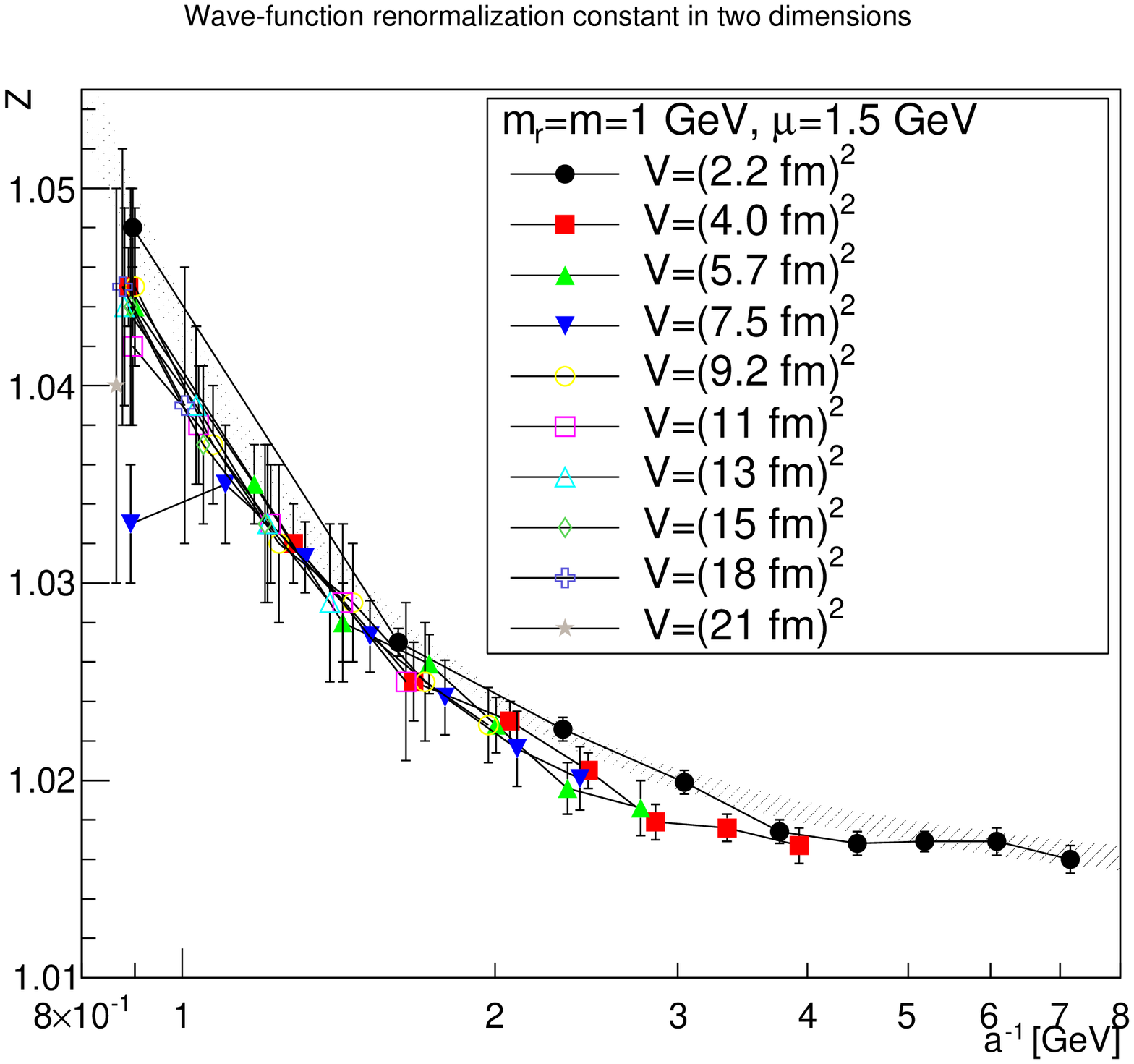}\includegraphics[width=0.475\linewidth]{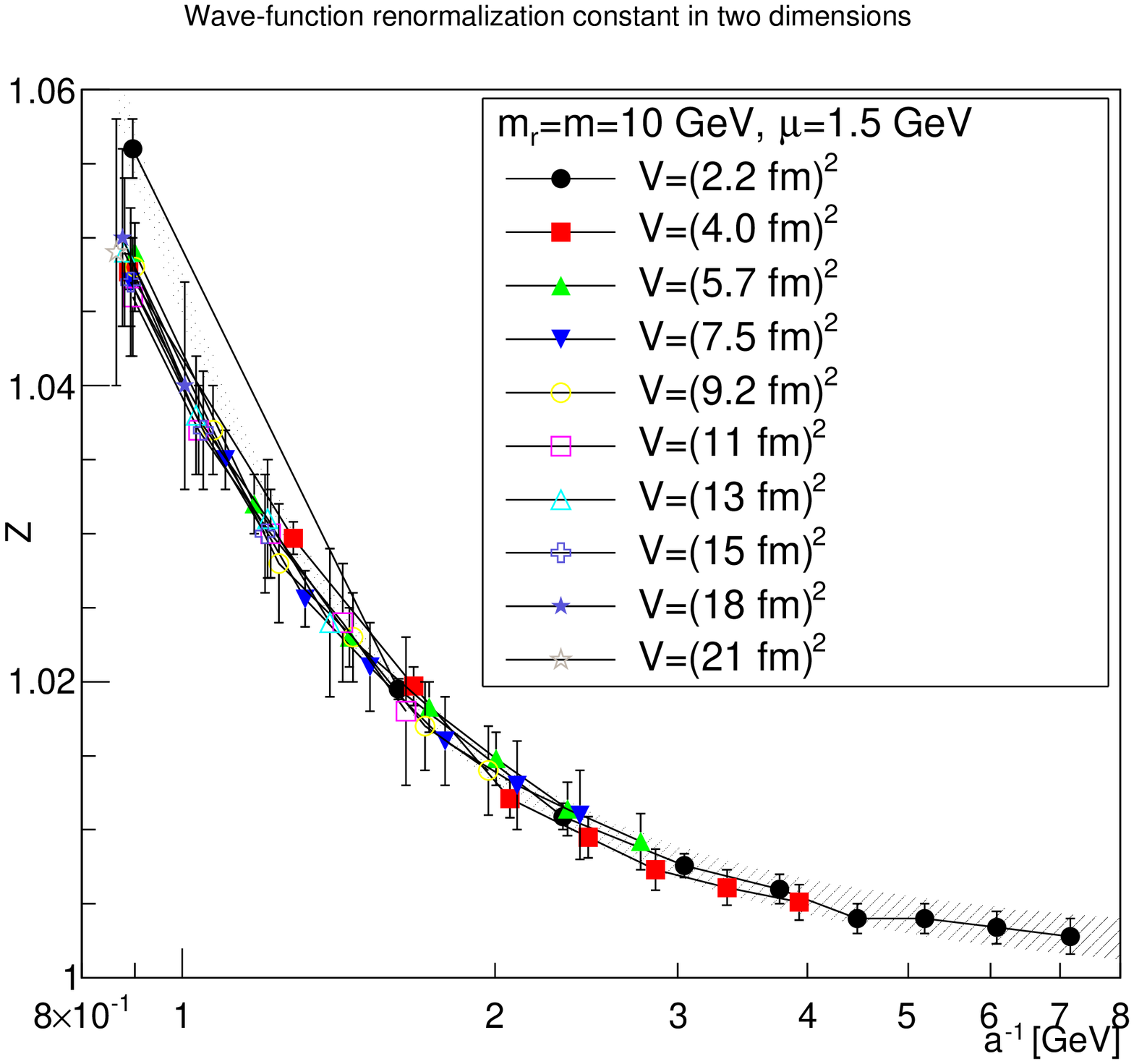}
\caption{\label{fig:z2}The wave-function renormalization constant as a function of the lattice cutoff and the lattice volume in two dimensions for $\mu=1.5$ GeV. The top-left panel shows the case of $m=m_r=0$ GeV, the top-right panel of $m=m_r=0.1$ GeV, the bottom-left panel of $m=m_r=1$ GeV, and the bottom-right panel of $m=m_r=10$ GeV. The hatched band is the fit \pref{zfit} with the parameters given in table \ref{fitsz}.}
\end{figure}

\begin{figure}[!htb]
\includegraphics[width=0.475\linewidth]{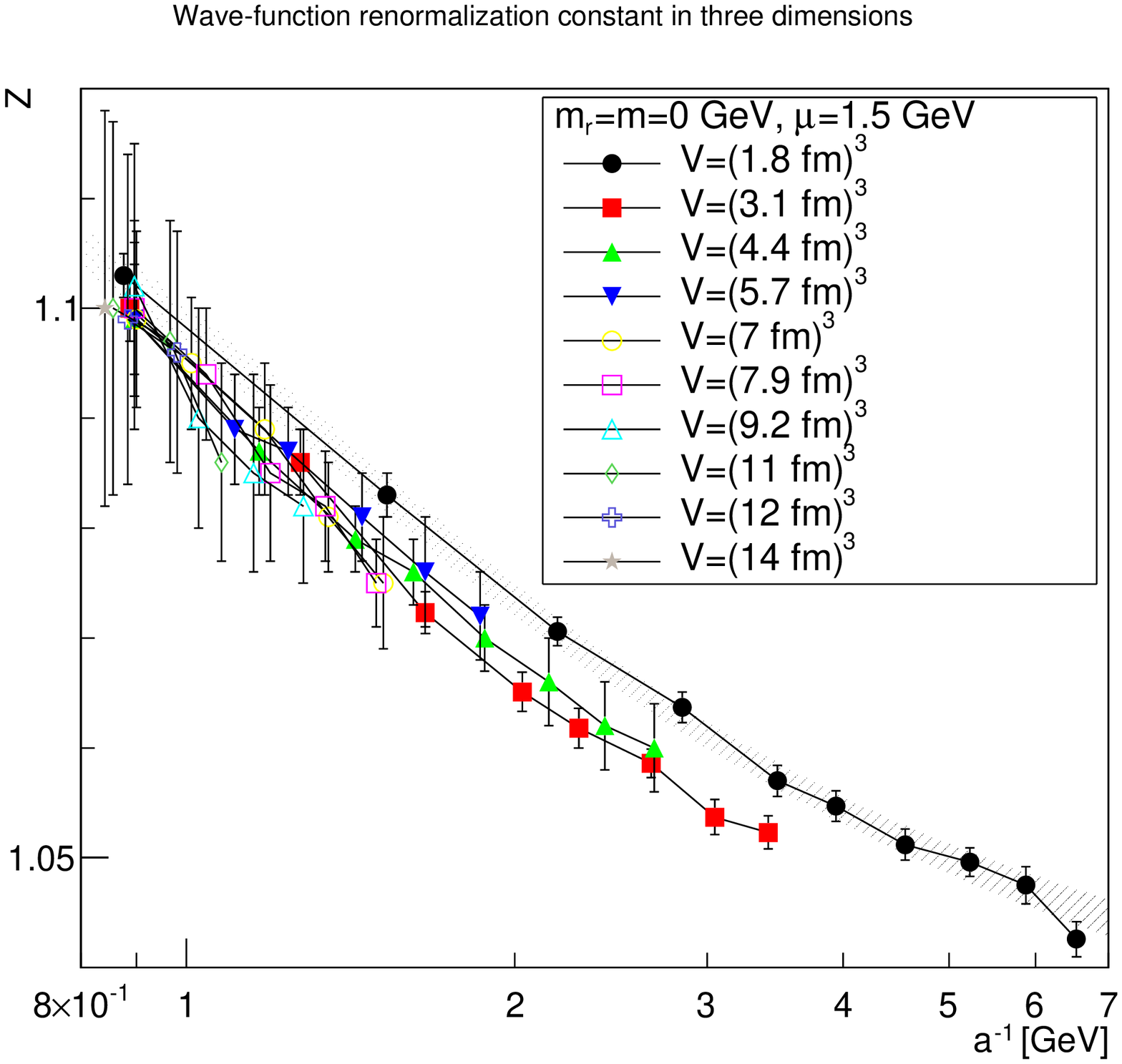}\includegraphics[width=0.475\linewidth]{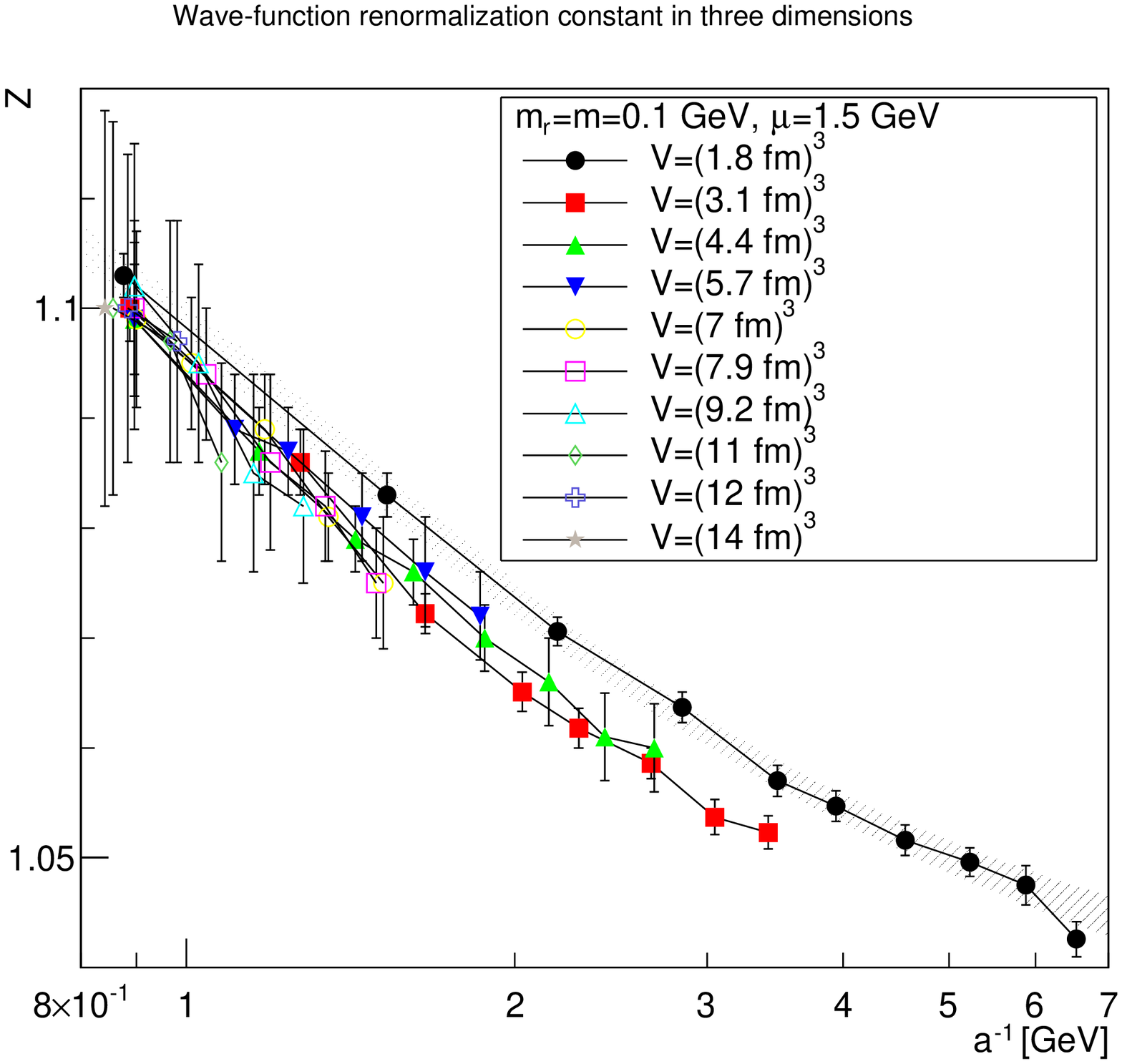}\\
\includegraphics[width=0.475\linewidth]{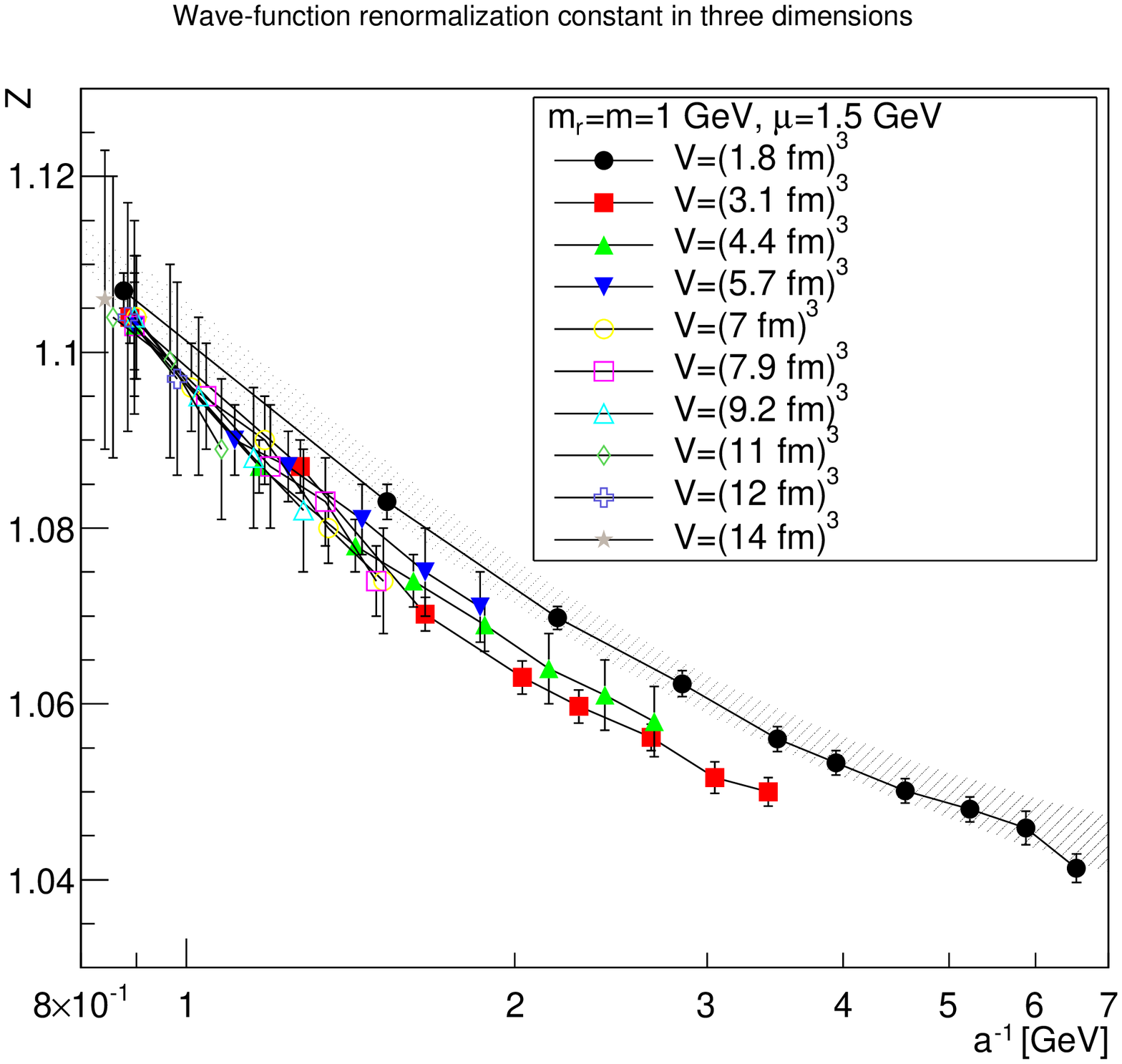}\includegraphics[width=0.475\linewidth]{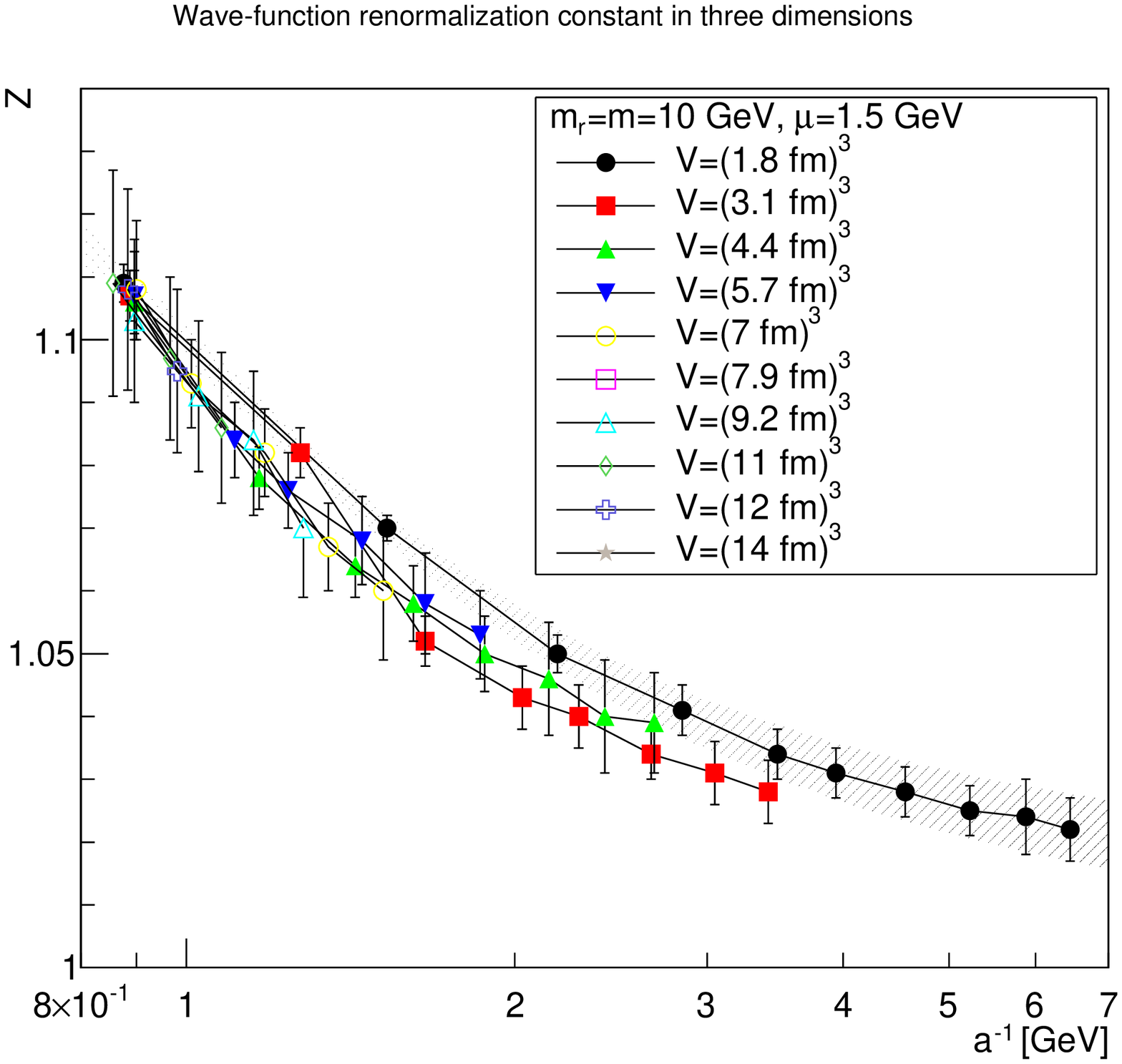}
\caption{\label{fig:z3}The wave-function renormalization constant as a function of the lattice cutoff and the lattice volume in three dimensions for $\mu=1.5$ GeV. The top-left panel shows the case of $m=m_r=0$ GeV, the top-right panel of $m=m_r=0.1$ GeV, the bottom-left panel of $m=m_r=1$ GeV, and the bottom-right panel of $m=m_r=10$ GeV. The hatched band is the fit \pref{zfit} with the parameters given in table \ref{fitsz}.}
\end{figure}

\begin{figure}[!htb]
\includegraphics[width=0.475\linewidth]{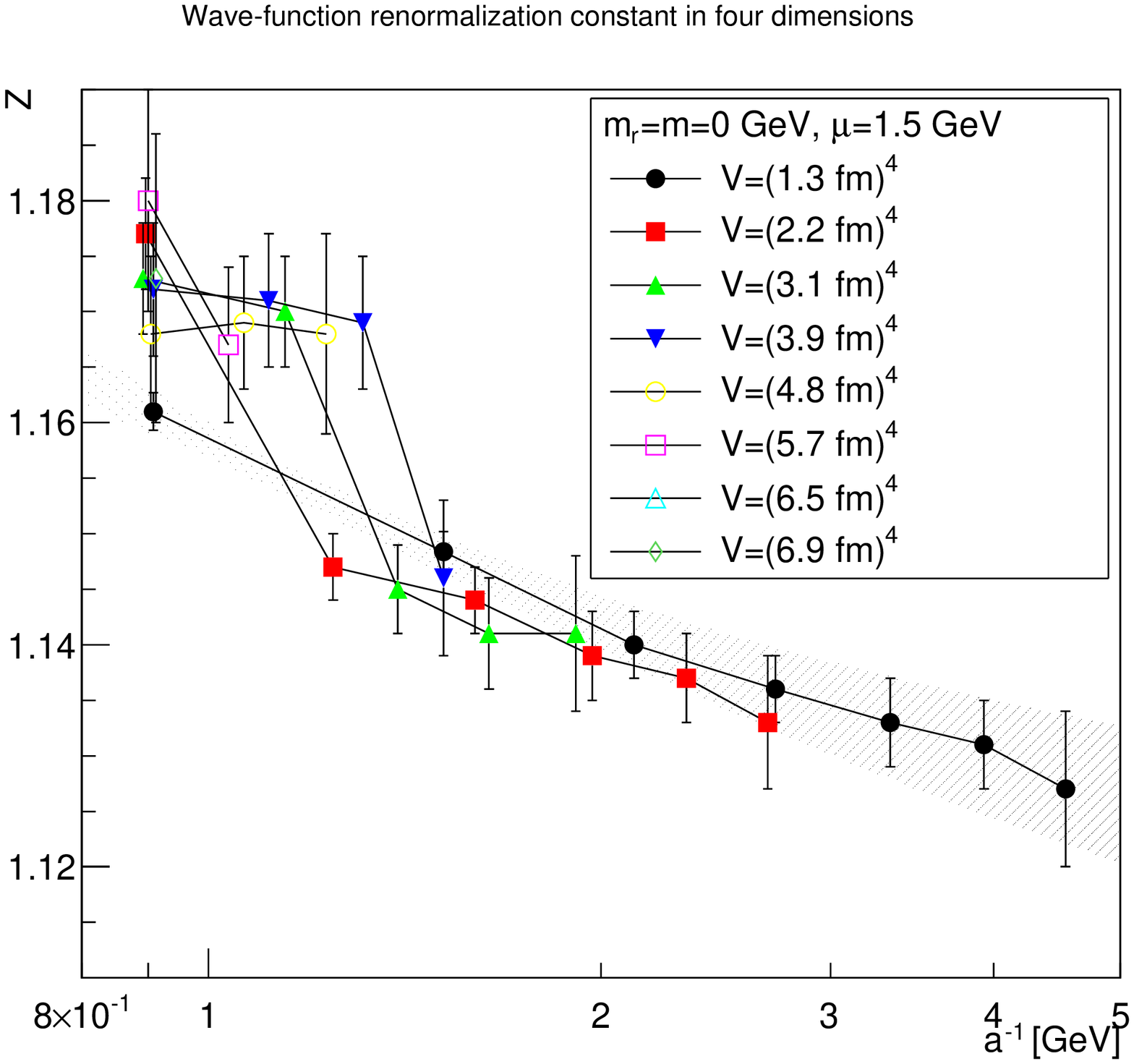}\includegraphics[width=0.475\linewidth]{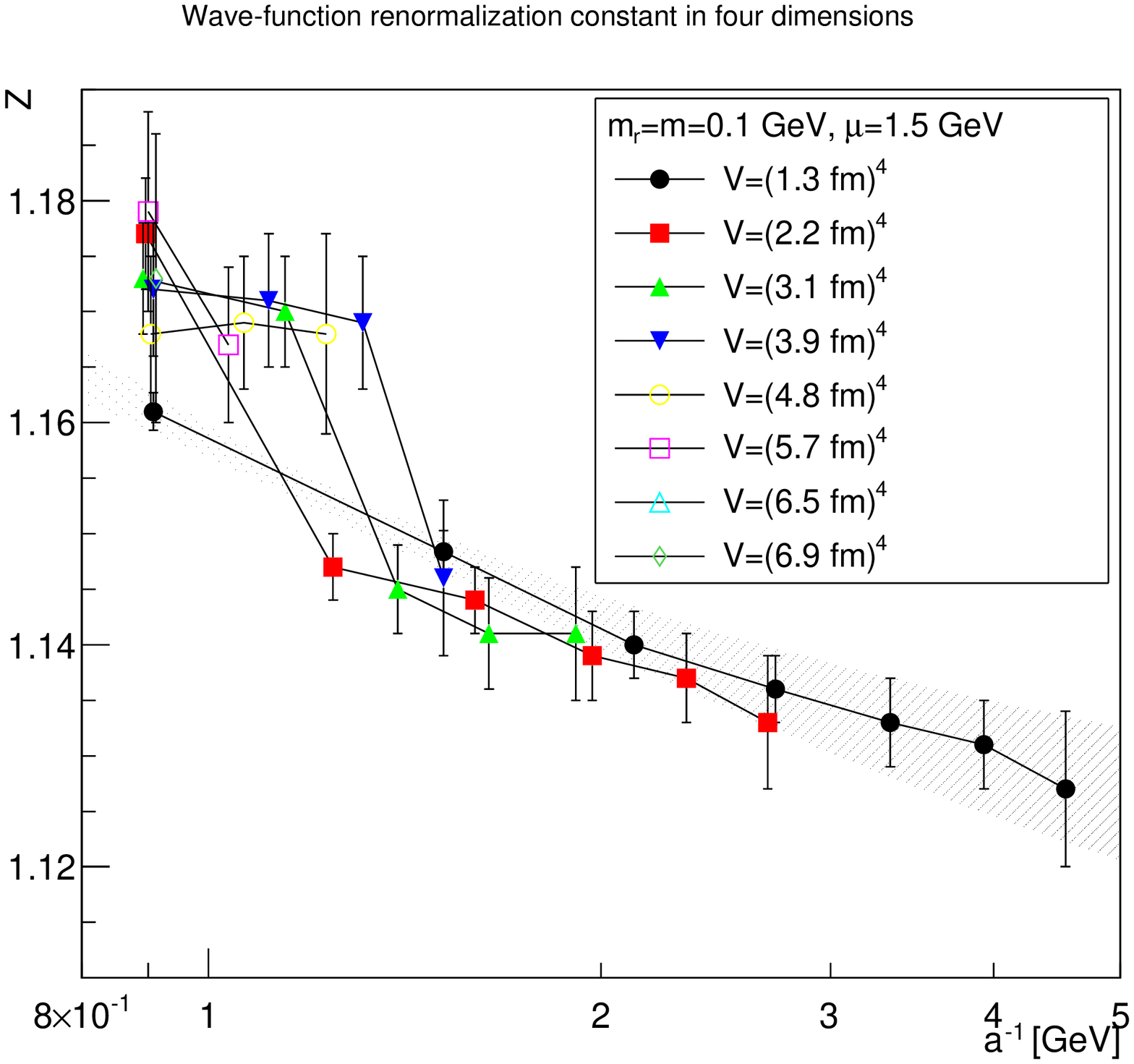}\\
\includegraphics[width=0.475\linewidth]{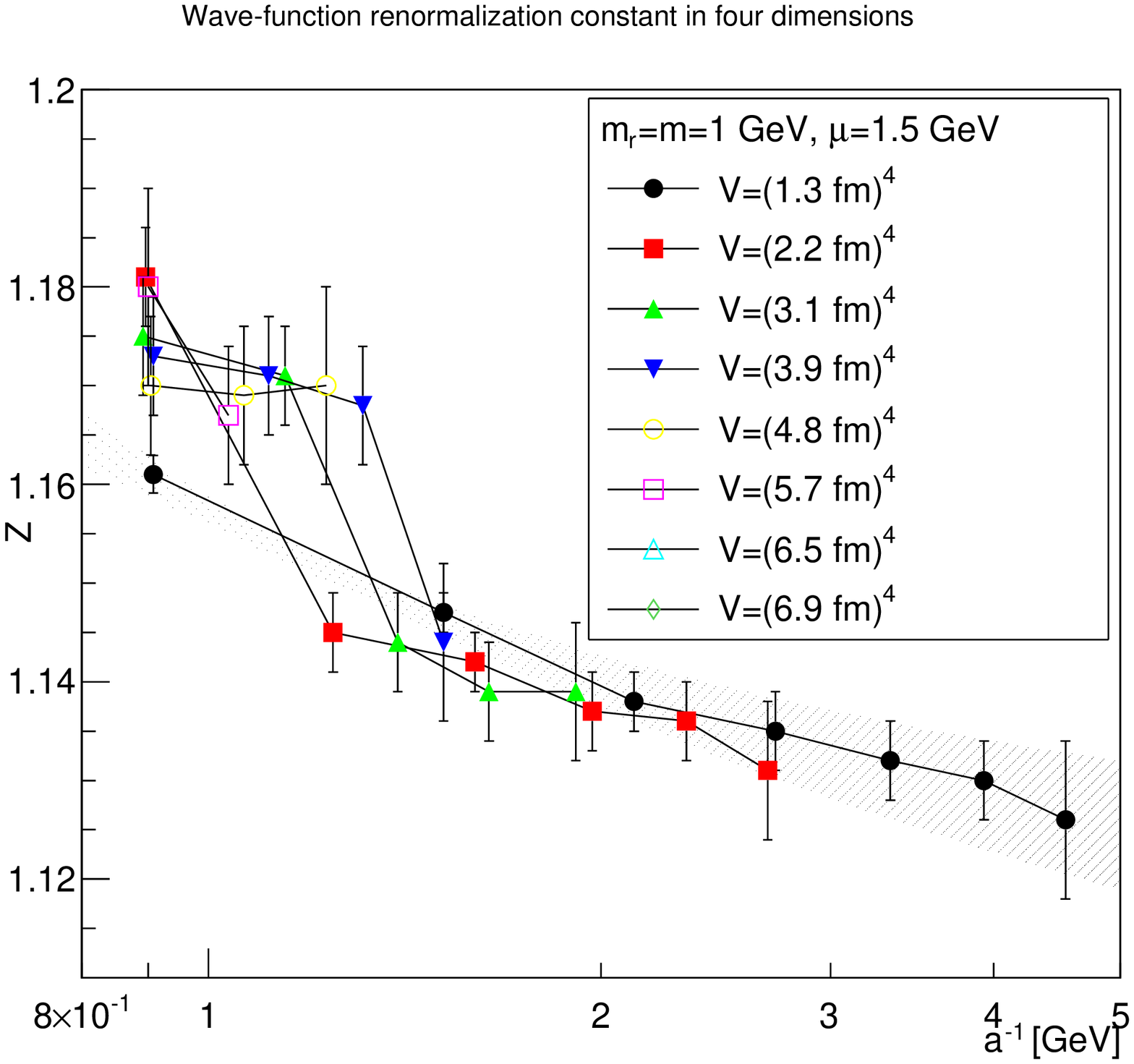}\includegraphics[width=0.475\linewidth]{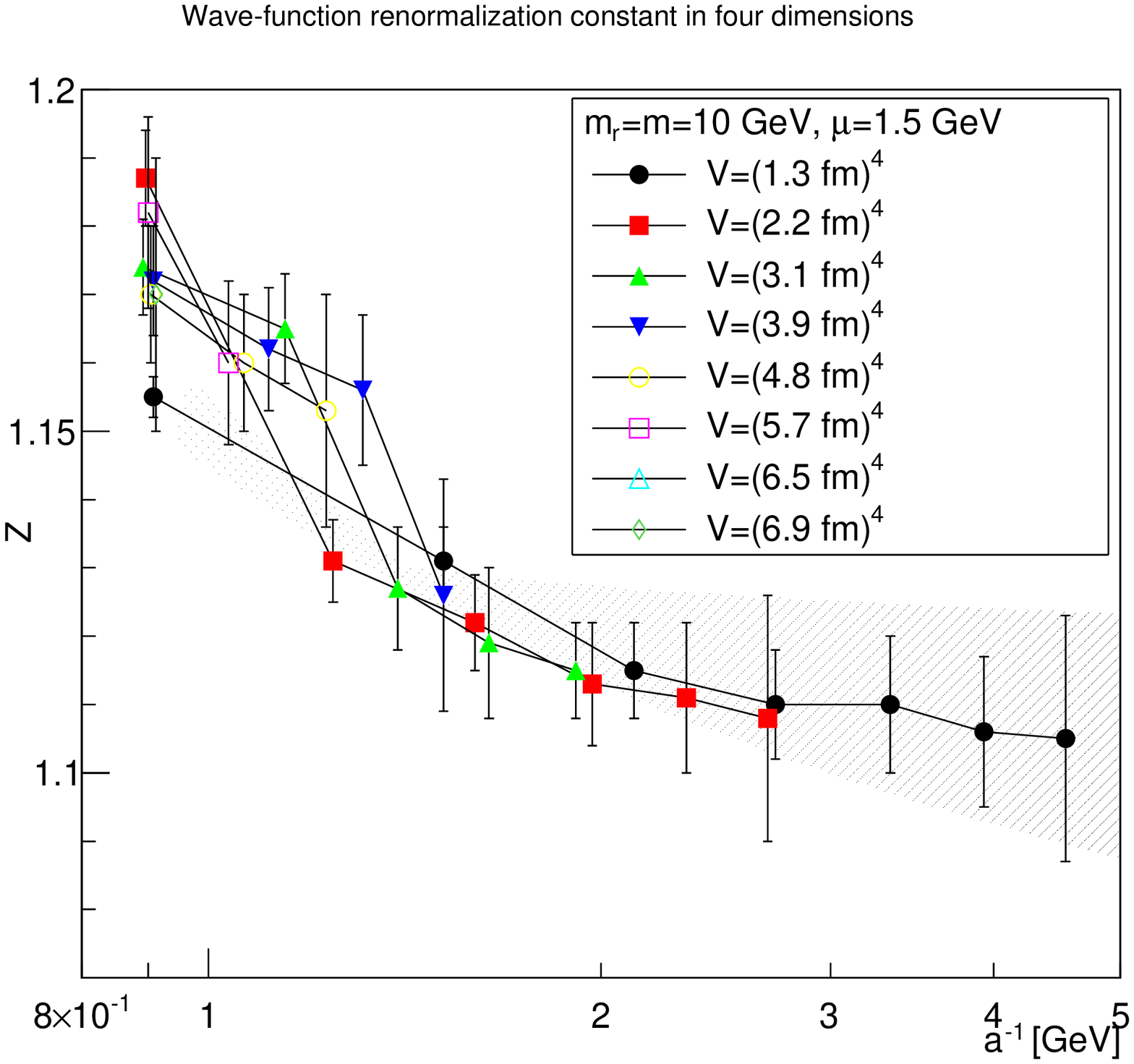}
\caption{\label{fig:z4}The wave-function renormalization constant as a function of the lattice cutoff and the lattice volume in four dimensions for $\mu=1.5$ GeV. The top-left panel shows the case of $m=m_r=0$ GeV, the top-right panel of $m=m_r=0.1$ GeV, the bottom-left panel of $m=m_r=1$ GeV, and the bottom-right panel of $m=m_r=10$ GeV. The hatched band is the fit \pref{zfit} with the parameters given in table \ref{fitsz}.}
\end{figure}

The results for the wave-function renormalization is shown in figures \ref{fig:z2}-\ref{fig:z4} for two, three, and four dimensions, respectively. All cases show a wave-function renormalization $Z(1/a)$ which decays with decreasing lattice spacing $a$. Volume effects are small, and especially do not affect the qualitative behavior, though there is some drift for volumes smaller than (2-3 fm)$^d$. The wave-function renormalization is essentially a continuous function of $a$ for two and three dimensions, while there is a pronounced jump in four dimensions around $a^{-1}=1-1.5$ GeV, which occurs later for larger volumes. However, this effect is still rather small, and only slightly affects the quantitative behavior.

\begin{longtable}[!H]{|c|c|c|c|c|c|}
\caption{\label{fitsz}Fit parameters of \pref{zfit} for the wave-function renormalization constants at $\mu=1.5$ GeV. A value of 0 for $Z_\infty$ indicates that no stable fit with a non-zero value for $Z_\infty$ could be found.}\cr
\hline
$d$	& m [GeV]	& $Z_\infty$	& $\Lambda$ [GeV]	& $\epsilon$ 	& $c$\cr
\hline	\endfirsthead
\hline
\multicolumn{6}{|l|}{Table \ref{fitsz} continued}\cr
\hline
$d$	& m [GeV]	& $Z_\infty$	& $\Lambda$ [GeV]	& $\epsilon$ 	& $c$\cr
\hline\endhead
\hline
\multicolumn{6}{|r|}{Continued on next page}\cr
\hline\endfoot
\endlastfoot
\hline
2	& 10		& 0.9991(16)	& 1.103(5)		& 1.58(7)	& 0.0325(9) \cr
\hline
2	& 1		& 1.0116(11)	& 1.040(4)		& 1.16(7)	& 0.0214(2) \cr
\hline
2	& 0.1		& 1.0157(8)	& 2.2(4)		& 5.2$\pm$1.5	& 0.4$_{-0.3}^{+1.9}$ \cr
\hline
2	& 0		& 1.0157(8)	& 2.1(5)		& 5.1$\pm$1.7	& 0.4$_{-0.3}^{+2.1}$ \cr
\hline
\hline
3	& 10		& 0.999(8)	& 1.39(5)		& 1.16(3)	& 0.110(11) \cr
\hline
3	& 1		& 1.00(19)	& 1.4(3)		& 0.7(6)	& 0.105$_{-0.004}^{+0.164}$ \cr
\hline
3	& 0.1		& 1.00(6)	& 1.48(15)		& 0.6(4)	& 0.11(4) \cr
\hline
3	& 0		& 1.00(4)	& 1.49(13)		& 0.7(3)	& 0.11(3) \cr
\hline
\hline
4	& 10		& 0		& 0.8(5)		& 1.129(7)	& 0.021(17) \cr
\hline
4	& 1		& 0		& 0.99(19)		& 0.018(7)	& 1.150(3) \cr
\hline
4	& 0.1		& 0		& 1.10(18)		& 0.020(7)	& 1.153(3) \cr
\hline
4	& 0		& 0		& 1.10(19)		& 0.020(7)	& 1.153(3) \cr
\hline
\end{longtable}

The slow evolution already suggests a logarithmic behavior. Indeed, the data can be fitted rather well using the fit form
\be
Z(a)=Z_\infty+c\left(\ln\left(\frac{\Lambda^2+\frac{1}{a^2}}{(1\text{ GeV})^2}\right)\right)^\epsilon\label{zfit}.
\ee
\no The resulting parameters are shown for the smallest volume in table \ref{fitsz}. They are also indicated as the hatched band in figures \ref{fig:z2}-\ref{fig:z4}.

Although the smallest volume shows some quantitative deviation from the trend of larger volumes, it allows to reach larger cutoffs $1/a$, and therefore more stable fits of the logarithmic tail. However, even then the values for the anomalous dimensions have large uncertainties. The remaining fit parameters show the expected behavior. The scale $\Lambda$ is of the typical scale of 1 GeV. In two and three dimensions, the renormalization constants tend to a constant (in fact essentially to 1) at infinite cutoff, showing that no wave function renormalization is required. This is the expected behavior from perturbation theory. In four dimensions no stable fit with a finite value of $Z_\infty$ was possible, which is also expected from perturbation theory \cite{Bohm:2001yx}.

Incidentally, this is already a first hint that the fundamental scalar, at least in four dimensions, is not a physical particle, due to the \"Ohme-Zimmermann superconvergence relation \cite{Oehme:1979ai}.

\begin{figure}[htb]
\includegraphics[width=0.475\linewidth]{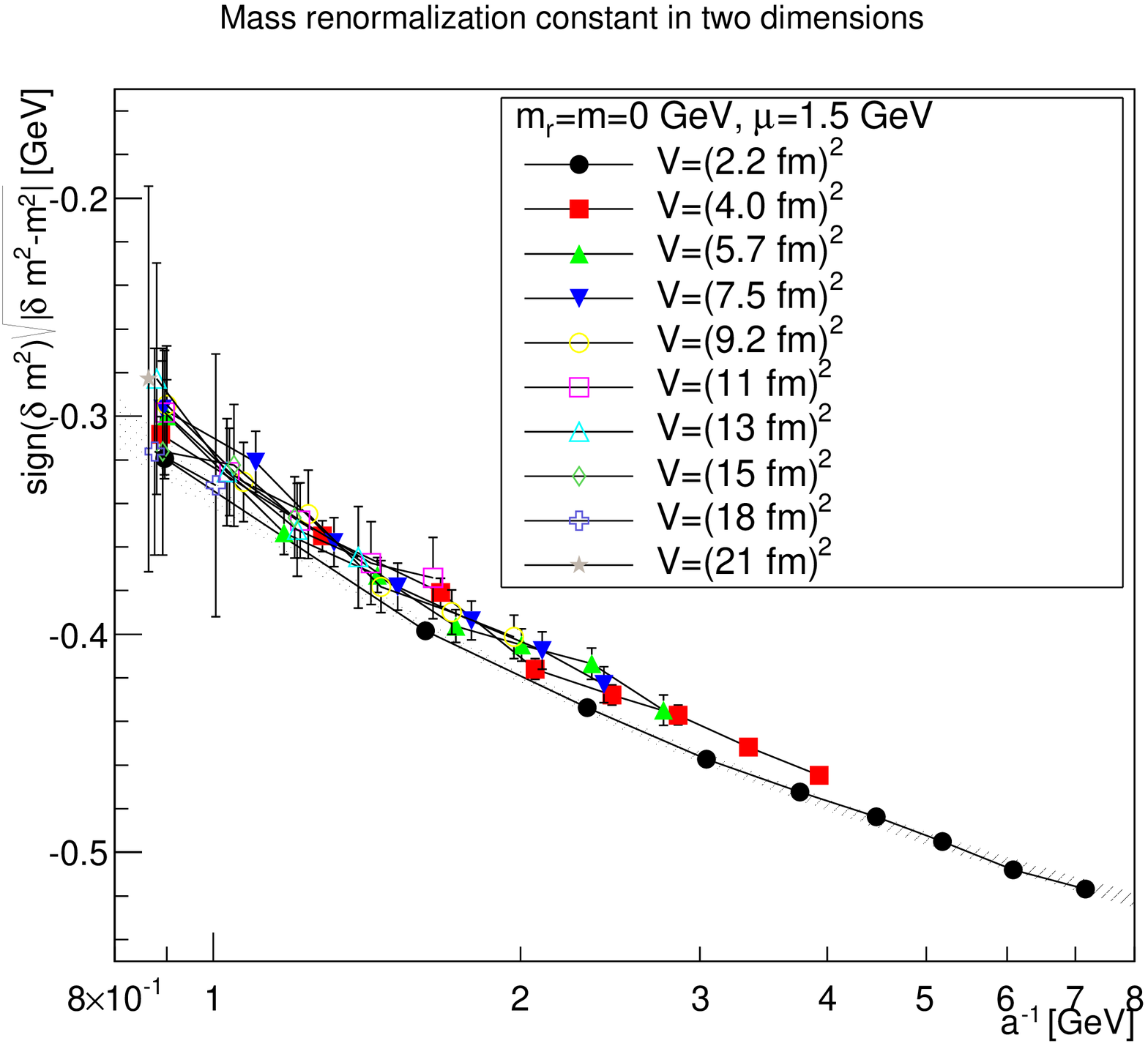}\includegraphics[width=0.475\linewidth]{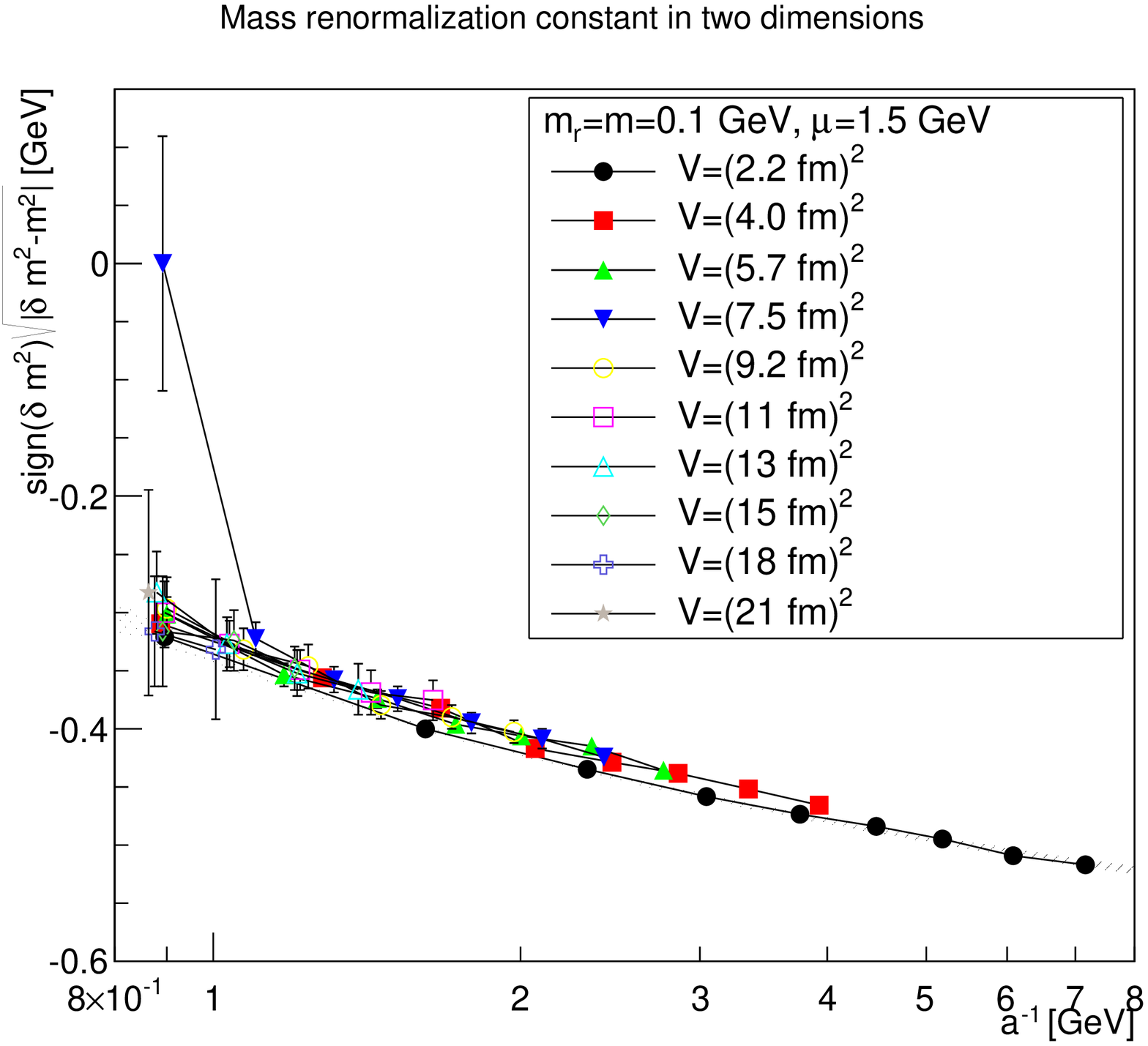}\\
\includegraphics[width=0.475\linewidth]{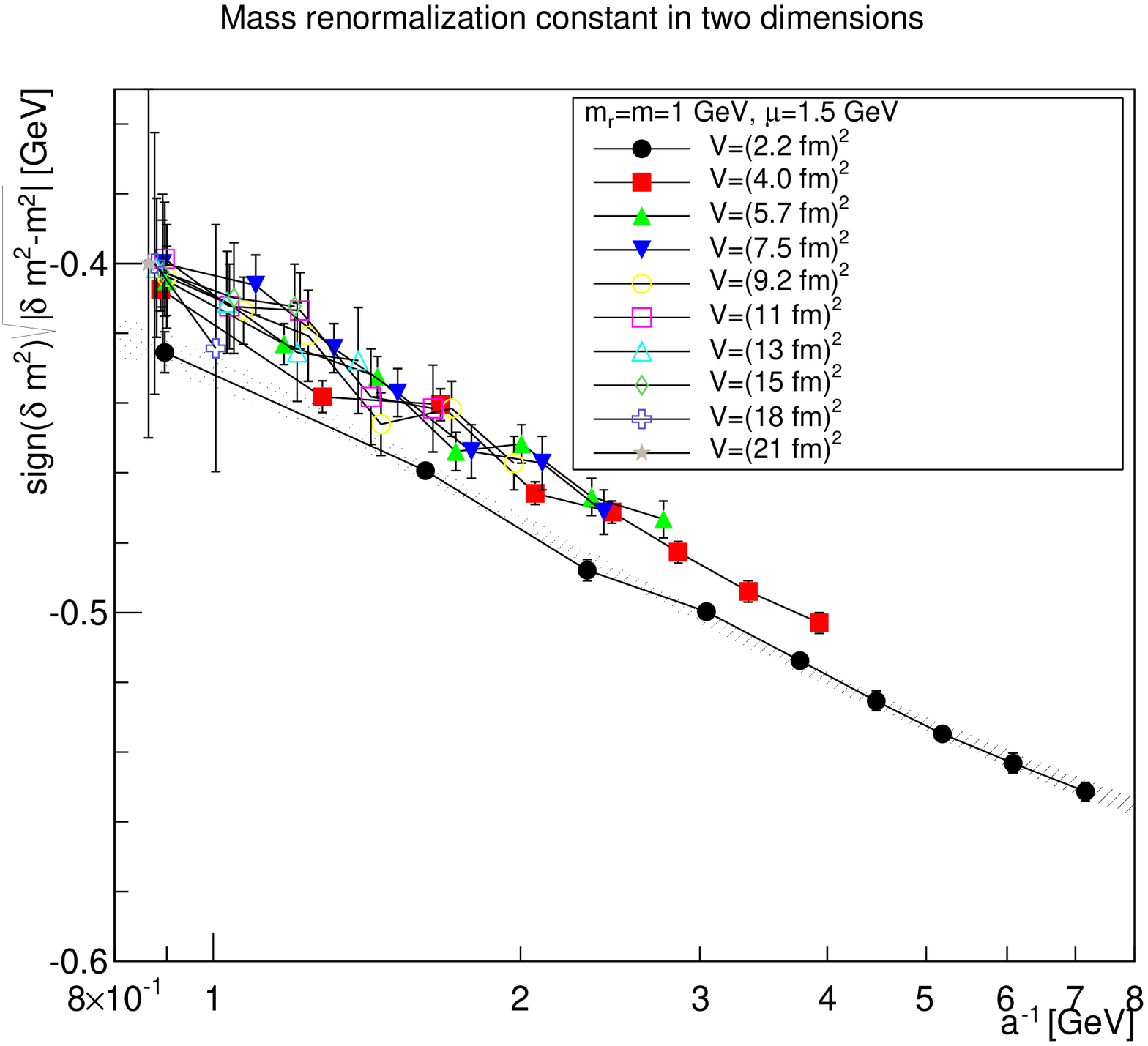}\includegraphics[width=0.475\linewidth]{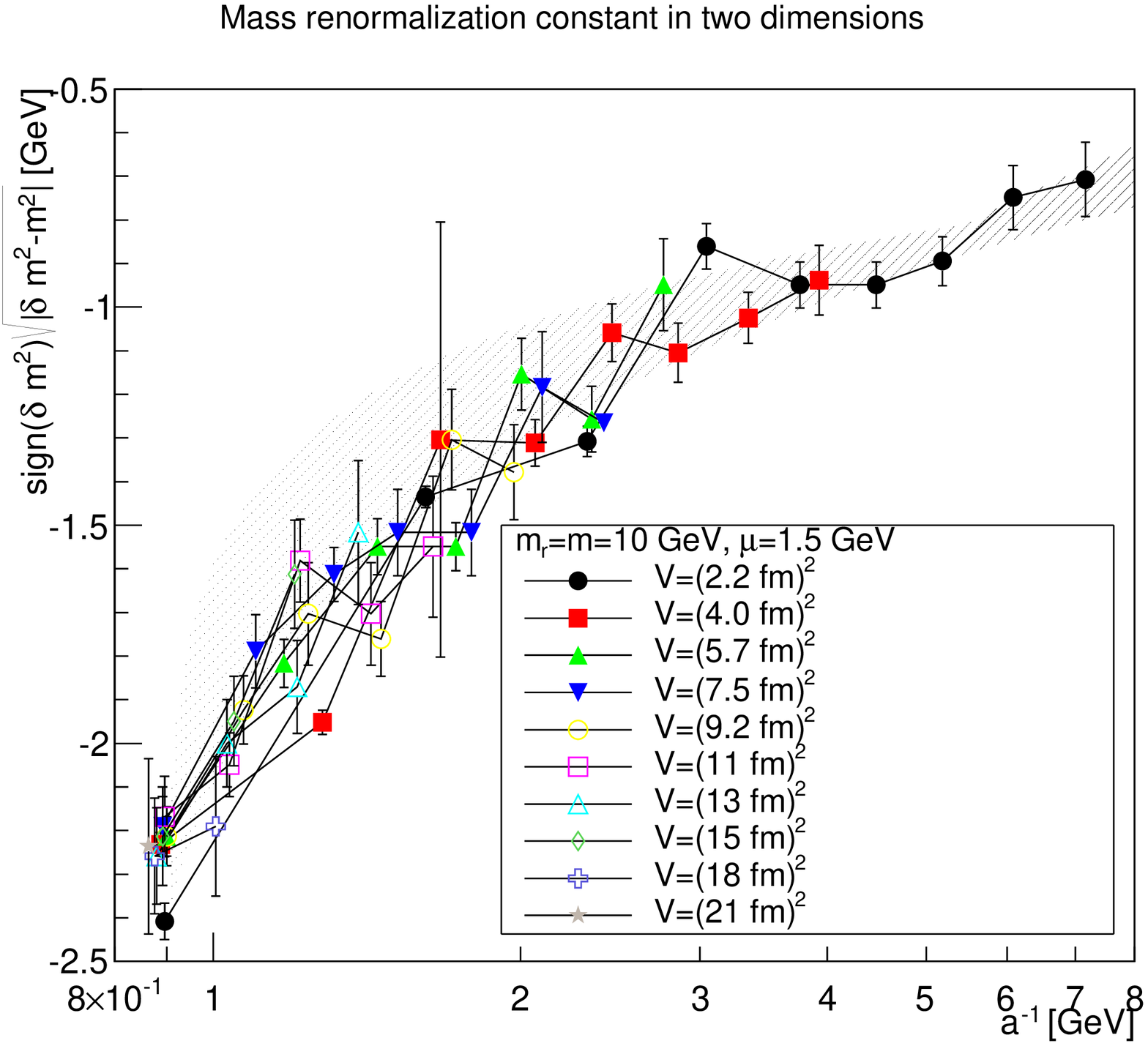}
\caption{\label{fig:m2}The mass renormalization constant as a function of the lattice cutoff and the lattice volume in two dimensions for $\mu=1.5$ GeV. The top-left panel shows the case of $m=m_r=0$ GeV, the top-right panel of $m=m_r=0.1$ GeV, the bottom-left panel of $m=m_r=1$ GeV, and the bottom-right panel of $m=m_r=10$ GeV. The hatched band is the fit \pref{mfit} with the parameters given in table \ref{fitsm}.}
\end{figure}

\begin{figure}[htb]
\includegraphics[width=0.475\linewidth]{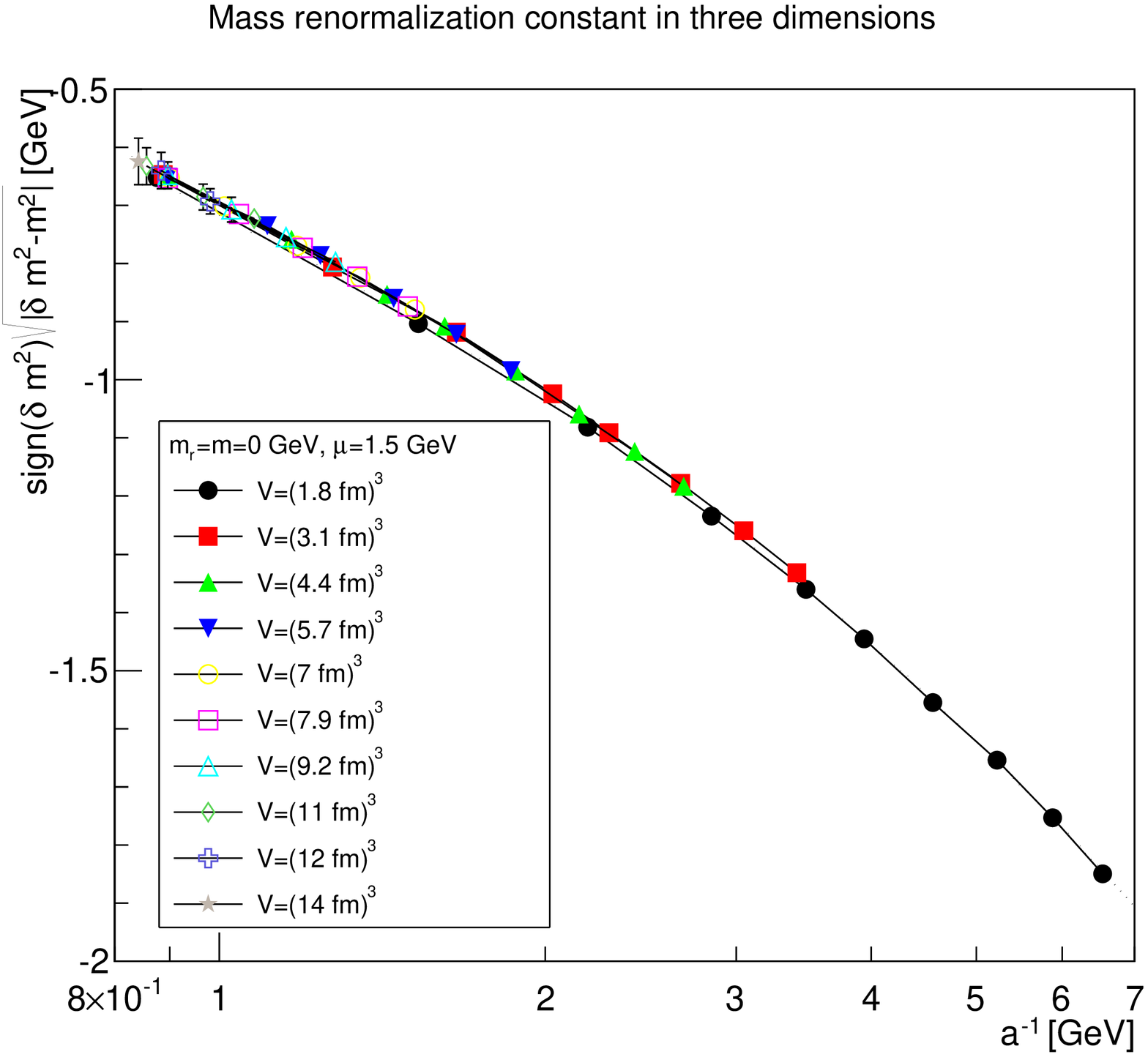}\includegraphics[width=0.475\linewidth]{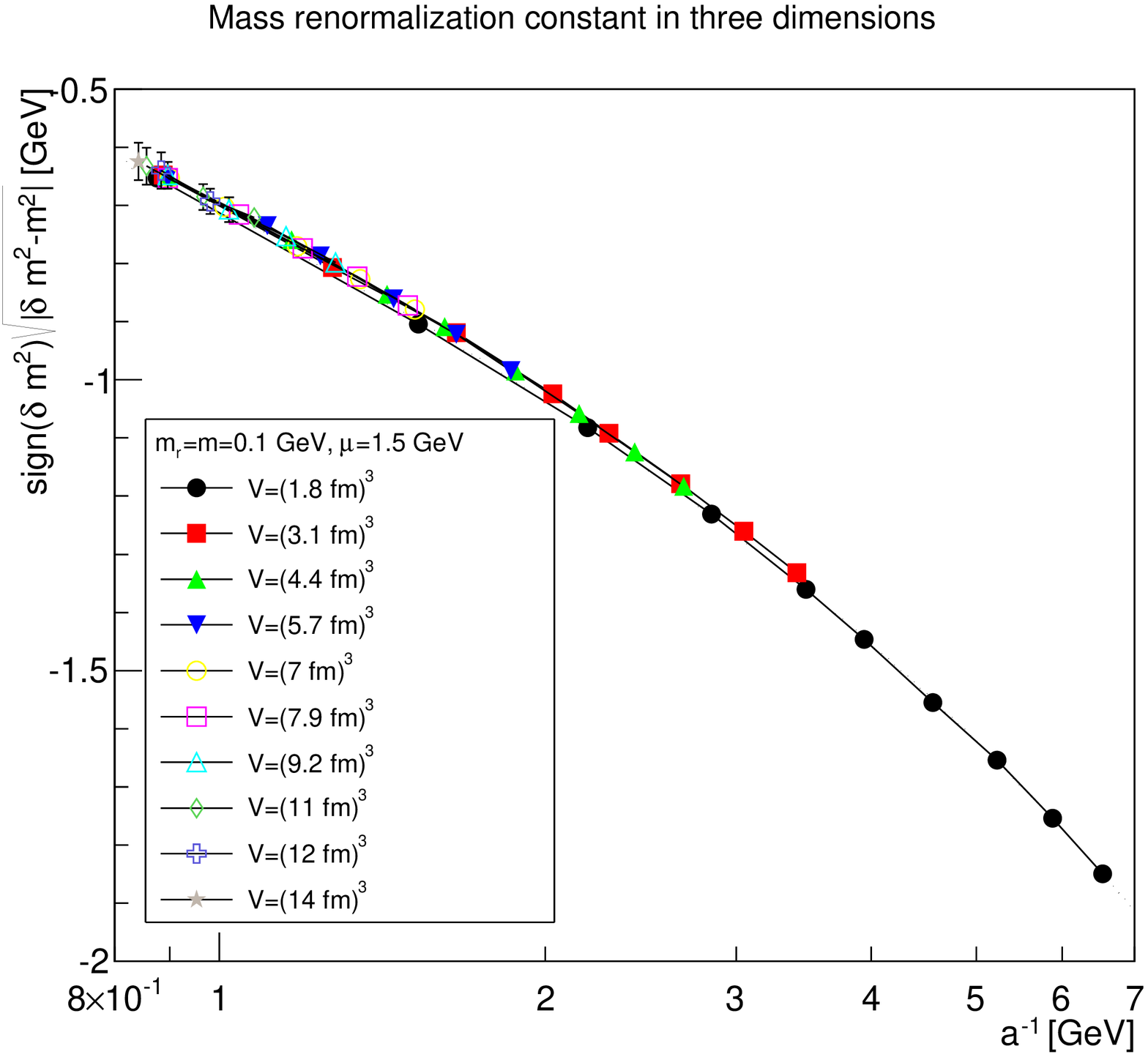}\\
\includegraphics[width=0.475\linewidth]{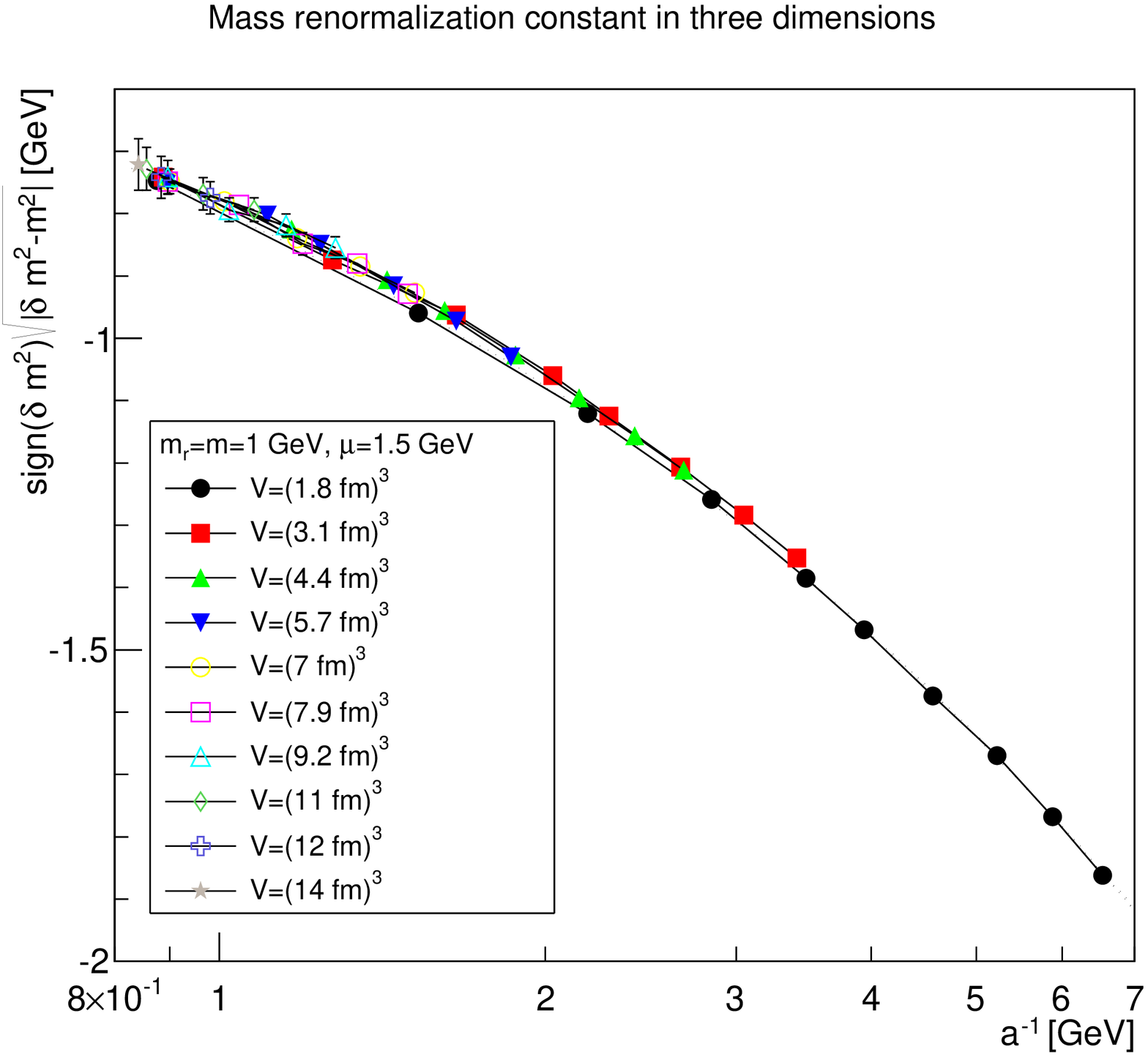}\includegraphics[width=0.475\linewidth]{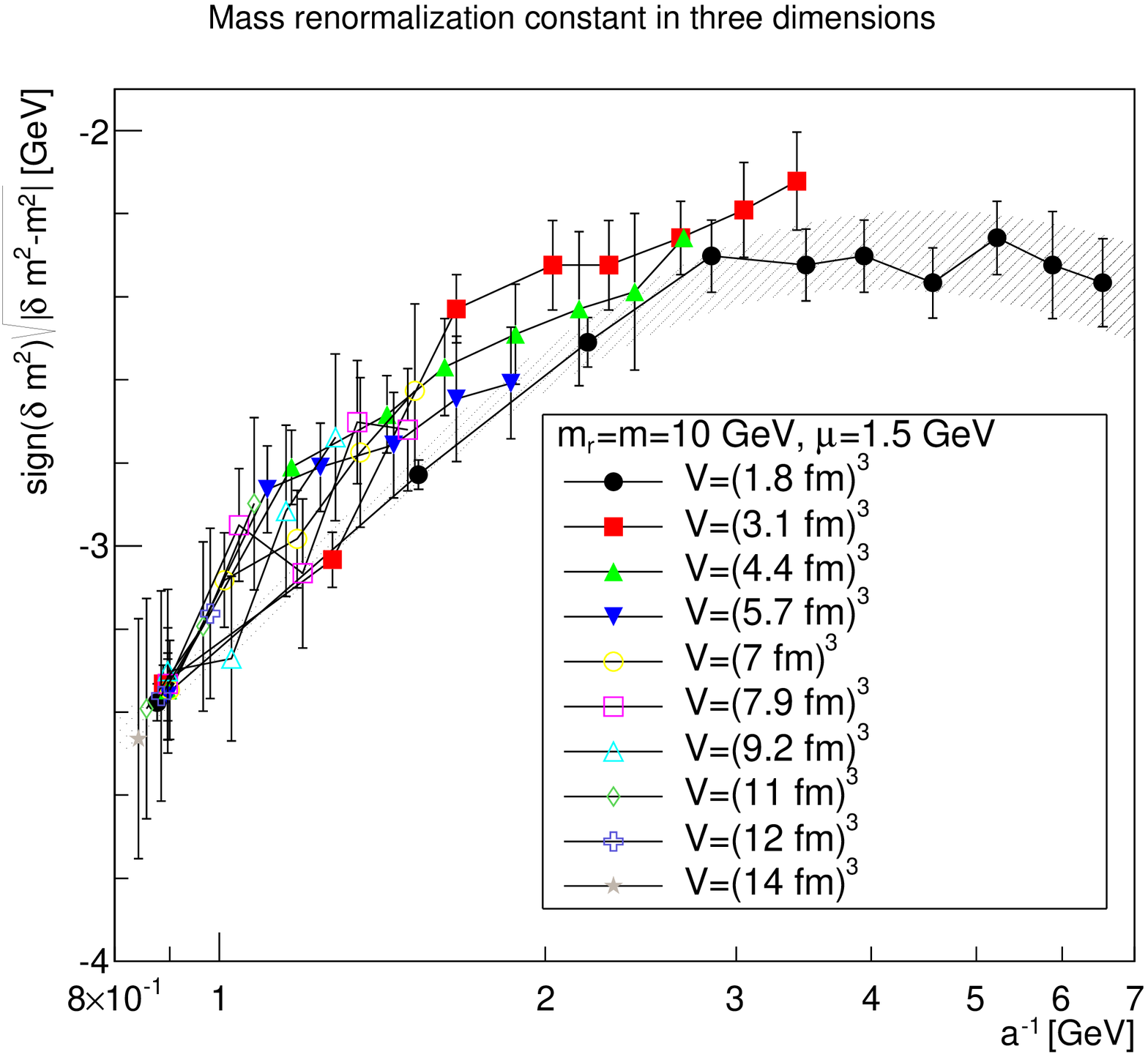}
\caption{\label{fig:m3}The mass renormalization constant as a function of the lattice cutoff and the lattice volume in three dimensions for $\mu=1.5$ GeV. The top-left panel shows the case of $m=m_r=0$ GeV, the top-right panel of $m=m_r=0.1$ GeV, the bottom-left panel of $m=m_r=1$ GeV, and the bottom-right panel of $m=m_r=10$ GeV. The hatched band is the fit \pref{mfit} with the parameters given in table \ref{fitsm}. Note that the hatched band can be as narrow as the lines, and therefore not be visible.}
\end{figure}

\begin{figure}[htb]
\includegraphics[width=0.475\linewidth]{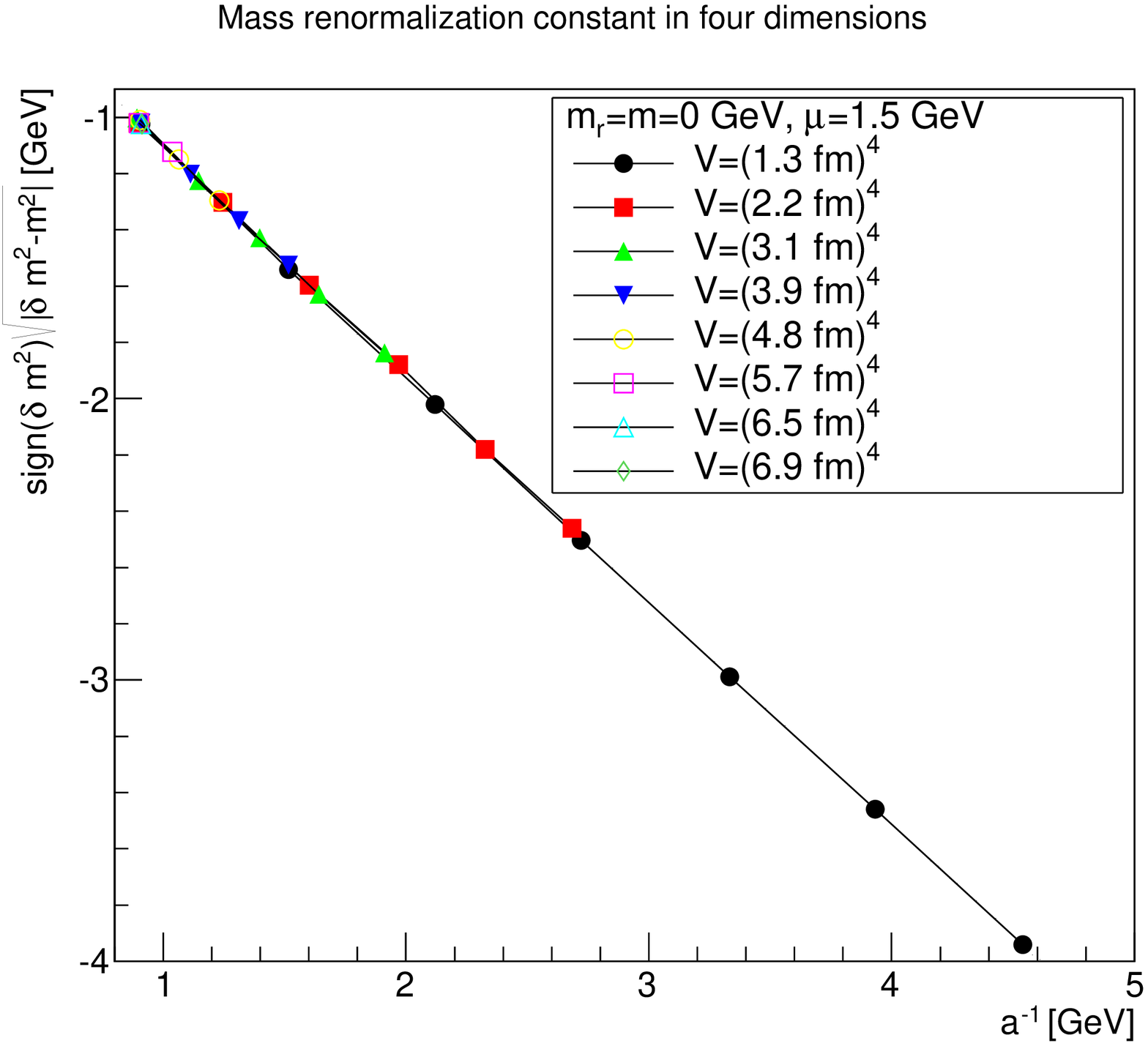}\includegraphics[width=0.475\linewidth]{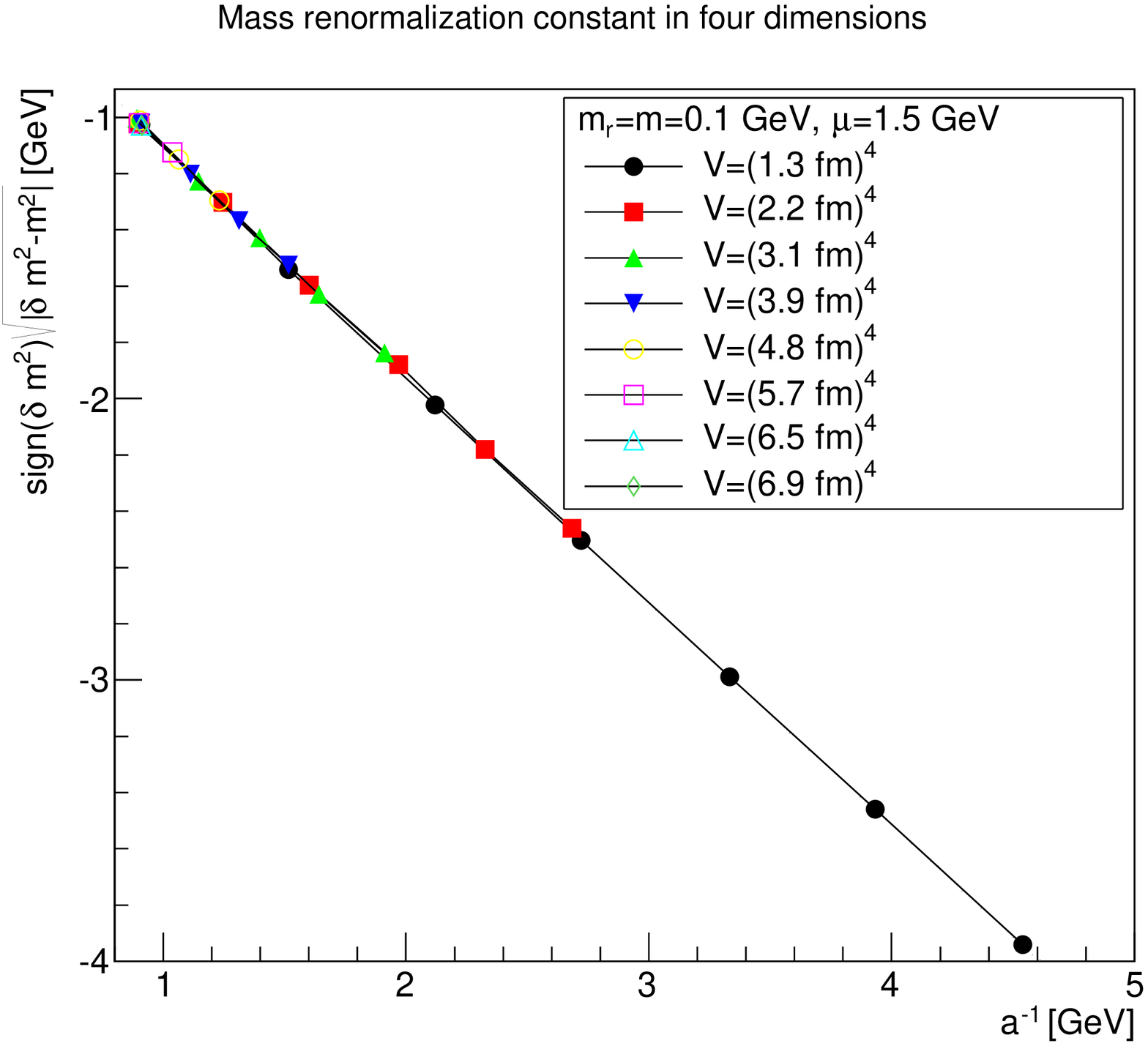}\\
\includegraphics[width=0.475\linewidth]{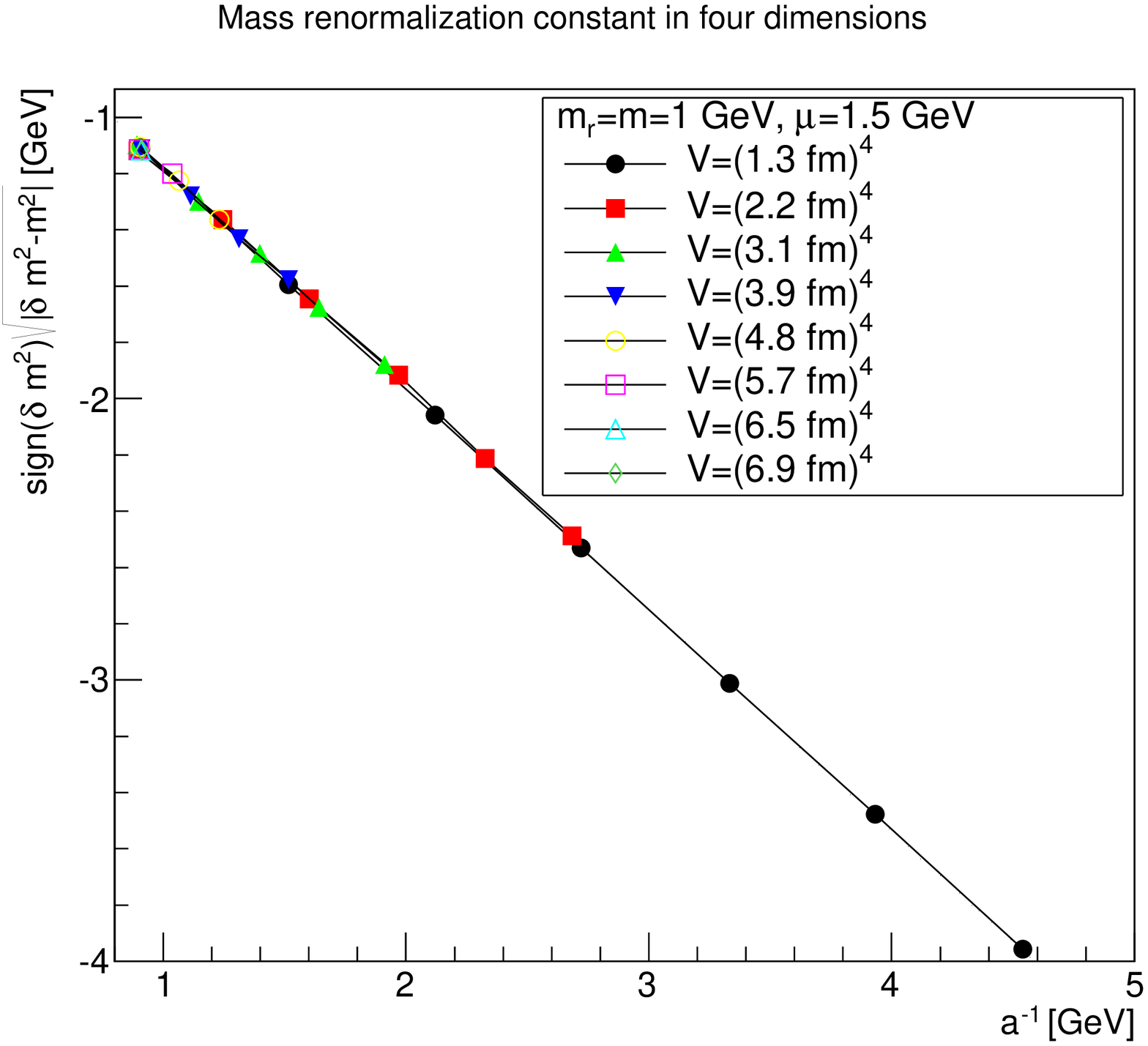}\includegraphics[width=0.475\linewidth]{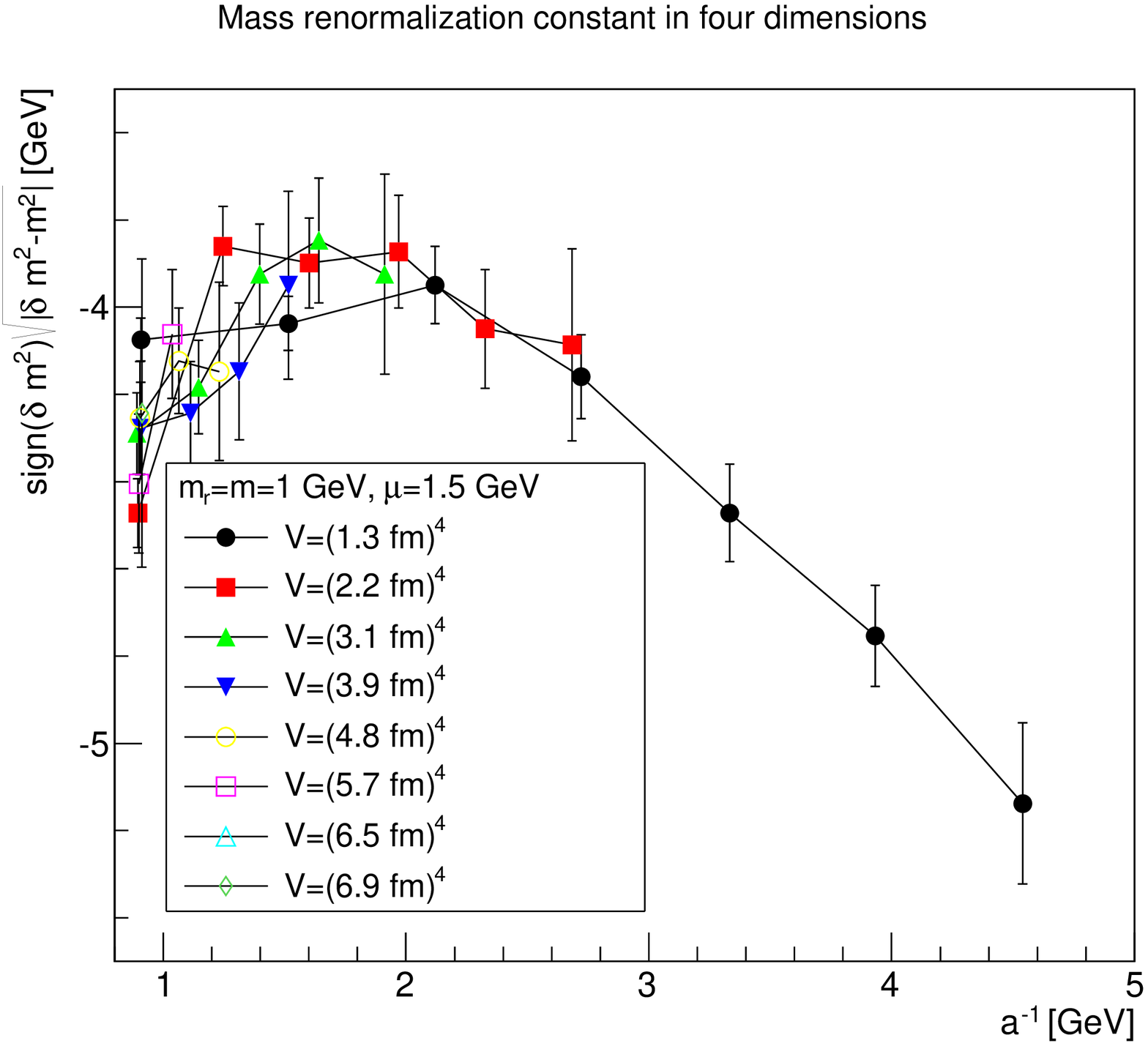}
\caption{\label{fig:m4}The mass renormalization constant as a function of the lattice cutoff and the lattice volume in four dimensions for $\mu=1.5$ GeV. The top-left panel shows the case of $m=m_r=0$ GeV, the top-right panel of $m=m_r=0.1$ GeV, the bottom-left panel of $m=m_r=1$ GeV, and the bottom-right panel of $m=m_r=10$ GeV. The hatched band is the fit \pref{mfit} with the parameters given in table \ref{fitsm}. Note that the hatched band can be as narrow as the lines, and therefore not be visible. No fit was possible using this fit form for $m=m_r=10$ GeV.}
\end{figure}

The situation for the mass renormalization constant, more precisely for $(|\delta m^2-m^2|)^{1/2}$, is shown in figures \ref{fig:m2}-\ref{fig:m4}. 

\begin{longtable}{|c|c|c|c|c|c|}
\caption{\label{fitsm}Fit parameters of \pref{mfit} for the mass renormalization constants at $\mu=1.5$ GeV. Note that a fit for $m=10$ GeV in four dimensions was not possible using this fit form.}\\
\hline
$d$	& m [GeV]	& $\Lambda$ [GeV]	& $\epsilon$ 	& $\delta$	& $c$ [GeV]	\cr
\hline\endfirsthead
\hline
\multicolumn{6}{|l|}{Table \ref{fitsm} continued}\\
\hline
$d$	& m [GeV]	&$\Lambda$ [GeV]	& $\epsilon$ 	& $\delta$	& $c$ [GeV$^{-\delta}$]	\cr
\hline\endhead
\multicolumn{6}{|r|}{Continued on next page}\\
\hline\endfoot
\endlastfoot
\hline
2	& 10	& 0.62(18)	& 0.06(4)	& 0	& 8.5(4)	\cr
\hline
2	& 1	& 1.47(9)	& 0.065(3)	& 0	& -1.4179(16)	\cr
\hline
2	& 0.1	& 0.91(6)	& 0.182(3)	& 0	& -0.4792(11)	\cr
\hline
2	& 0	& 0.89(6)	& 0.223(3)	& 0	& -0.3793(11)	\cr
\hline
\hline
3	& 10	& 1.150(19)	& 0.383(11)	& -0.2409(7)	& 7.209(4)	\cr
\hline
3	& 1	& 1.142(4)	& 0.1916(8)	& 0.4018(3)	& -1.734(2)	\cr
\hline
3	& 0.1	& 1.950(15)	& -0.267(16)	& 0.588(9)	& -0.906(5)	\cr
\hline
3	& 0	& 0.610(8)	& 0.034(3)	& 0.4603(16)	& -0.7428(15)	\cr
\hline
\hline
4	& 1	& 1.086(3)	& -0.596(11)	& 1.082(8)	& -1.885(3)	\cr
\hline
4	& 0.1	& 1.110(5)	& -0.319(11)	& 1.083(8)	& -1.1257(18)	\cr
\hline
4	& 0	& 1.139(6)	& -0.2722(12)	& 1.076(9)	& -1.0520(18)	\cr
\hline
\end{longtable}

The results can be fitted rather well for the three lighter cases by the form
\be
\frac{\delta m^2(a)-m^2}{(1\text{ GeV})^{2-\epsilon}}=-\left(m+ca^{-\delta}\ln\left(\Lambda^2+\frac{1}{a^2}\right)\right)^\epsilon\label{mfit}.
\ee
\no The fit parameters are listed in table \ref{fitsm}. As expected, the dependence on $a^{-1}$ is roughly logarithmic in two dimensions, linear in three dimensions, and quadratic in four dimensions. Again, slight modifications of the fit form work equally well, but the general trend remains the same. This form also shows that in the limit $a\to\infty$ the mass moves close to the classical mass, especially in three and four dimensions.

The situation for $m=10$ GeV is more involved. Replacing $m$ by $-m$, the fit form \pref{mfit} also works in two and three dimensions, but not in four dimensions. The reason can essentially be inferred from the comparison of the three dimensionalities. In all case, there appears to be a competition between two effects. One, which pushes the mass renormalization to zero, and one which pushes it to infinity. The latter wins out earlier the higher the dimension, and is the one with the expected dependence on $a$. It appears thus reasonable that this is the actual behavior of the renormalization constant, in agreement with expectations, while the other contribution is likely a lattice artifact. Just because the divergent part grows quicker the higher the dimension this effect wins for smaller $a$ in higher dimensions, while the bending-over has not yet been reached in two dimensions. In four dimensions, this is happening but the full behavior cannot be captured by the fit as in three dimensions. Fitting with values at large cutoff only would probably be possible, but not yet enough points are available to do this. This is also consistent with the expectation that lattice artifacts due to lattice spacings should be largest for the largest mass, and thus a similar effect for the lighter masses should be suppressed.

The comparison of the parameters in the tables \ref{fitsz} and \ref{fitsm} show for the light masses a slight dependence on the mass, as was to expected in the present mass-dependent scheme. On the other hand, the figures \ref{fig:z2}-\ref{fig:m4} show pretty clearly that once a certain minimal volume of a few fm$^d$ has been reached, the renormalization constants are essentially volume-independent. Thus, small-volume high-statistic runs can be used to get already a reasonable result for the renormalization constants, if need would be.

With this, the renormalization of the propagators appears sufficiently well under control that their analytic structure can be investigated next.

\section{Analytic structure}\label{s:ana}

\subsection{Momentum space properties}\label{ss:mom}

Since the results of the previous section, especially figure \ref{fig:ur}, strongly suggest that discretization artifacts are small in the following only the results for the finest lattices will be considered. In fact, checking all the lattice setups individually only shows an, almost statistically insignificant, tendency for the propagators to be a little more infrared enhanced the smaller $a$ is.

In addition to the propagator themselves also the dressing functions are interesting, which will be defined as
\be
H(p^2)=(p^2+m_r^2)D(p^2)\label{df},
\ee
\no and which therefore describe the deviation from the corresponding tree-level propagator, and hence the influence of quantum corrections.

\begin{figure}[!htb]
\includegraphics[width=\linewidth]{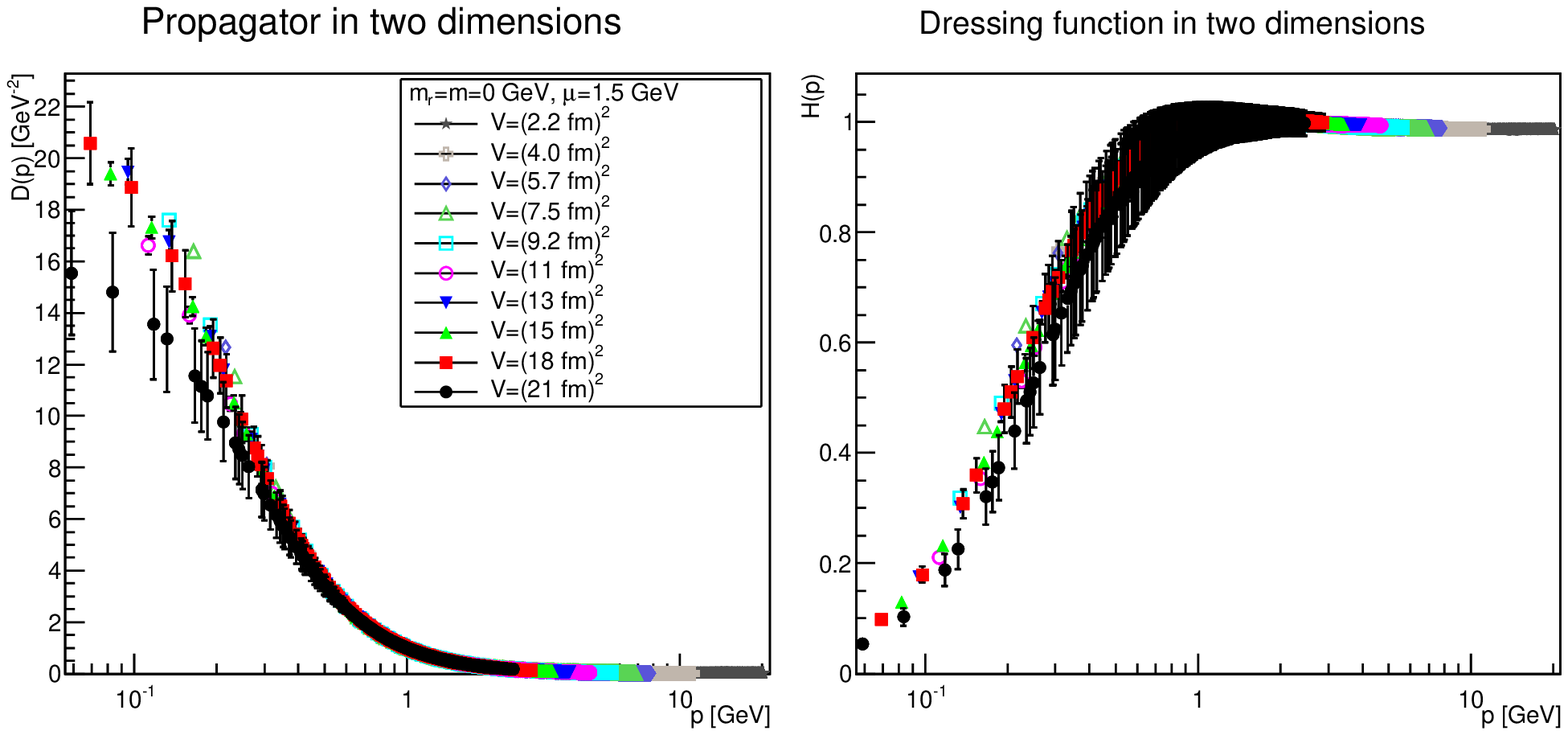}
\caption{\label{fig:d20}The propagator (left panel) and the dressing function \pref{df} (right panel) in two dimensions for $m=m_r=0$ GeV and $\mu=1.5$ GeV.}
\end{figure}

\begin{figure}[!htb]
\includegraphics[width=\linewidth]{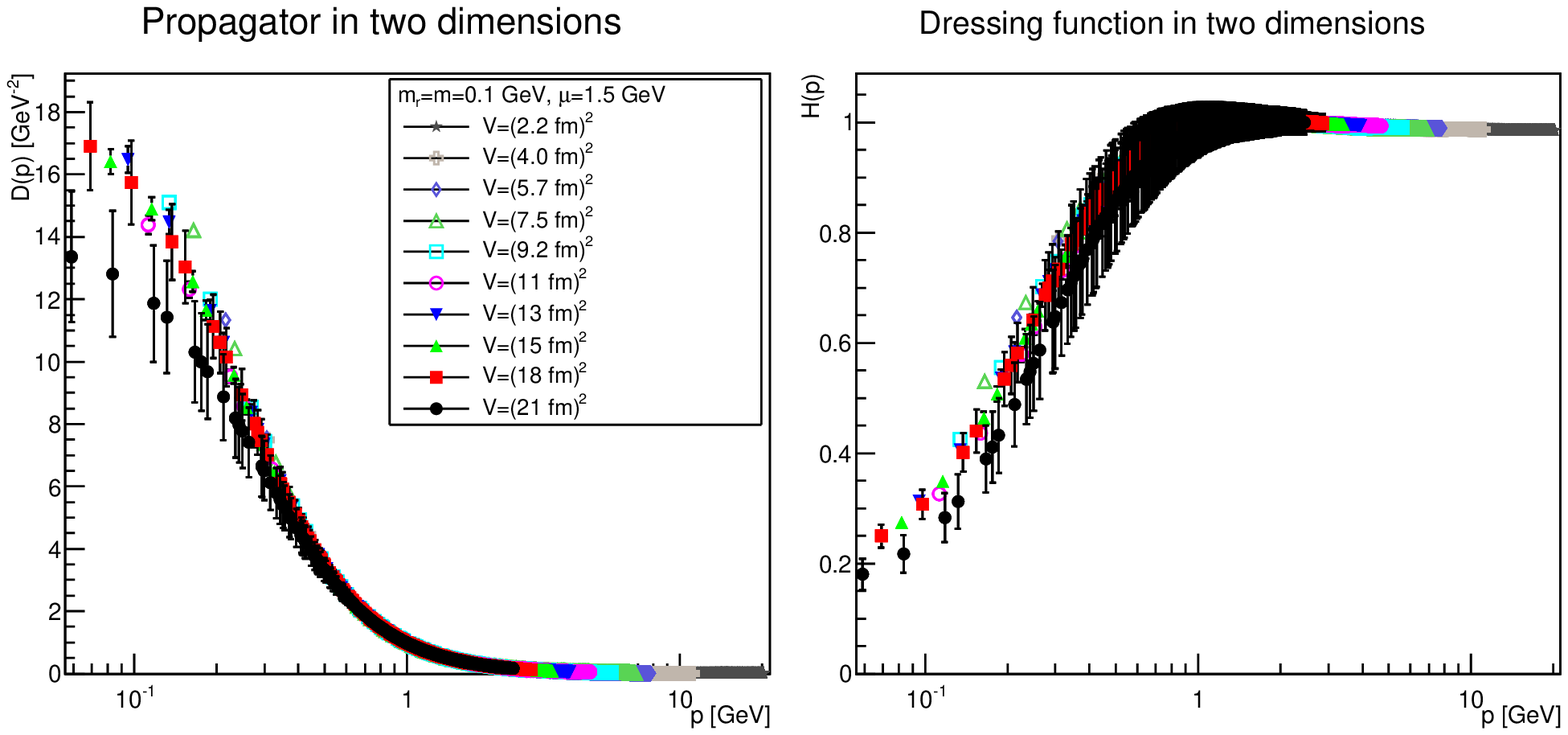}
\caption{\label{fig:d21}The propagator (left panel) and the dressing function \pref{df} (right panel) in two dimensions for $m=m_r=0.1$ GeV and $\mu=1.5$ GeV.}
\end{figure}

\begin{figure}[!htb]
\includegraphics[width=\linewidth]{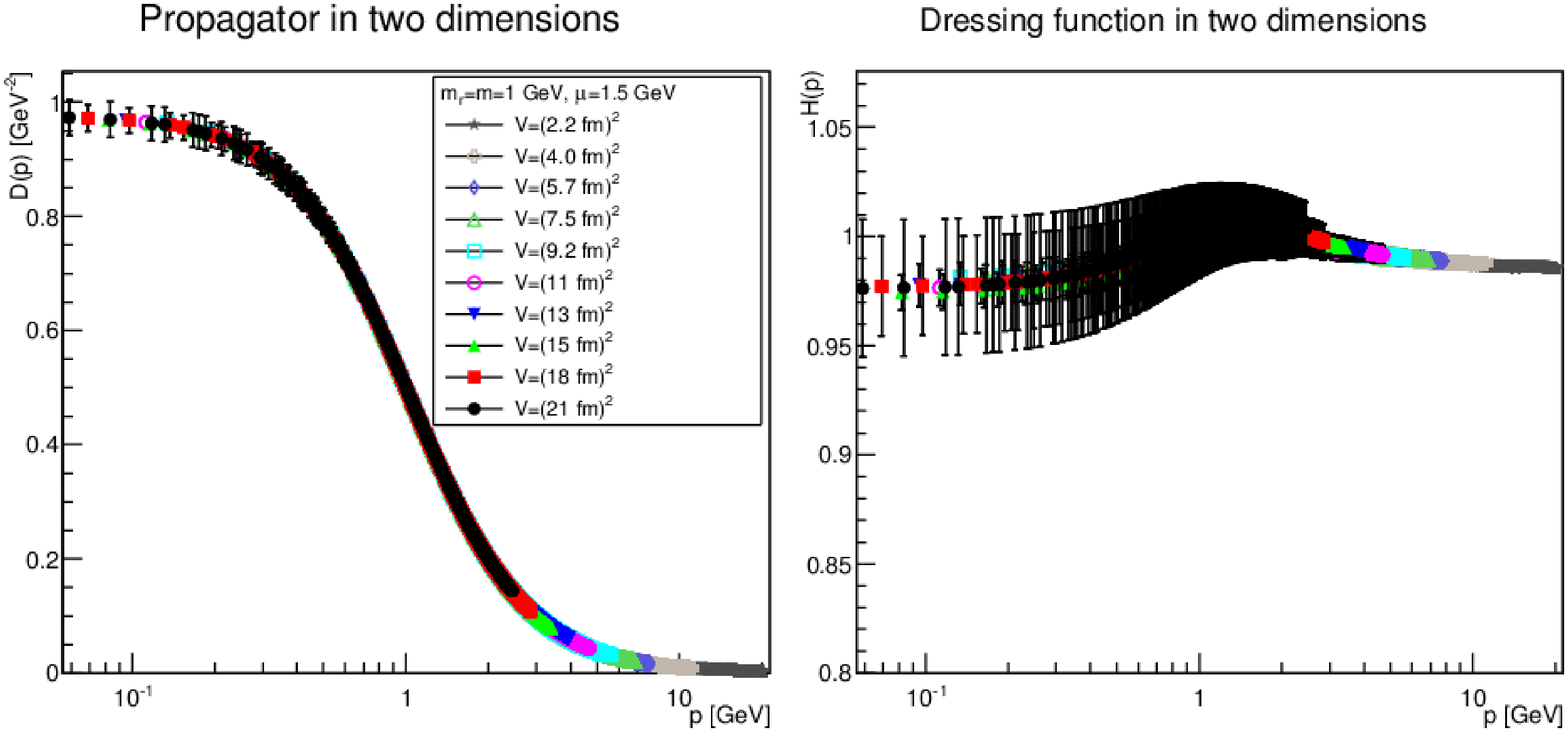}
\caption{\label{fig:d22}The propagator (left panel) and the dressing function \pref{df} (right panel) in two dimensions for $m=m_r=1$ GeV and $\mu=1.5$ GeV. Note the different scale in the right-hand panel compared to figures \ref{fig:d20} and \ref{fig:d21}.}
\end{figure}

\begin{figure}[!htb]
\includegraphics[width=\linewidth]{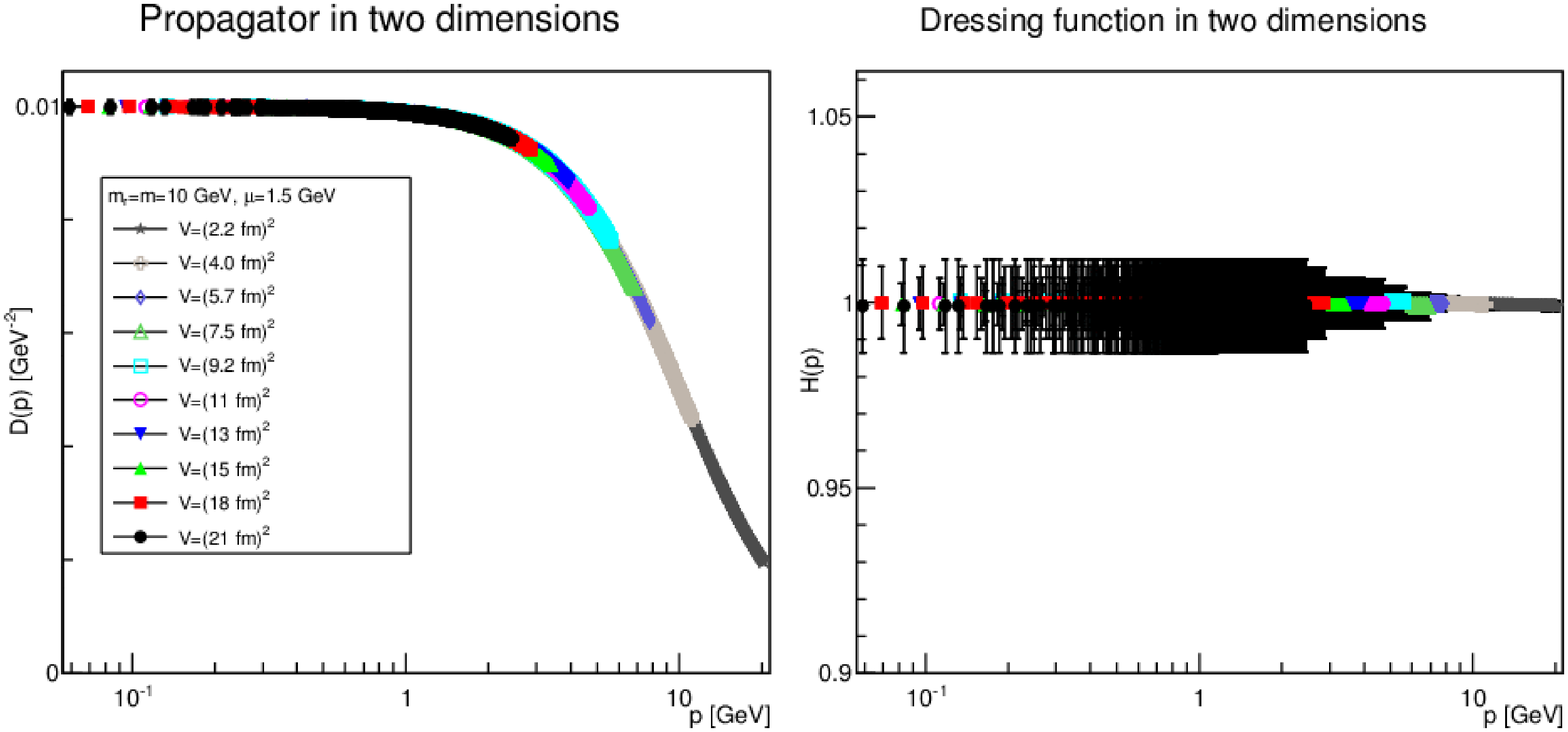}
\caption{\label{fig:d23}The propagator (left panel) and the dressing function \pref{df} (right panel) in two dimensions for $m=m_r=10$ GeV and $\mu=1.5$ GeV. Note the different scale in the right-hand panel compared to figures \ref{fig:d20} and \ref{fig:d21}.}
\end{figure}

\begin{figure}[!htb]
\includegraphics[width=\linewidth]{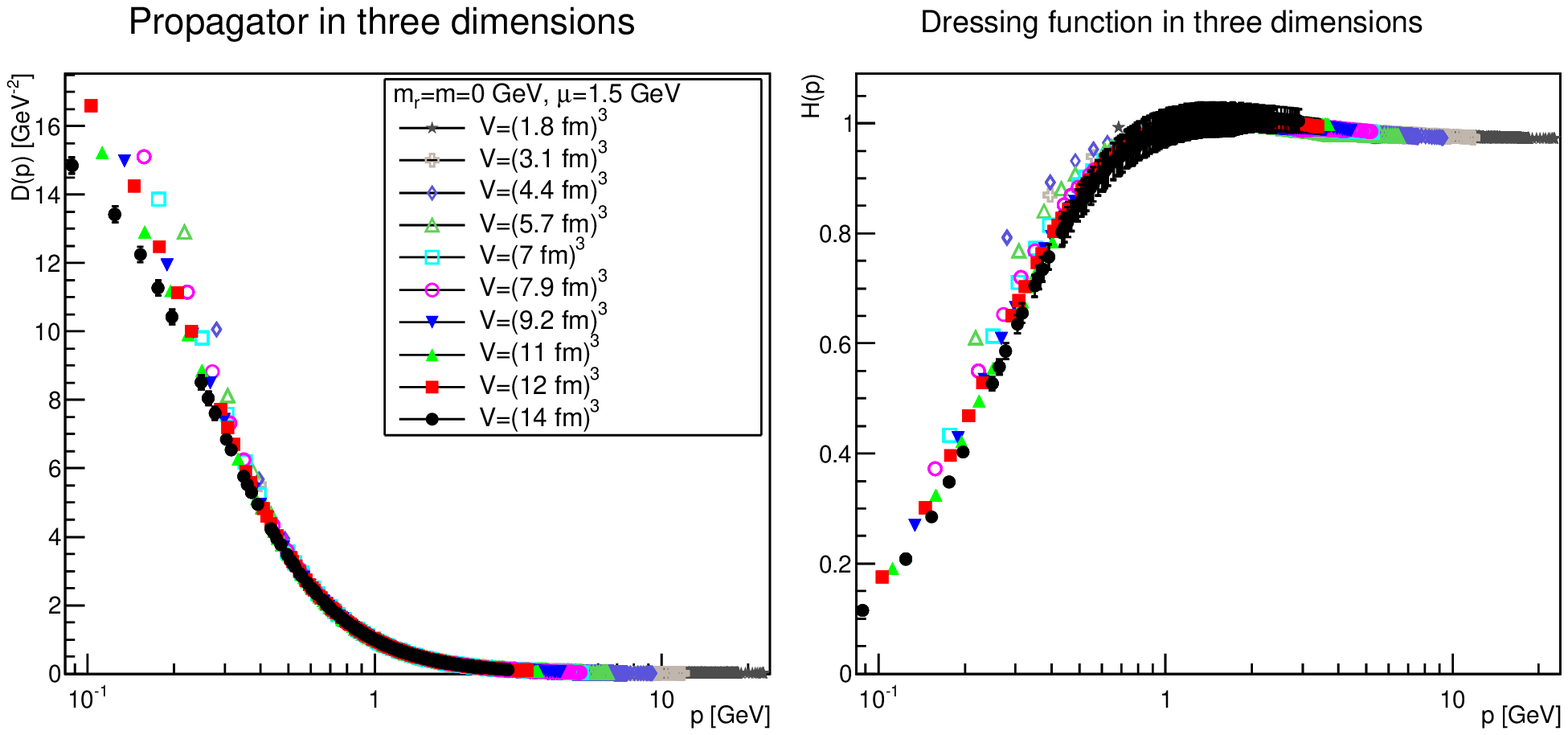}
\caption{\label{fig:d30}The propagator (left panel) and the dressing function \pref{df} (right panel) in three dimensions for $m=m_r=0$ GeV and $\mu=1.5$ GeV.}
\end{figure}

\begin{figure}[!htb]
\includegraphics[width=\linewidth]{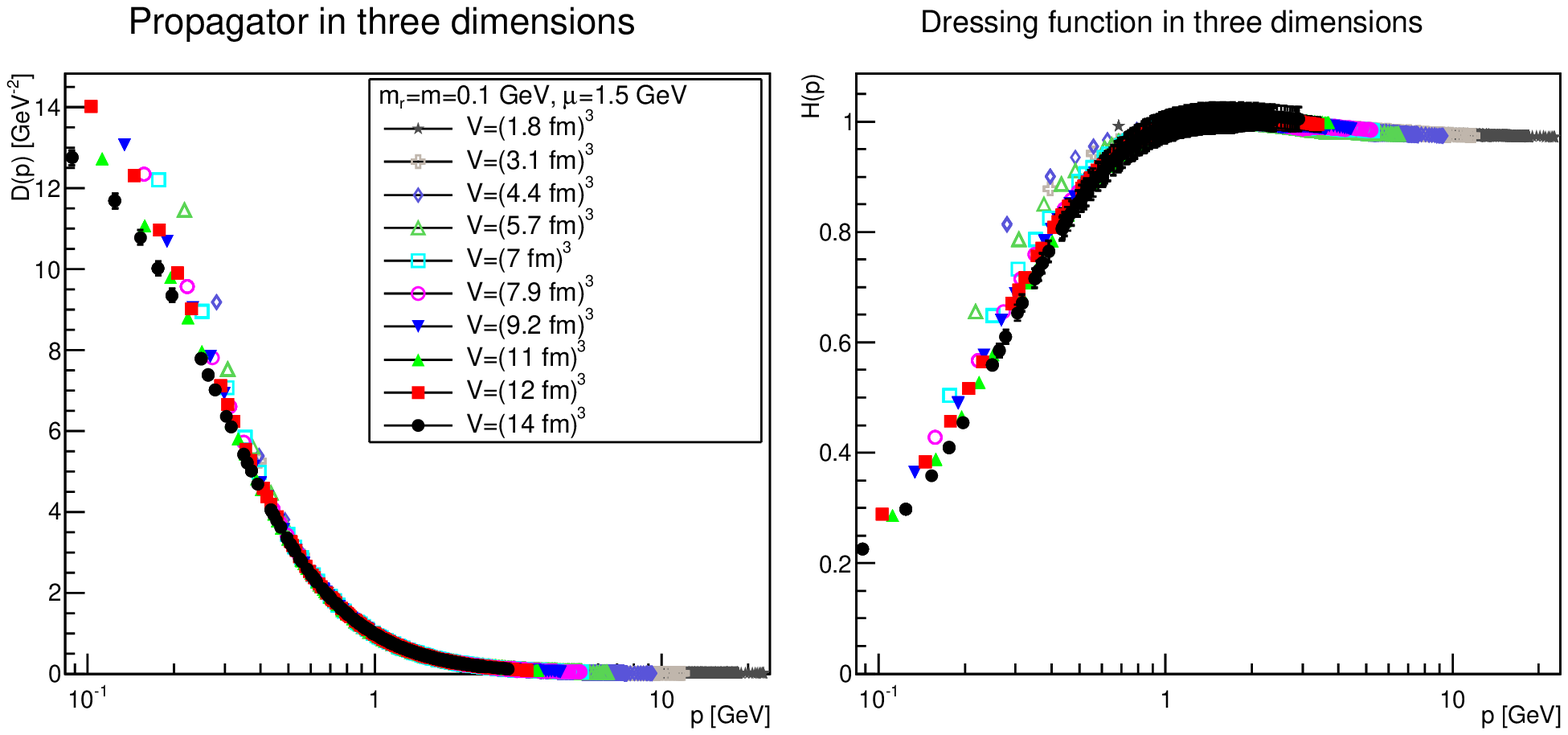}
\caption{\label{fig:d31}The propagator (left panel) and the dressing function \pref{df} (right panel) in three dimensions for $m=m_r=0.1$ GeV and $\mu=1.5$ GeV.}
\end{figure}

\begin{figure}[!htb]
\includegraphics[width=\linewidth]{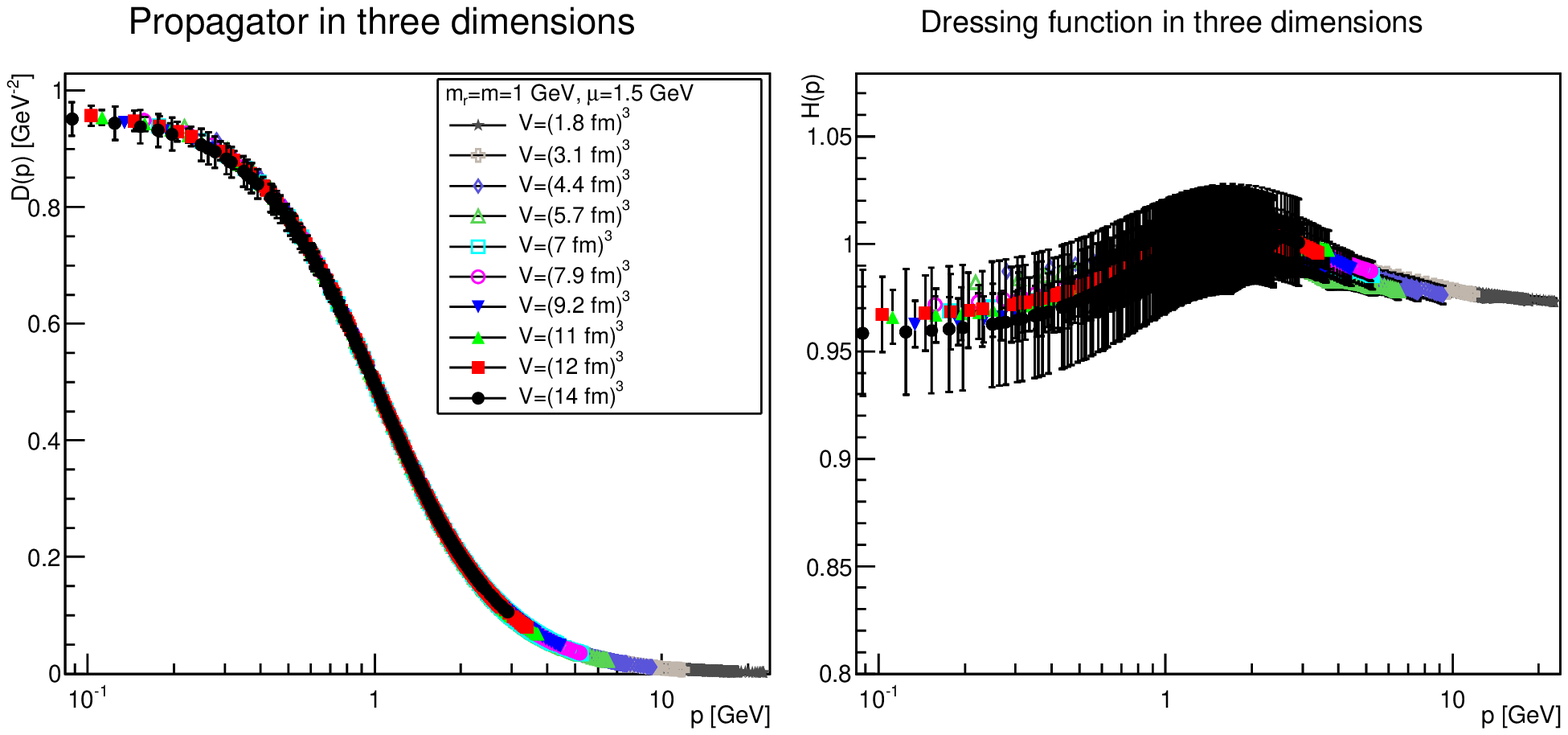}
\caption{\label{fig:d32}The propagator (left panel) and the dressing function \pref{df} (right panel) in three dimensions for $m=m_r=1$ GeV and $\mu=1.5$ GeV. Note the different scale in the right-hand panel compared to figures \ref{fig:d30} and \ref{fig:d31}.}
\end{figure}

\begin{figure}[!htb]
\includegraphics[width=\linewidth]{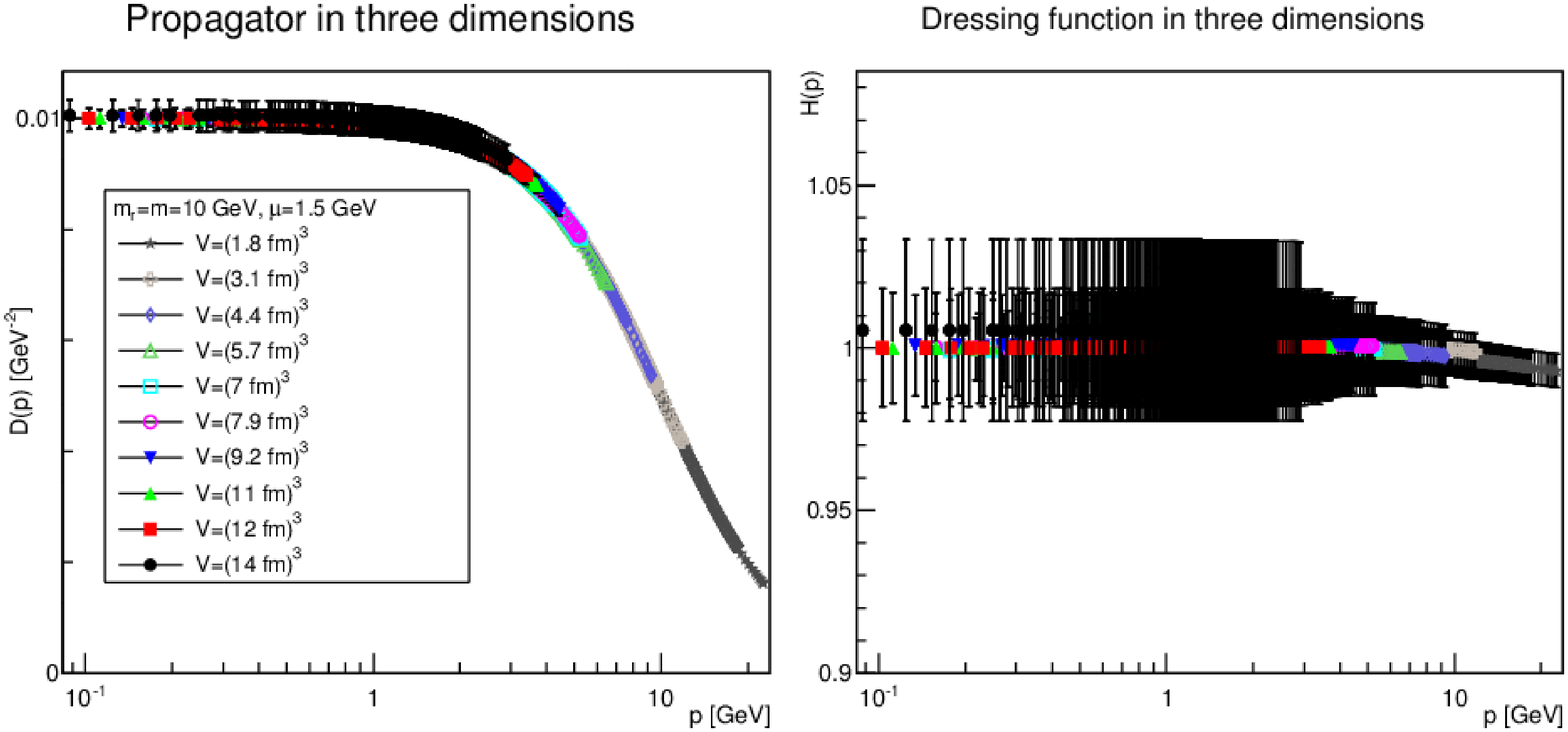}
\caption{\label{fig:d33}The propagator (left panel) and the dressing function \pref{df} (right panel) in three dimensions for $m=m_r=10$ GeV and $\mu=1.5$ GeV. Note the different scale in the right-hand panel compared to figures \ref{fig:d30} and \ref{fig:d31}.}
\end{figure}

\begin{figure}[!htb]
\includegraphics[width=\linewidth]{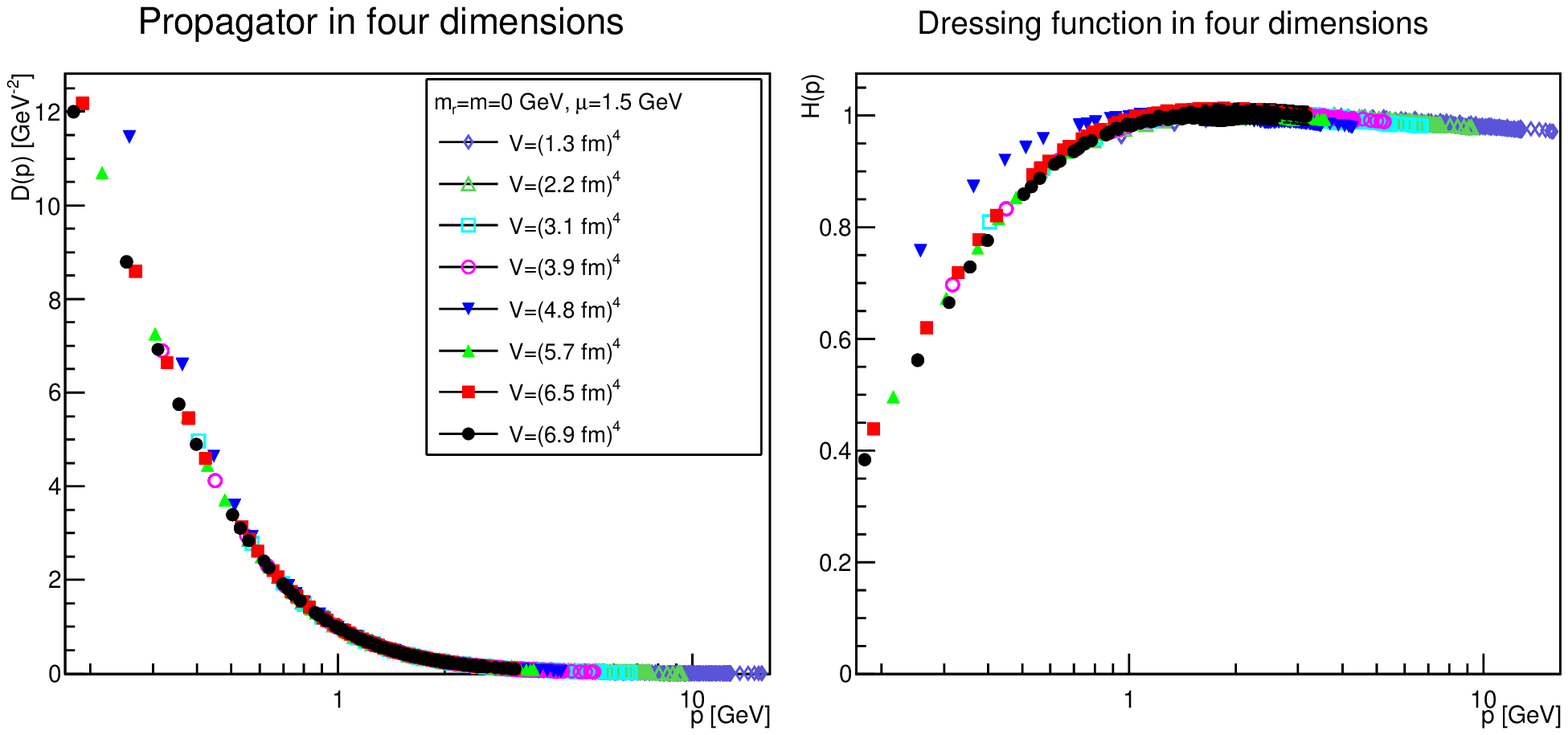}
\caption{\label{fig:d40}The propagator (left panel) and the dressing function \pref{df} (right panel) in four dimensions for $m=m_r=0$ GeV and $\mu=1.5$ GeV.}
\end{figure}

\begin{figure}[!htb]
\includegraphics[width=\linewidth]{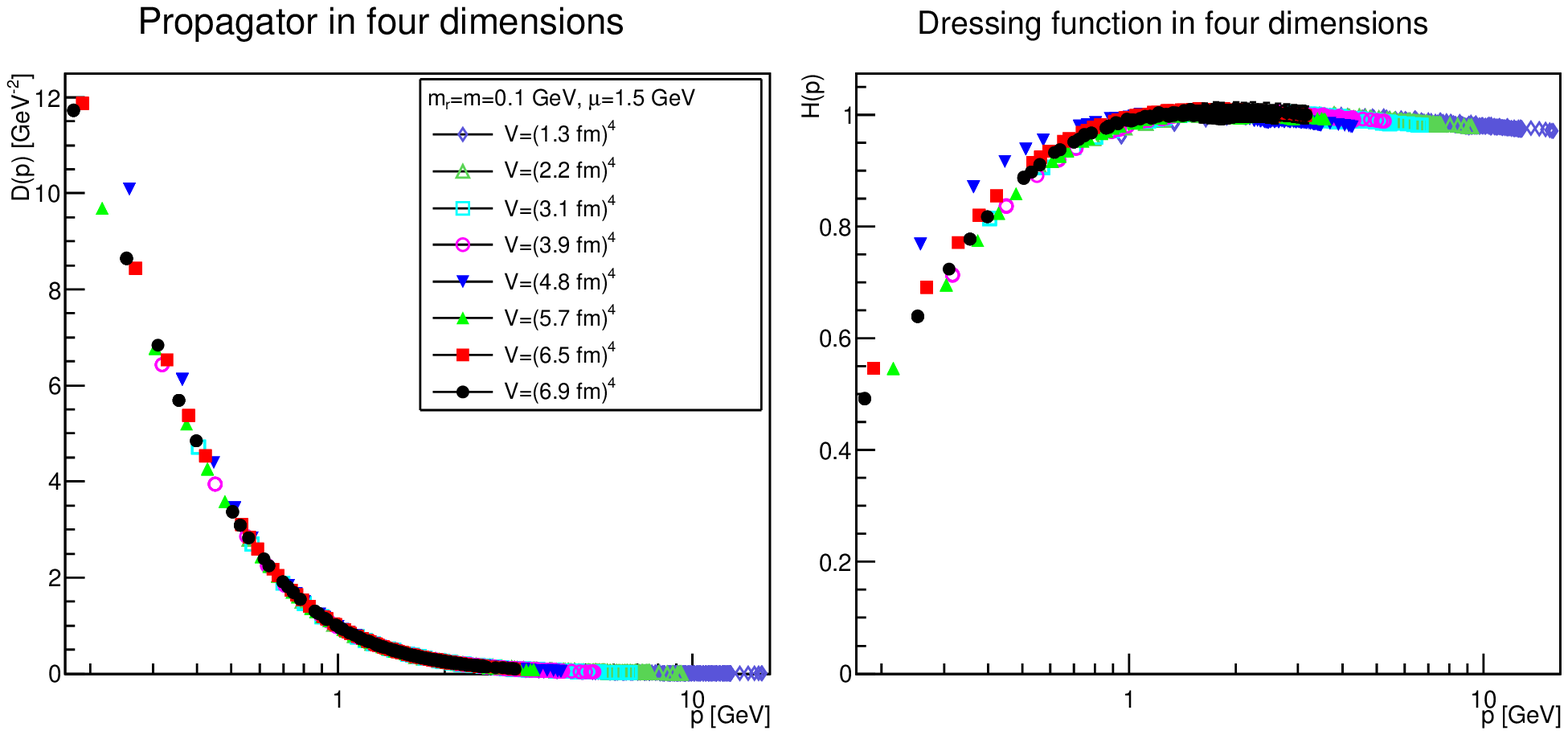}
\caption{\label{fig:d41}The propagator (left panel) and the dressing function \pref{df} (right panel) in four dimensions for $m=m_r=0.1$ GeV and $\mu=1.5$ GeV.}
\end{figure}

\begin{figure}[!htb]
\includegraphics[width=\linewidth]{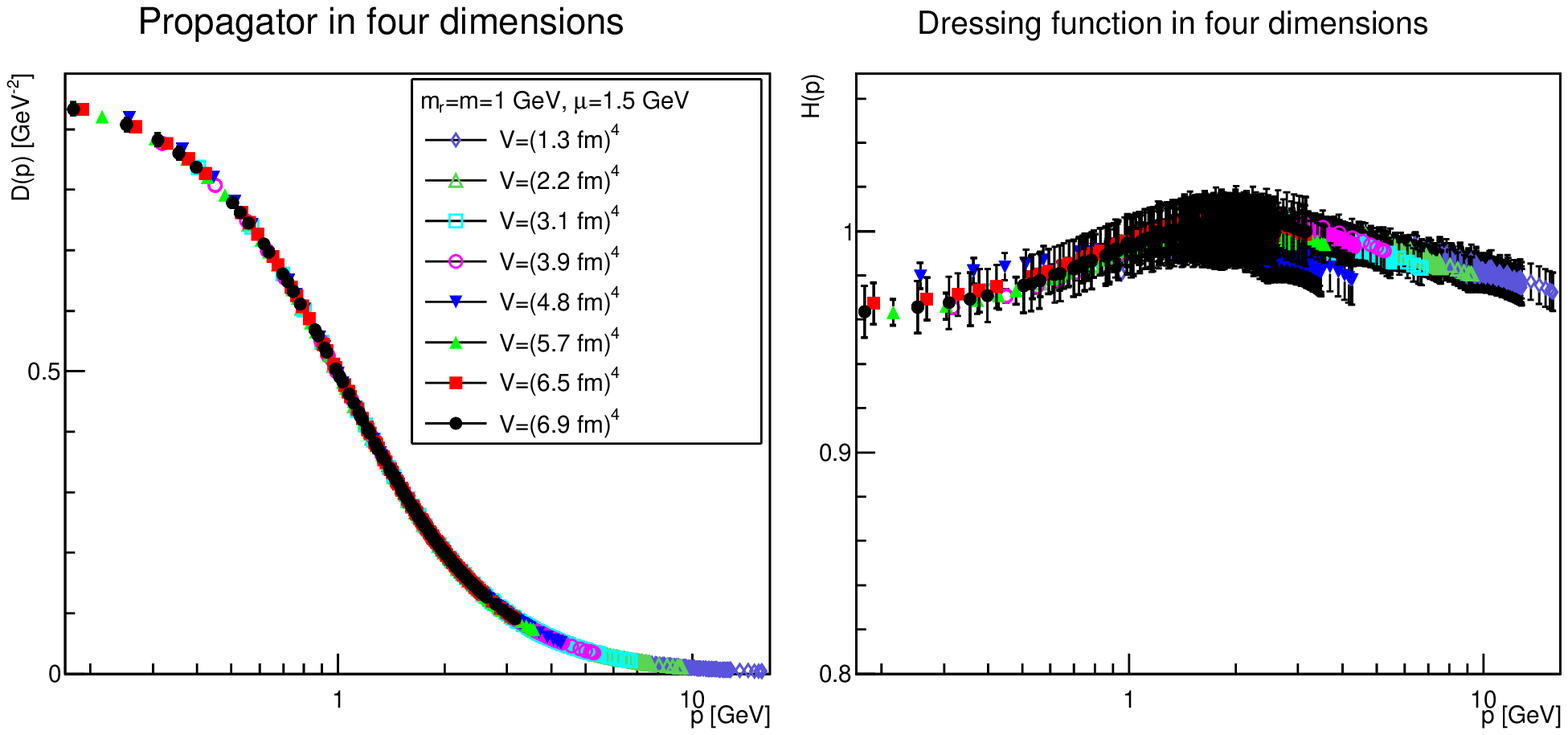}
\caption{\label{fig:d42}The propagator (left panel) and the dressing function \pref{df} (right panel) in four dimensions for $m=m_r=1$ GeV and $\mu=1.5$ GeV. Note the different scale in the right-hand panel compared to figures \ref{fig:d40} and \ref{fig:d41}.}
\end{figure}

\begin{figure}[!htb]
\includegraphics[width=\linewidth]{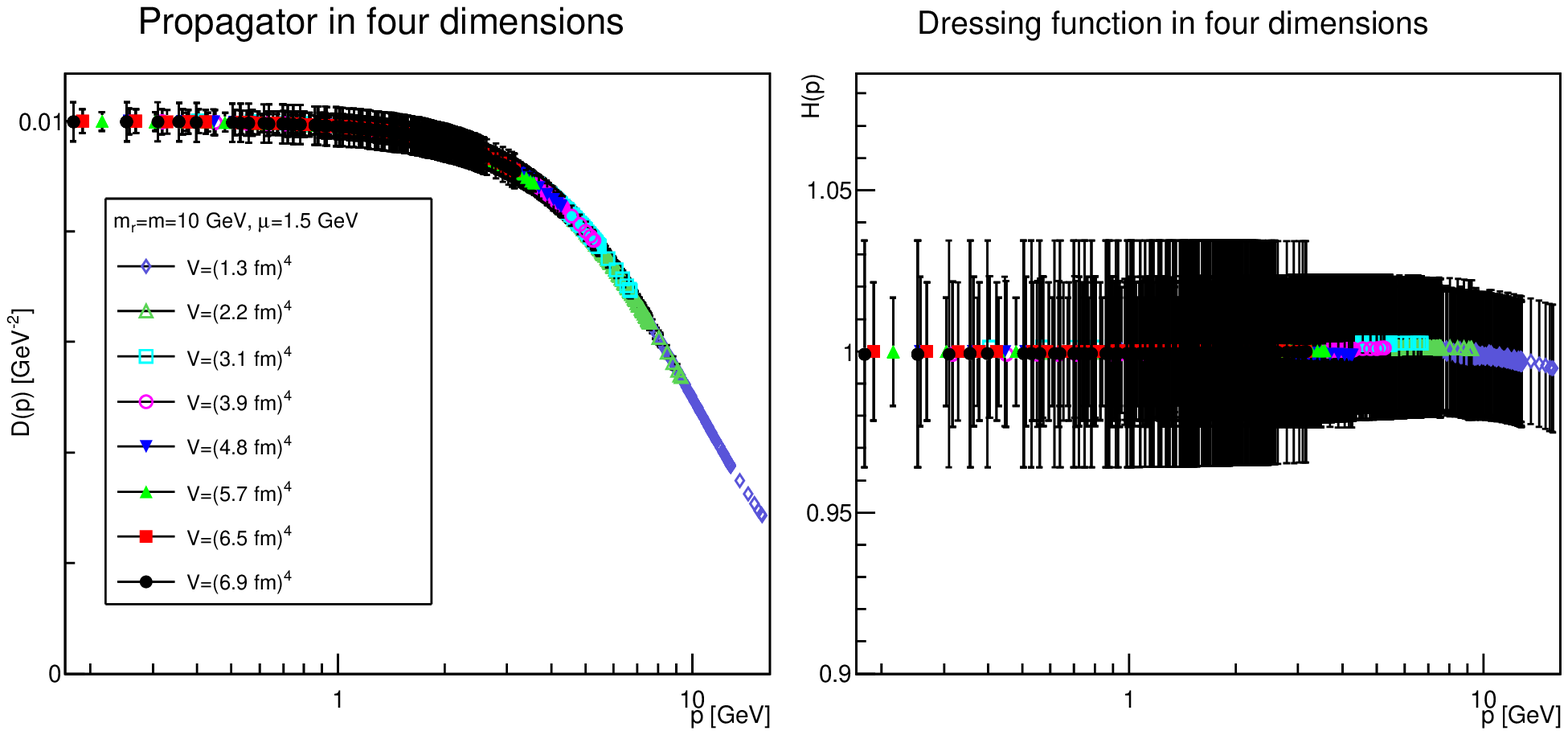}
\caption{\label{fig:d43}The propagator (left panel) and the dressing function \pref{df} (right panel) in four dimensions for $m=m_r=10$ GeV and $\mu=1.5$ GeV. Note the different scale in the right-hand panel compared to figures \ref{fig:d40} and \ref{fig:d41}.}
\end{figure}

The results are shown for two dimensions in figures \ref{fig:d20}-\ref{fig:d23}, for three dimensions in figures \ref{fig:d30}-\ref{fig:d33}, and for four dimensions in figures \ref{fig:d40}-\ref{fig:d43}. Of course, at $\mu$ all propagators have, by construction, a dressing function of 1.

\begin{figure}[!htb]
\includegraphics[width=\linewidth]{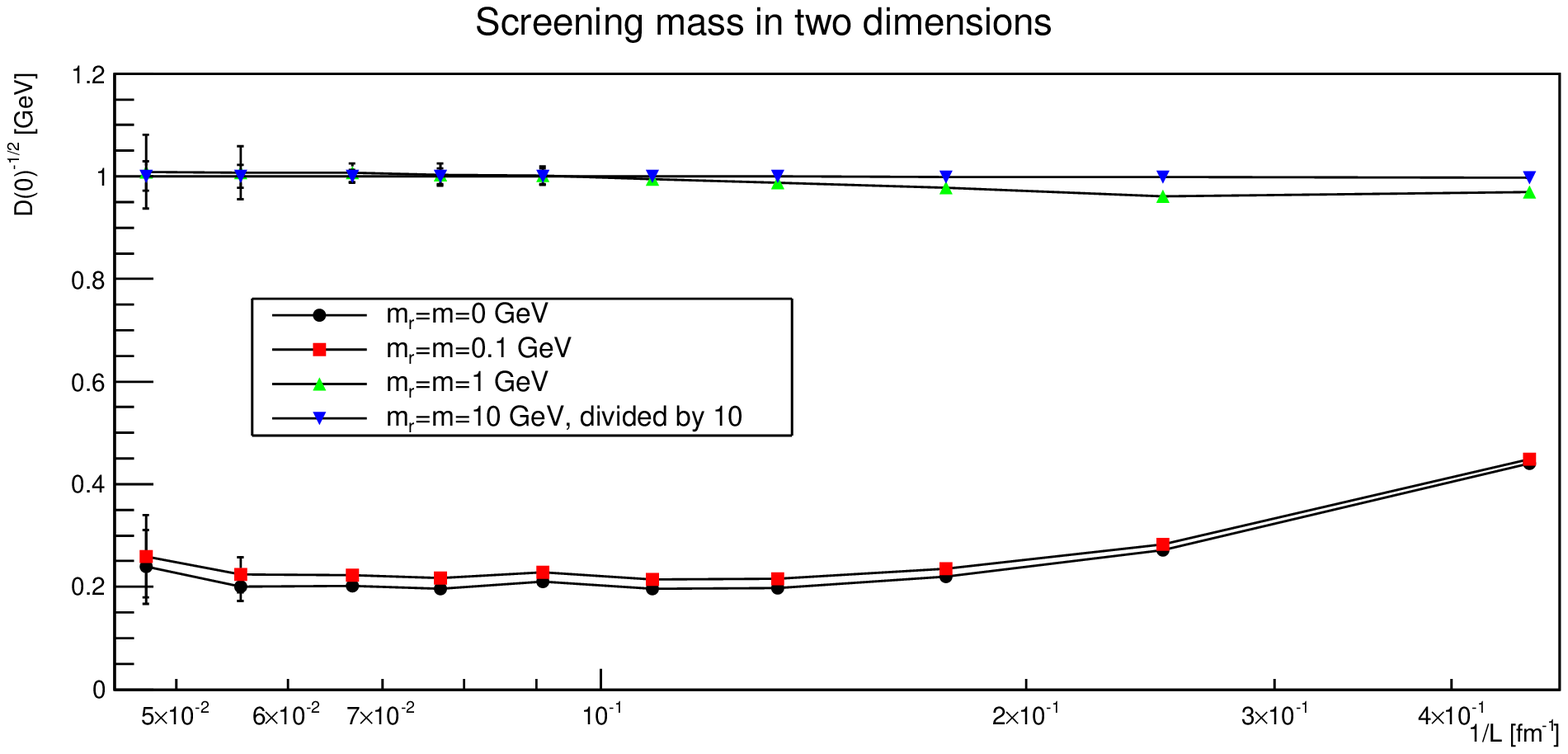}\\
\includegraphics[width=\linewidth]{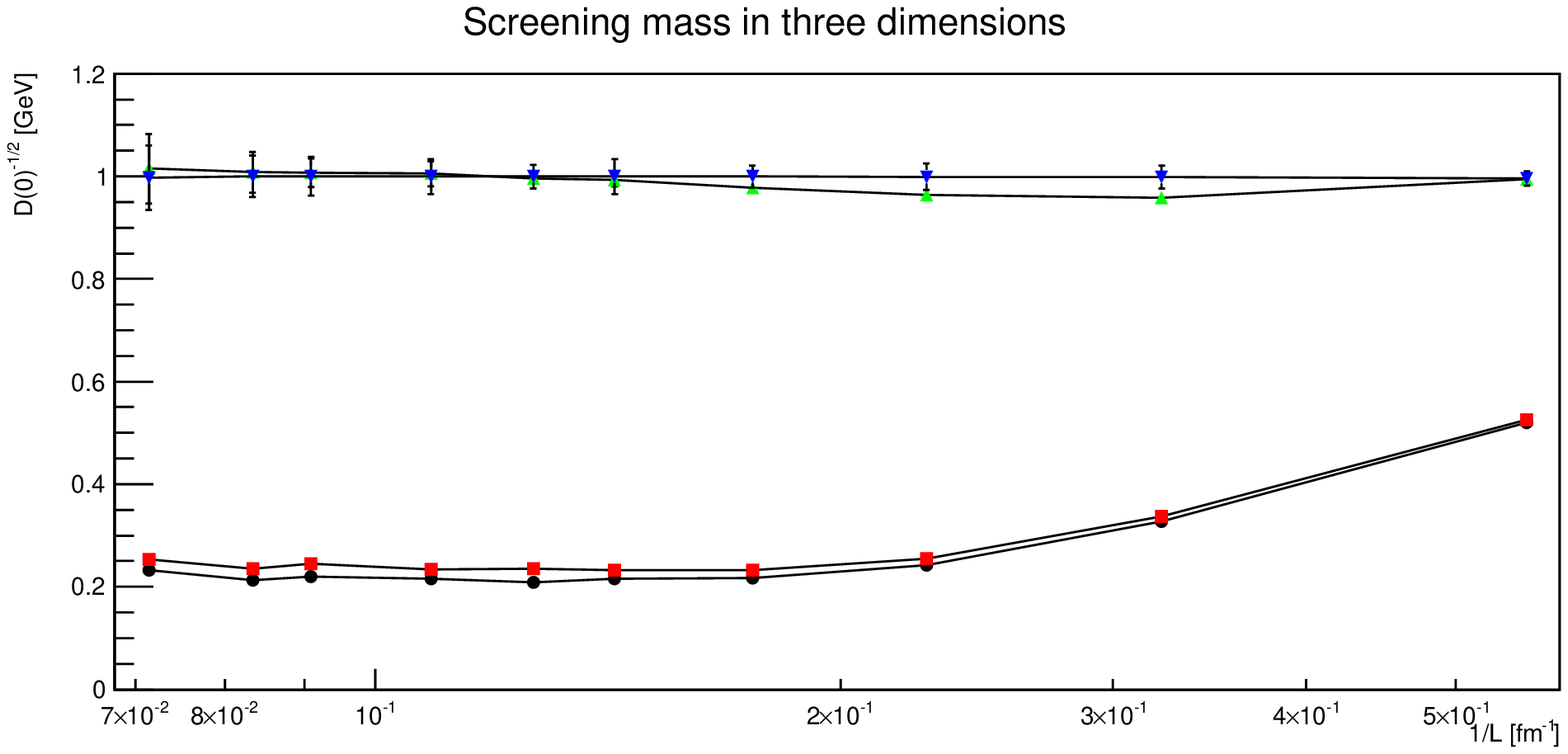}\\
\includegraphics[width=\linewidth]{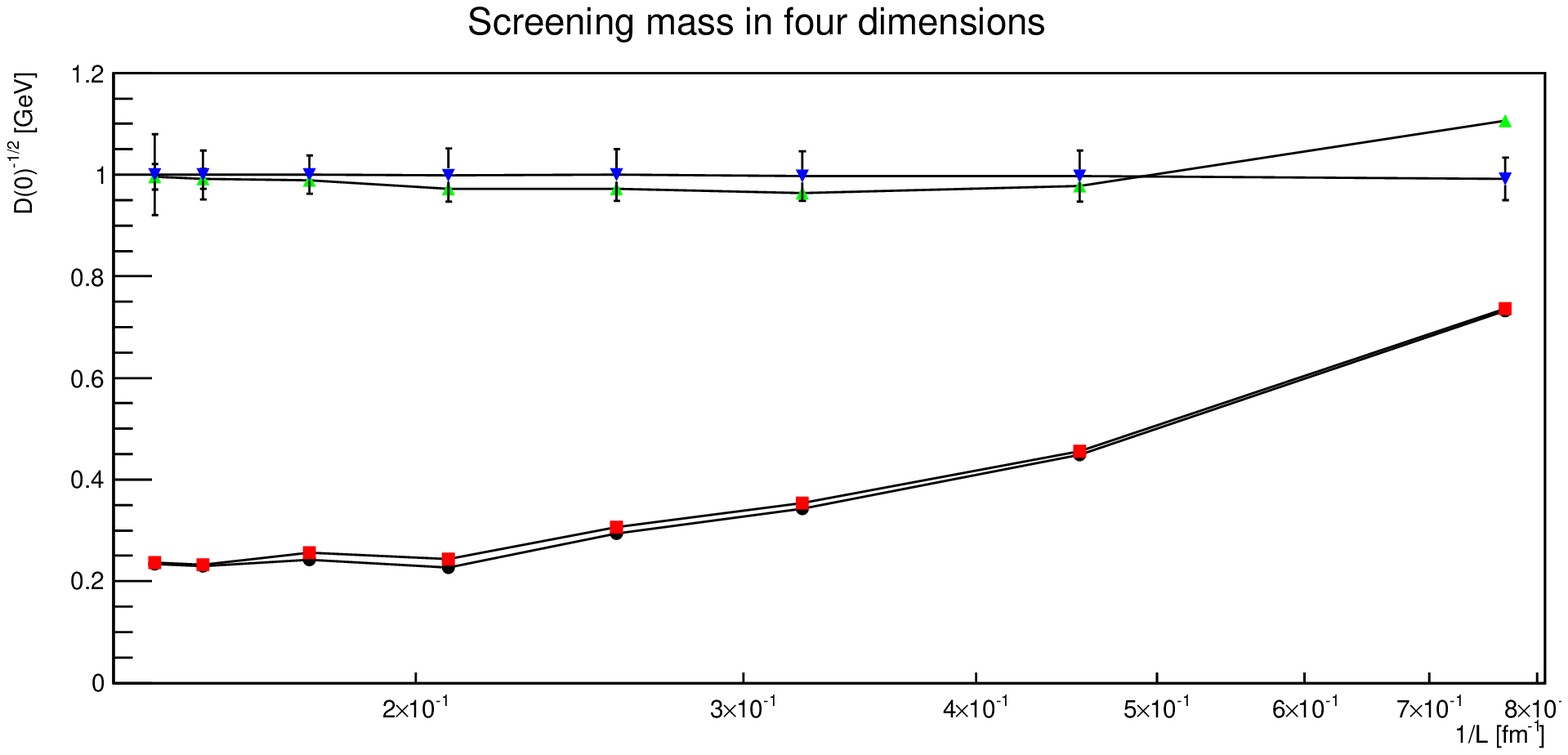}
\caption{\label{fig:d0}The value of the screening mass $D(0)^{-1/2}$ as a function of lattice extent in two (top panel), three (middle panel), and four dimensions (bottom panel).}
\end{figure}

More interesting is the observation that the propagator for $m=m_r=0$ GeV tends to a finite value in the infrared, which is also visible in the dressing functions, irrespective of the dimension. To emphasize this, the value of the propagators at zero momentum, or more precisely its screening mass $D(0)^{-1/2}$, is shown as a function of lattice extension in figure \ref{fig:d0}. Note that since the propagator has not been evaluated there, this value is obtained by a linear extrapolation of the propagator at the two lowest non-zero momenta.

This is interesting as it implies that even if the renormalization condition forces the propagators to behave like a massless one at $\mu$, the actual propagator in the infrared develops a non-zero screening mass, and thus a finite value for the propagator. Hence, even though the renormalization scheme and the tree-level mass enforce a massless behavior, the dynamics induce a screening or gaping at small momenta.

The value of this additional screening mass is somewhat erratically changing from volume to volume. This indicates that a mixture of discretization and volume effects as well as systematic effects in the renormalization process affect the actual value.

A similar effect is also seen for larger masses. In all cases, the renormalized propagator is below the tree-level value at small momenta, though the effect becomes smaller and smaller the larger the mass. In the case of $m=m_r=10$ GeV it is essentially gone. Indeed, the values of the propagators at small masses suggest a screening mass of the order of 200-250 MeV, which is surprisingly similar to the screening mass observed for quarks in the fundamental representation in the chiral and quenched limit \cite{Roberts:1994dr,Roberts:2015lja,Alkofer:2000wg,Fischer:2006ub}. This may indicate that the generation of screening masses is actual similar for bosons and fermions. However, in contrast to fermions the contribution from the screening mass seems to diminish with increasing renormalized mass, though the additive shift in the mass may overlay this to some extent.

The behavior of the propagators at higher momenta is then following more or less the expected pattern. At high mass the propagators also start to deviate again from the tree-level one. In four dimensions, this follows from the logarithmic decay due to the renormalization effects. In lower dimensions, this is somewhat unexpected, and in contrast to the gauge propagators \cite{Maas:2011se}. This is, however, likely due to the additional wave-function renormalization, which compensates partly for a self-energy contribution, and this discrepancy yields the observed effect: Due to asymptotic freedom, at large momenta all propagators in two and three dimensions tend to $D=1/(Zp^2)$, yielding $H(p)=1/Z$, rather than unity.

\subsection{Schwinger function and effective mass}\label{ss:ssda}

The Schwinger function
\be
\Delta(t)=\frac{1}{\pi}\int_0^\infty dp_0\cos(tp_0)D(p_0^2)=\frac{1}{a\pi}\frac{1}{N_t}\sum_{P_0=0}^{N_t-1}\cos\left(\frac{2\pi tP_0}{N_t}\right)D(P_0^2)\nn,
\ee
\no essentially the temporal correlator, is obtained from the renormalized propagator. The calculation is straightforward in principle, though requires obtaining the removed value at zero momentum. As above, this is obtained by a linear extrapolation of the propagator at the two lowest momenta. Especially for the larger masses this is relatively accurate, but induces some systematic error for the smaller masses on small physical volumes. However, this only adds a constant to the Schwinger function, which yields a negative contribution to the effective mass. This offset vanishes as a function of the physical volume.

These correlation functions have a very simple behavior, a sum of exponentials, for gauge-invariant Euclidean correlation functions on any finite lattice \cite{Seiler:1982pw}. The situation is quite different for gauge-dependent correlation functions \cite{Maas:2011se,Alkofer:2000wg,Alkofer:2003jj}: There are no general constraints.

From the Schwinger function the effective (time-dependent) mass
\be
m_\text{eff}=-\ln\frac{\Delta(t+a)}{\Delta(t)}\nn,
\ee
\no can be derived, which in the case of a simple exponential decay coincides with the usual mass. On a finite lattice, for any physical particle with a positive spectral function this effective mass is a monotonously decreasing function for $t\le L/2$, and corresponds to the effective mass at a given time. Eventually, at sufficiently long time, it is just the mass of the ground state.

If the effective mass is non-monotonous decreasing, the spectral function has necessarily negative contributions. Therefore it does not describe a physical particle.

\begin{figure}[!ht]
\includegraphics[width=\linewidth]{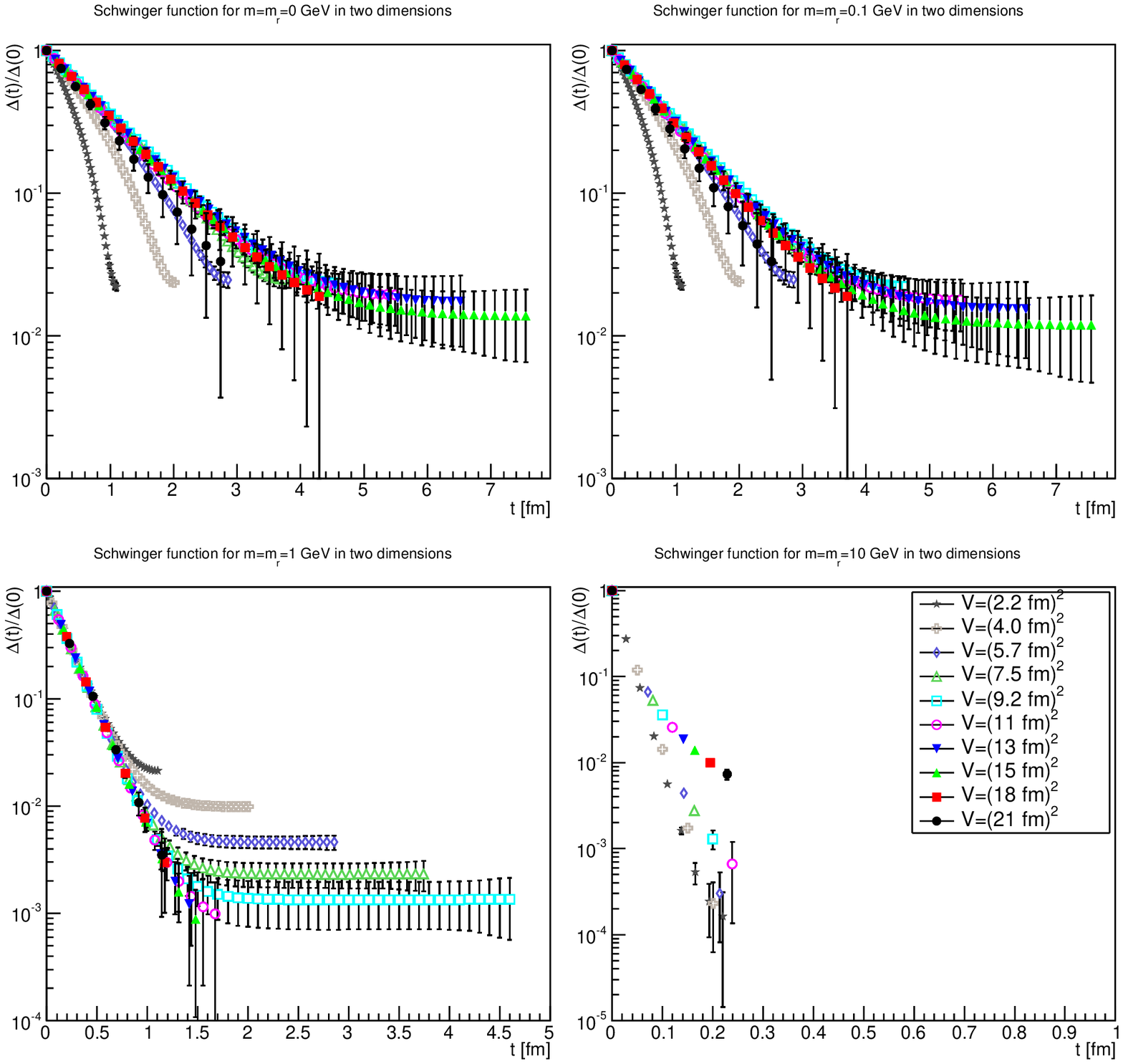}
\caption{\label{fig:a2d}The Schwinger function in two dimensions for $m=m_r=0$ GeV (top-left panel), $m=m_r=0.1$ GeV (top-right panel), $m=m_r=1$ GeV (bottom-left panel), and $m=m_r=10$ GeV (bottom-right panel). Points with a relative error larger than 100\% have been omitted.}
\end{figure}

\begin{figure}[!ht]
\includegraphics[width=\linewidth]{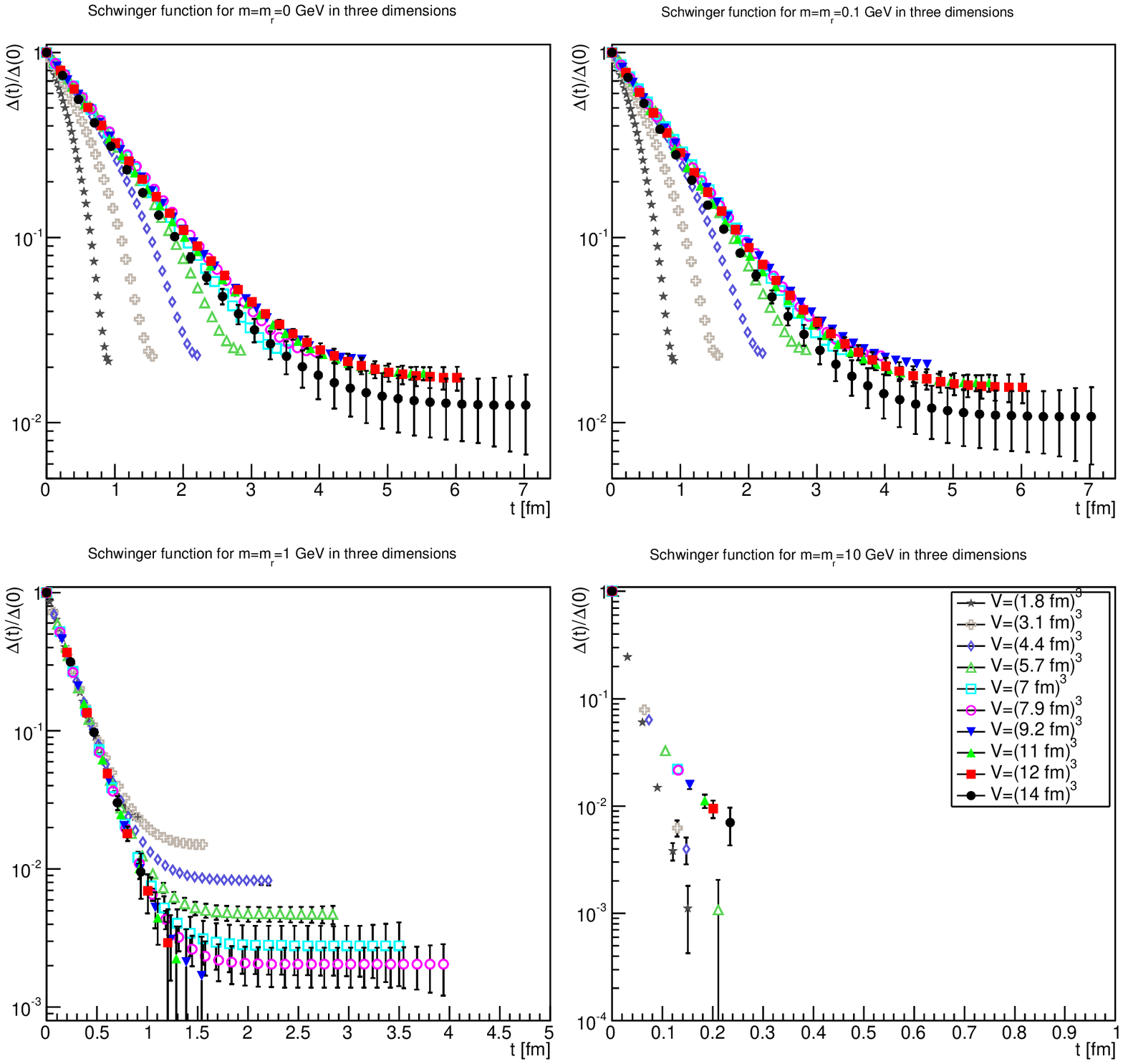}
\caption{\label{fig:a3d}The Schwinger function in three dimensions for $m=m_r=0$ GeV (top-left panel), $m=m_r=0.1$ GeV (top-right panel), $m=m_r=1$ GeV (bottom-left panel), and $m=m_r=10$ GeV (bottom-right panel). Points with a relative error larger than 100\% have been omitted.}
\end{figure}

\begin{figure}[!ht]
\includegraphics[width=\linewidth]{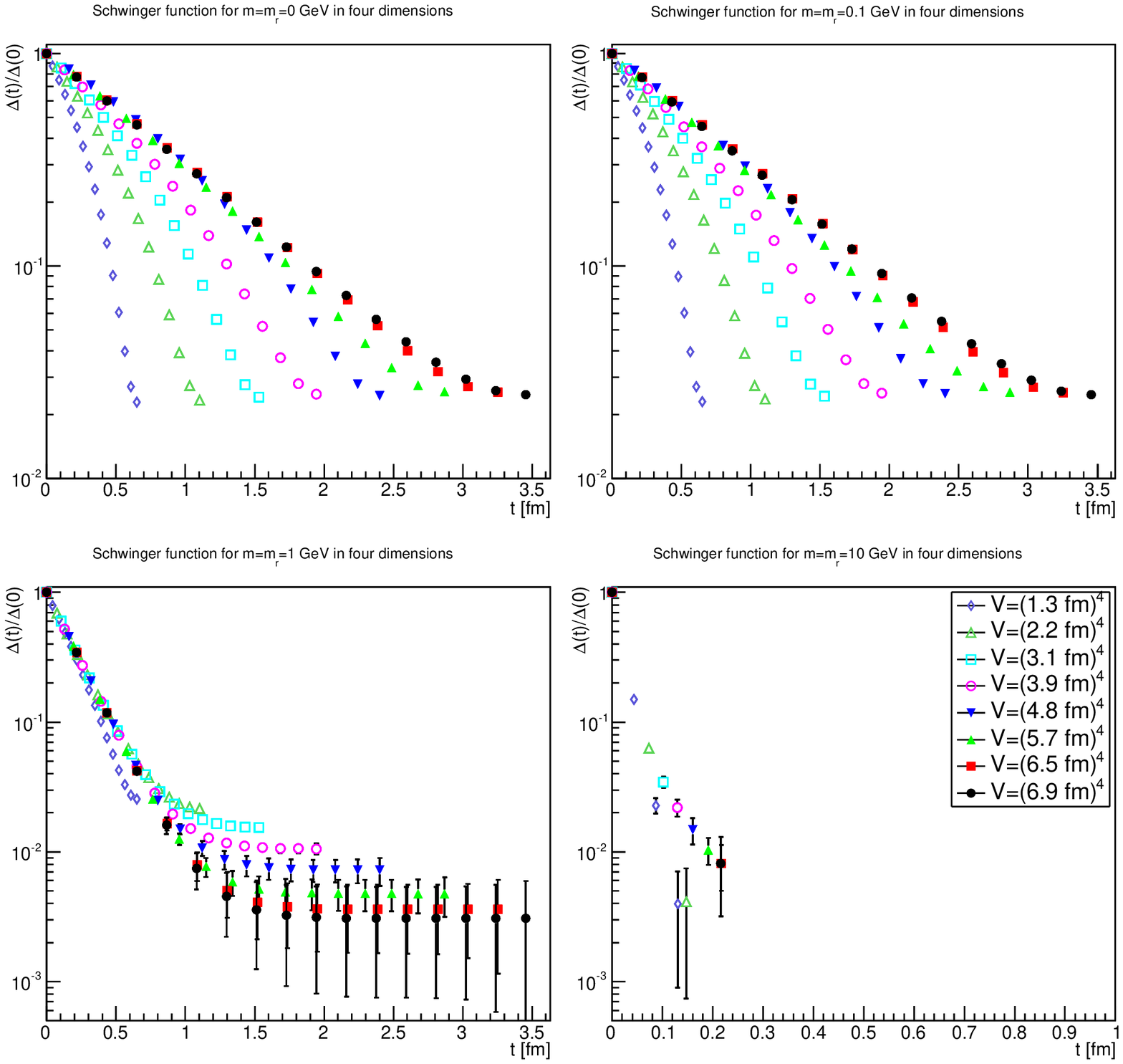}
\caption{\label{fig:a4d}The Schwinger function in four dimensions for $m=m_r=0$ GeV (top-left panel), $m=m_r=0.1$ GeV (top-right panel), $m=m_r=1$ GeV (bottom-left panel), and $m=m_r=10$ GeV (bottom-right panel). Points with a relative error larger than 100\% have been omitted.}
\end{figure}

\begin{figure}[!ht]
\includegraphics[width=\linewidth]{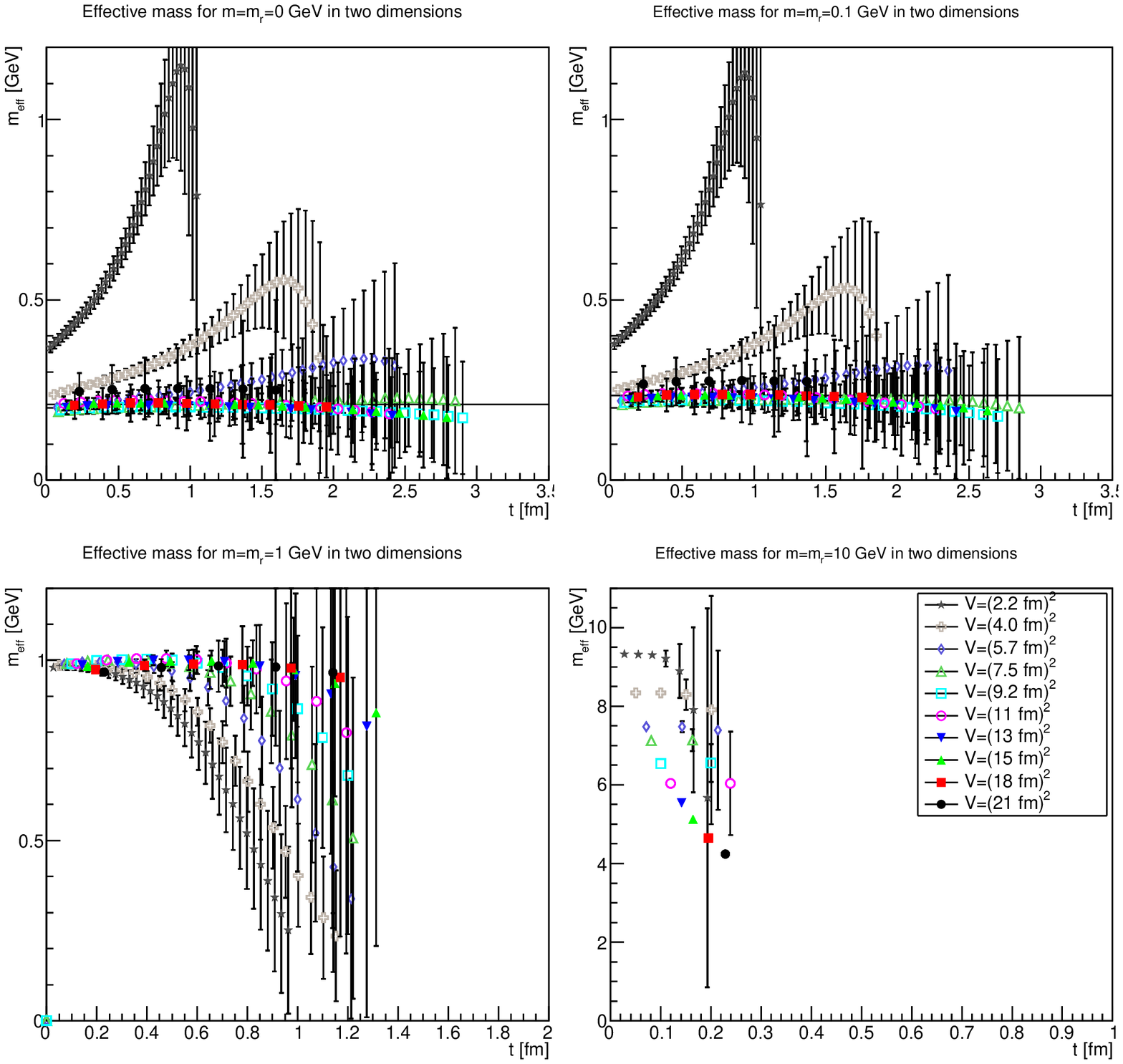}
\caption{\label{fig:m2d}The effective mass in two dimensions for $m=m_r=0$ GeV (top-left panel), $m=m_r=0.1$ GeV (top-right panel), $m=m_r=1$ GeV (bottom-left panel), and $m=m_r=10$ GeV (bottom-right panel). Points with a relative error larger than 100\% have been omitted. The lines in the top-left panel and the top-right panel correspond to 210 and 235 MeV, respectively.}
\end{figure}

\begin{figure}[!ht]
\includegraphics[width=\linewidth]{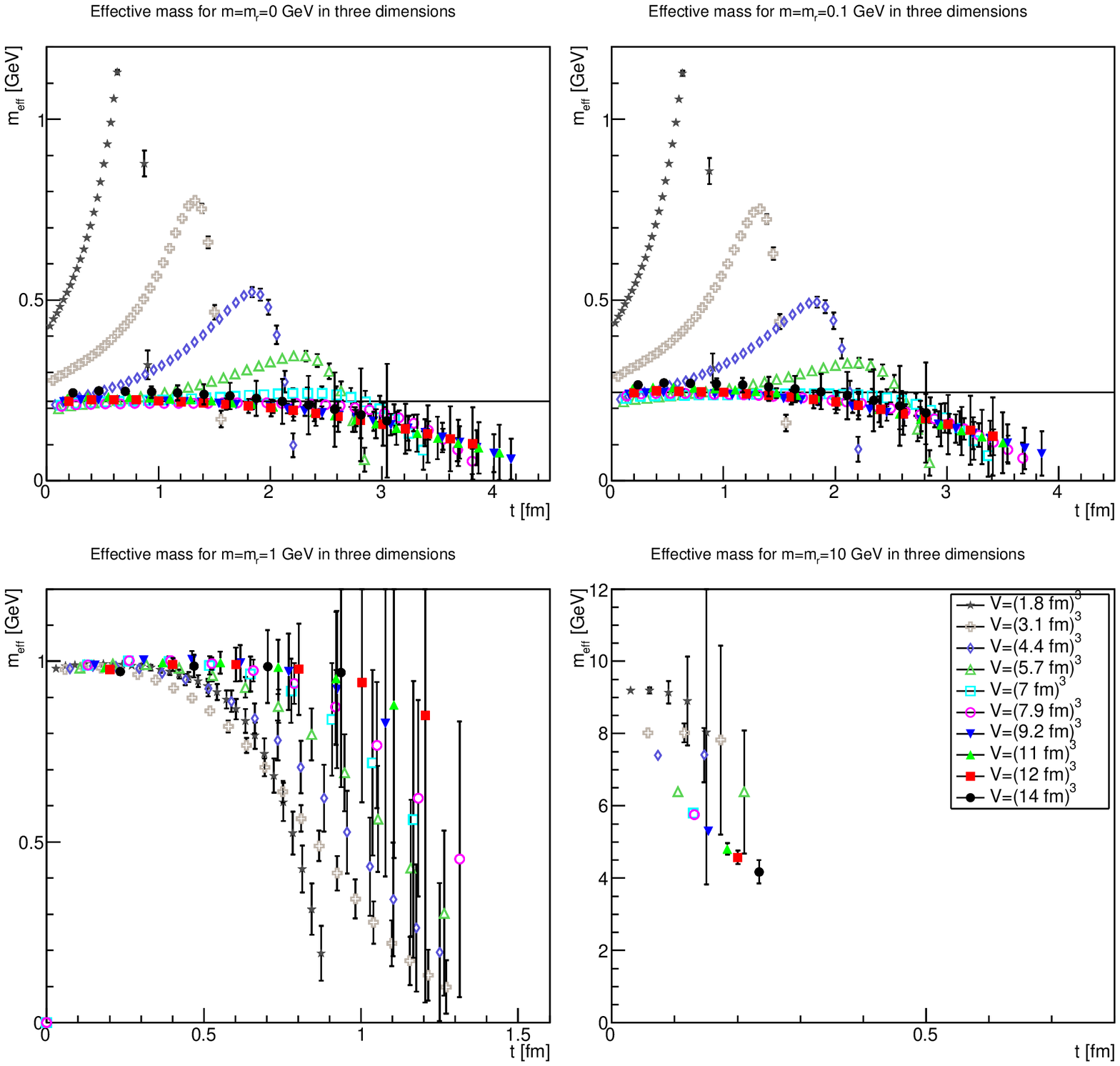}
\caption{\label{fig:m3d}The effective mass in three dimensions for $m=m_r=0$ GeV (top-left panel), $m=m_r=0.1$ GeV (top-right panel), $m=m_r=1$ GeV (bottom-left panel), and $m=m_r=10$ GeV (bottom-right panel). Points with a relative error larger than 100\% have been omitted. The lines in the top-left panel and the top-right panel correspond to 220 and 245 MeV, respectively.}
\end{figure}

\begin{figure}[!ht]
\includegraphics[width=\linewidth]{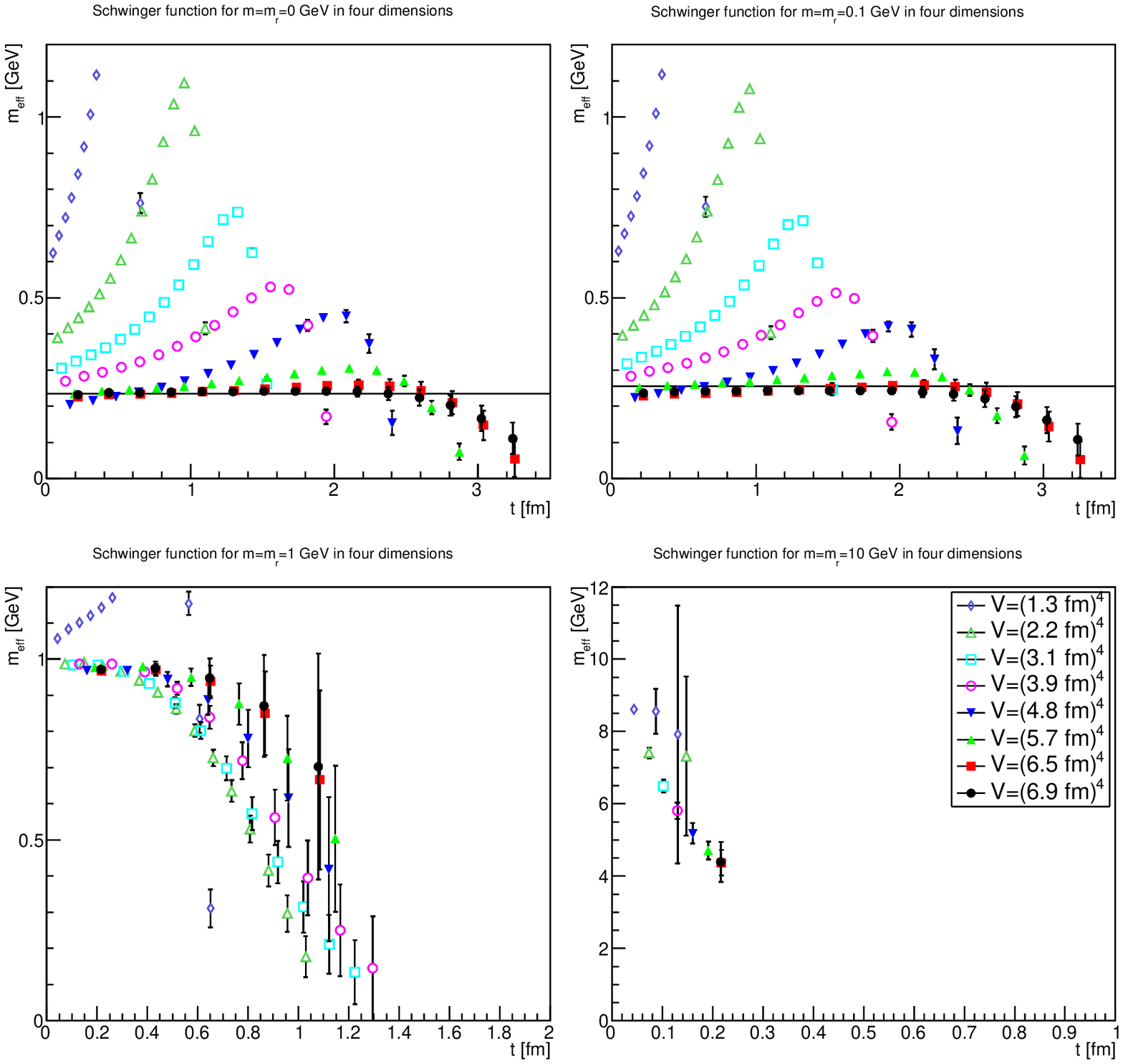}
\caption{\label{fig:m4d}The effective mass in four dimensions for $m=m_r=0$ GeV (top-left panel), $m=m_r=0.1$ GeV (top-right panel), $m=m_r=1$ GeV (bottom-left panel), and $m=m_r=10$ GeV (bottom-right panel). Points with a relative error larger than 100\% have been omitted. The lines in the top-left panel and the top-right panel correspond to 235 and 255 MeV, respectively.}
\end{figure}

The Schwinger function and the corresponding effective masses are shown in figures \ref{fig:a2d}-\ref{fig:a4d} and \ref{fig:m2d}-\ref{fig:m4d}, respectively.

The Schwinger function themselves show a number of interesting features. First of all, the results are quite similar, independent of dimensions. Secondly, finite-volume effects are visible, especially for small physical volumes. Since the physical volumes become smaller with increasing dimension, this effect is emphasized with increasing dimension. This effect is also much stronger for the smaller renormalized masses. In the case of 10 GeV, the drop is so sharp that for almost no relevant time distance a reasonable statement can be made before the signal drowns in noise.

The third observation is that the Schwinger function is positive for all times. This is in stark contrast to the gluon propagator \cite{Maas:2011se} and the (quenched) adjoint quark propagator \cite{August:2013jia}, which show such violations of positivity on a time scale of order 1 fm. Most interesting would be a comparison to the fundamental quark propagator, as both are in the same representation and both are affected in the same way by the Wilson confinement criterion. But here the situation is yet undecided \cite{Alkofer:2003jj}, though the indications also suggest a violation of positivity. It is also unclear whether a violation of positivity is seen for the unquenched fundamental scalar \cite{Maas:2013aia} for QCD-like regions of the phase diagram. For Higgs-like regions of the phase diagram there is no indication of a positivity violation.

Thus, the absence of an explicit positivity violation in the quenched case is an intriguing result, as it is so far the first correlator to show this behavior in a sufficiently reliable way. However, whether this has something to do with the absence of string-breaking in the quenched case is at this point at best speculation.

Though there is no such explicit violation of positivity, this by no means implies that the corresponding spectral density is positive. This is already visible by eye for small volumes, where the bending of the Schwinger function is clearly wrong for the two lightest masses. However, with increasing physical volume this type of wrong bending is reduced. There is also no obvious sign of it at the larger mass of 1 GeV.

Turning the view to the effective masses in figure \ref{fig:m2d}-\ref{fig:m4d} provides some more intriguing results. First of all, again the dependence is very similar in all dimensions, and thus no indication of a difference between dynamical and geometrical confinement.

Secondly, there is a very strong volume dependence. For small physical volumes, the effective mass bends upwards, which is not compatible with the properties of a physical particle. This behavior is substantially reduced the larger the volume up to the point where the behavior looks like that of a physical particle, with an essentially correctly curved effective mass. In fact, any deviation from the physical curvature becomes so small that statistical and systematic errors are too large to see if it remains. Such a behavior is not observed for either the gluon nor the adjoint quark propagator \cite{Maas:2011se,August:2013jia}, and was thus not expected.

The third interesting observation is that for the lightest two masses the so obtained effective mass is about 200-300 MeV, despite the renormalization condition attempts to force them to much smaller values. For the renormalized mass of 1 GeV, there is a small deviation downwards of the effective mass, before a strong suppression sets in. For the largest renormalized mass of 10 GeV, there are too few points to make any firm statement, but the reliable points are at large volumes clearly incompatible with an effective mass as large as the renormalized mass.

An investigation of the lattice spacing dependence shows that for the three lightest masses the lattice spacing has essentially no impact, but the effective mass for the largest mass substantially increases. Thus, this undershooting is most probably one of the expected discretization artifacts for such a large renormalized mass.

These observations, especially for the lightest mass, are very good in line with the observations for the propagators in momentum space in section \ref{ss:mom}, which also showed a stronger screening for the lighter two masses than just from the influence of the renormalized mass. Especially, it fits well with the results for the screening mass in figure \ref{fig:d0}. This suggests that there is an intrinsic mass scale in the infrared.

\subsection{Renormalization effects on the effective mass}\label{ss:sren}

To understand whether the mass scale, as well as the other properties of the effective mass, are something which could be interpreted as a feature of physics, it is necessary to understand the dependence on the necessary renormalization. Of course, even if this would signal any kind of interesting feature, this would still require to check the gauge-dependence of the results, which is beyond the scope of the present work.

\begin{figure}[!ht]
\includegraphics[width=\linewidth]{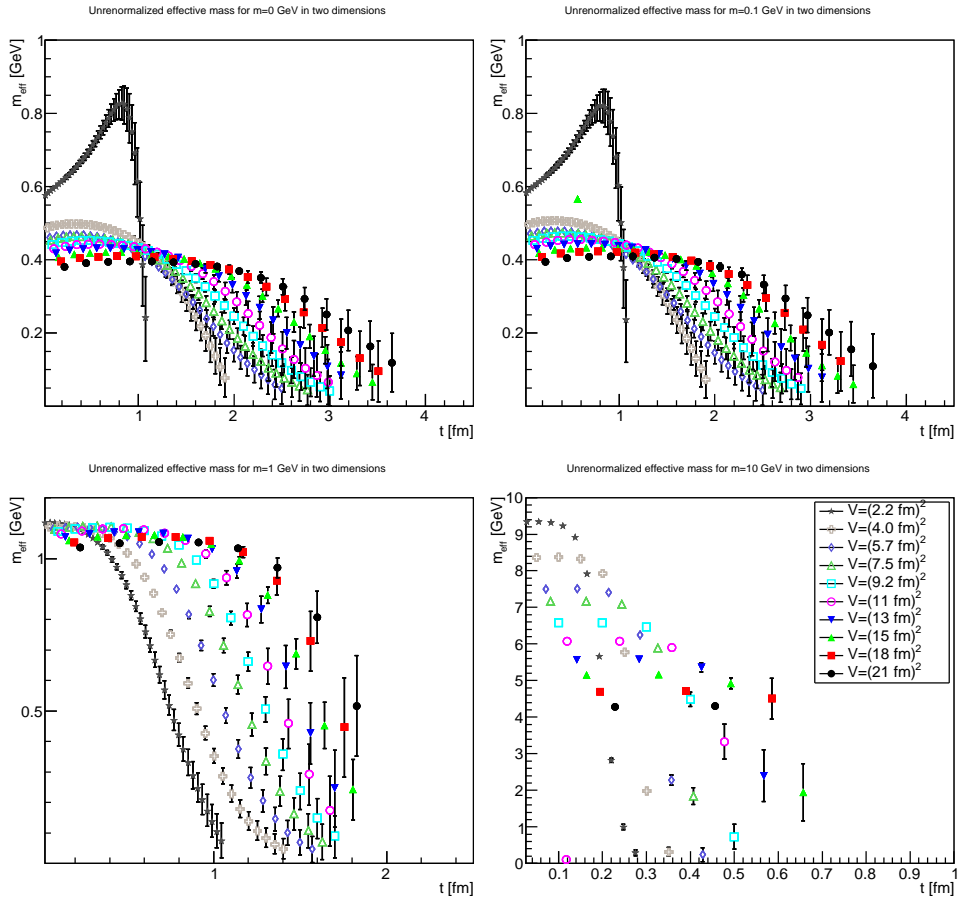}
\caption{\label{fig:um2d}The unrenormalized effective mass in two dimensions for $m=0$ GeV (top-left panel), $m=0.1$ GeV (top-right panel), $m=1$ GeV (bottom-left panel), and $m=10$ GeV (bottom-right panel). Points with a relative error larger than 100\% have been omitted.}
\end{figure}

\begin{figure}[!ht]
\includegraphics[width=\linewidth]{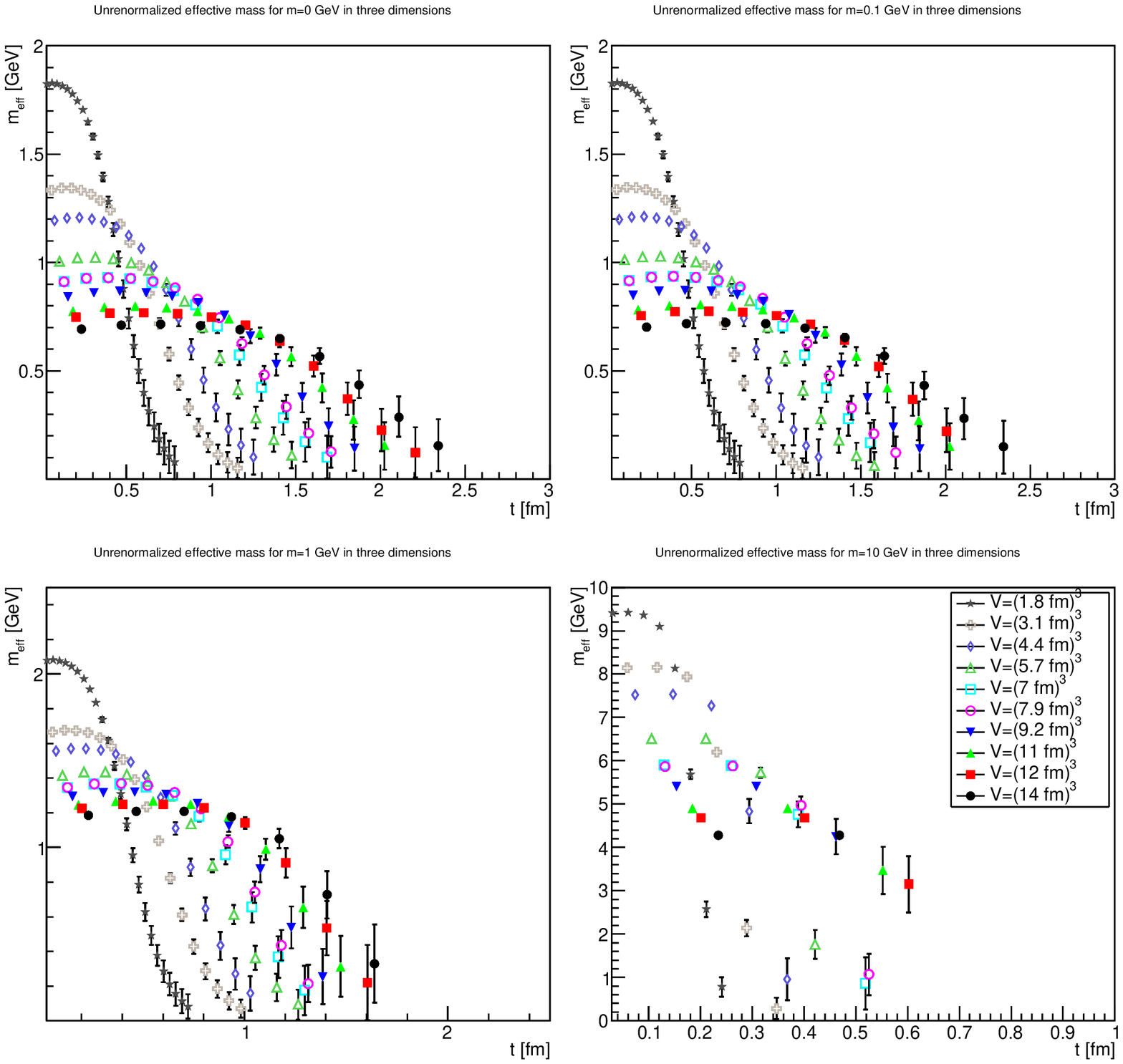}
\caption{\label{fig:um3d}The unrenormalized effective mass in three dimensions for $m=0$ GeV (top-left panel), $m=0.1$ GeV (top-right panel), $m=1$ GeV (bottom-left panel), and $m=10$ GeV (bottom-right panel). Points with a relative error larger than 100\% have been omitted.}
\end{figure}

\begin{figure}[!ht]
\includegraphics[width=\linewidth]{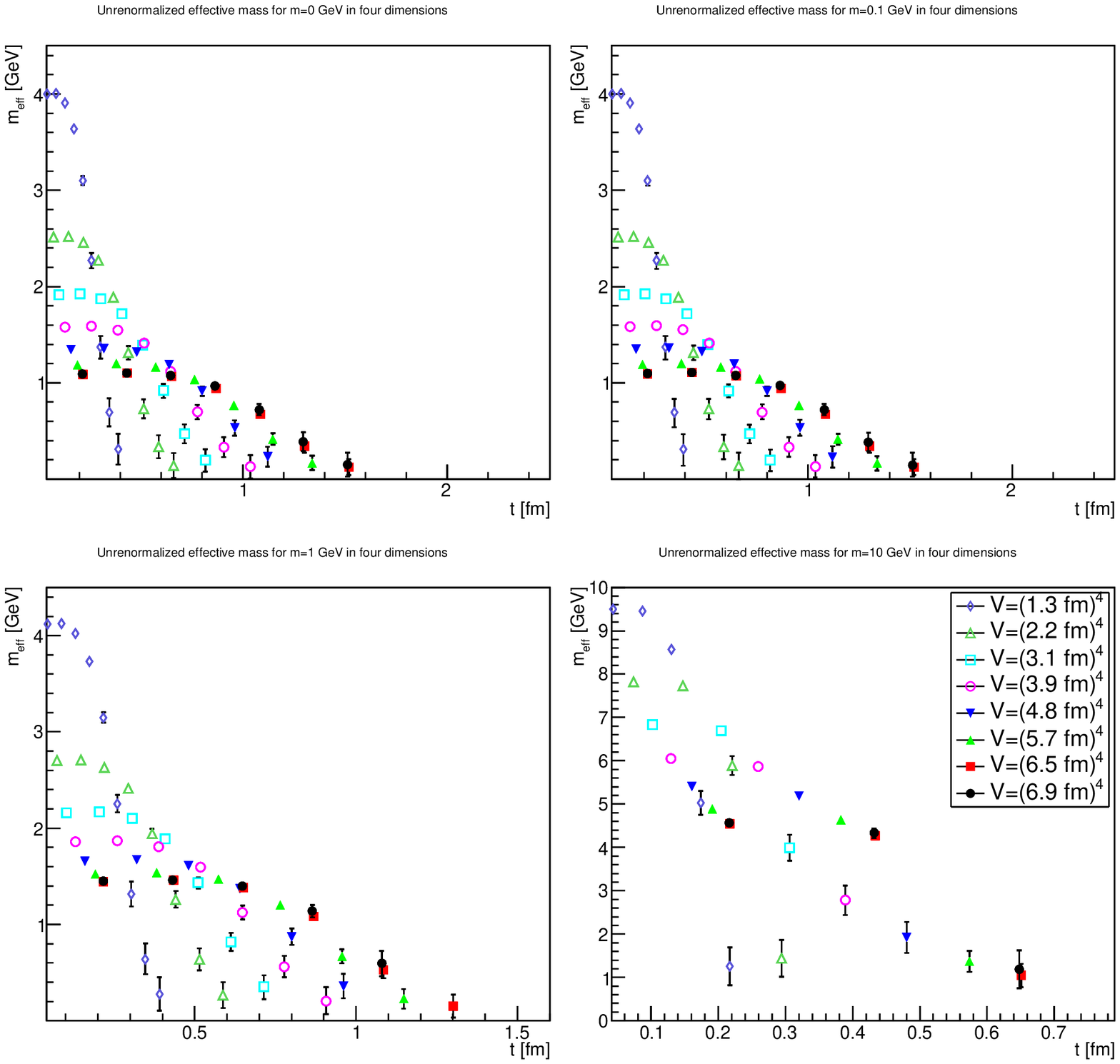}
\caption{\label{fig:um4d}The unrenormalized effective mass in four dimensions for $m=0$ GeV (top-left panel), $m=0.1$ GeV (top-right panel), $m=1$ GeV (bottom-left panel), and $m=10$ GeV (bottom-right panel). Points with a relative error larger than 100\% have been omitted.}
\end{figure}

If the effective mass should have a physical property it should be re\-norm\-al\-izati\-on-independent, and therefore already be a feature of the unrenormalized mass function. This possibility is investigated in figures\footnote{The fact that the errors are substantially smaller at long times than for the renormalized propagator is due to the error propagation from the mass renormalization, even though these errors are just at the few percent level. The error for the wave-function renormalization also contribute, but they are much smaller.} \ref{fig:um2d}-\ref{fig:um4d}. The results show a rather clear trend. First of all, in all cases it is now visible that the effective mass has an unphysical bending. Also, there is a substantial drift with the physical volume, which lets the mass drift towards zero. Both facts already suggest that the effective mass is not a physical quantity.

\begin{figure}[!ht]
\includegraphics[width=\linewidth]{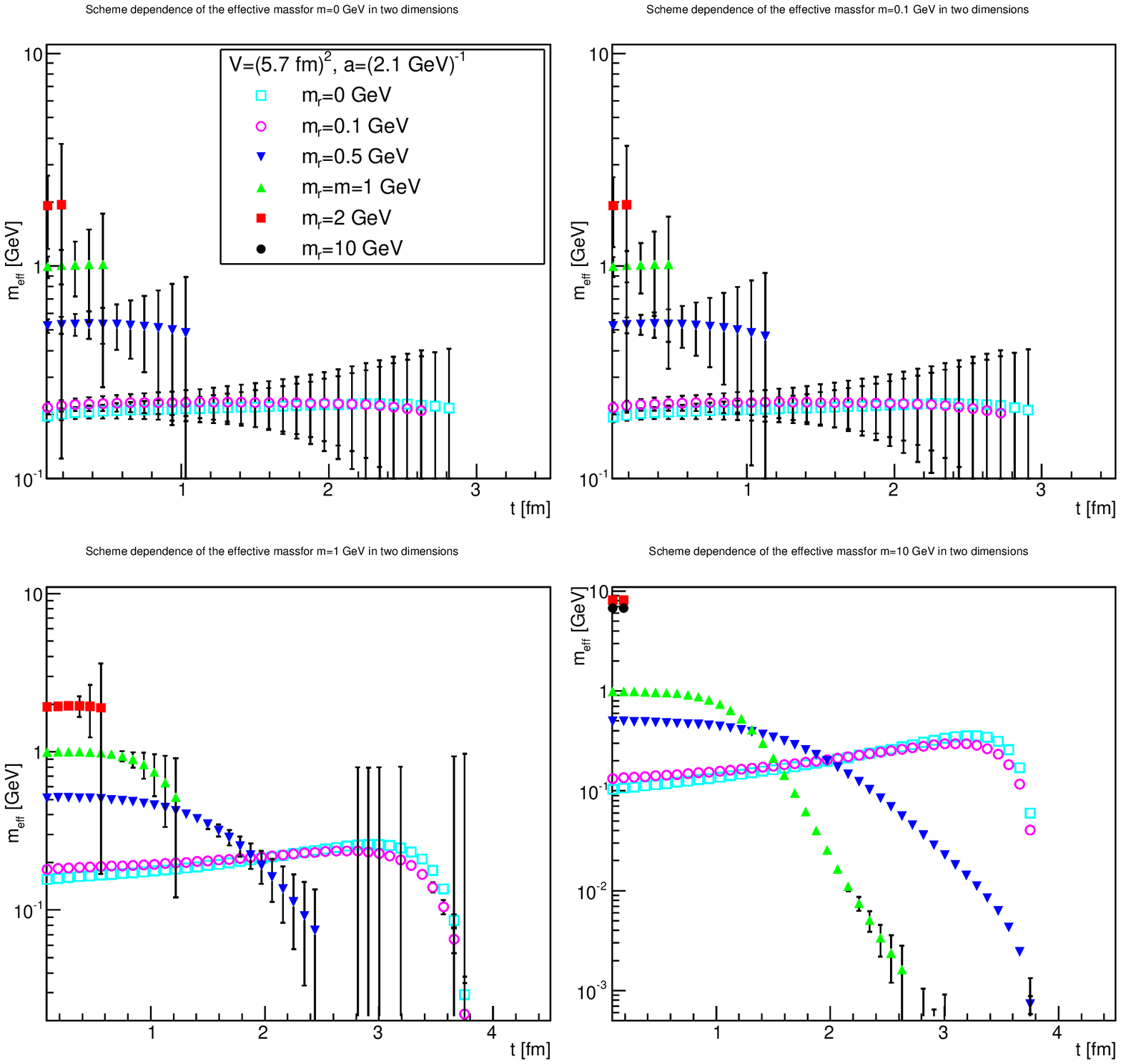}
\caption{\label{fig:sdm2d}The scheme-dependence of the effective mass in two dimensions for $m=0$ GeV (top-left panel), $m=0.1$ GeV (top-right panel), $m=1$ GeV (bottom-left panel), and $m=10$ GeV (bottom-right panel). Points with a relative error larger than 100\% have been omitted.}
\end{figure}

\begin{figure}[!ht]
\includegraphics[width=\linewidth]{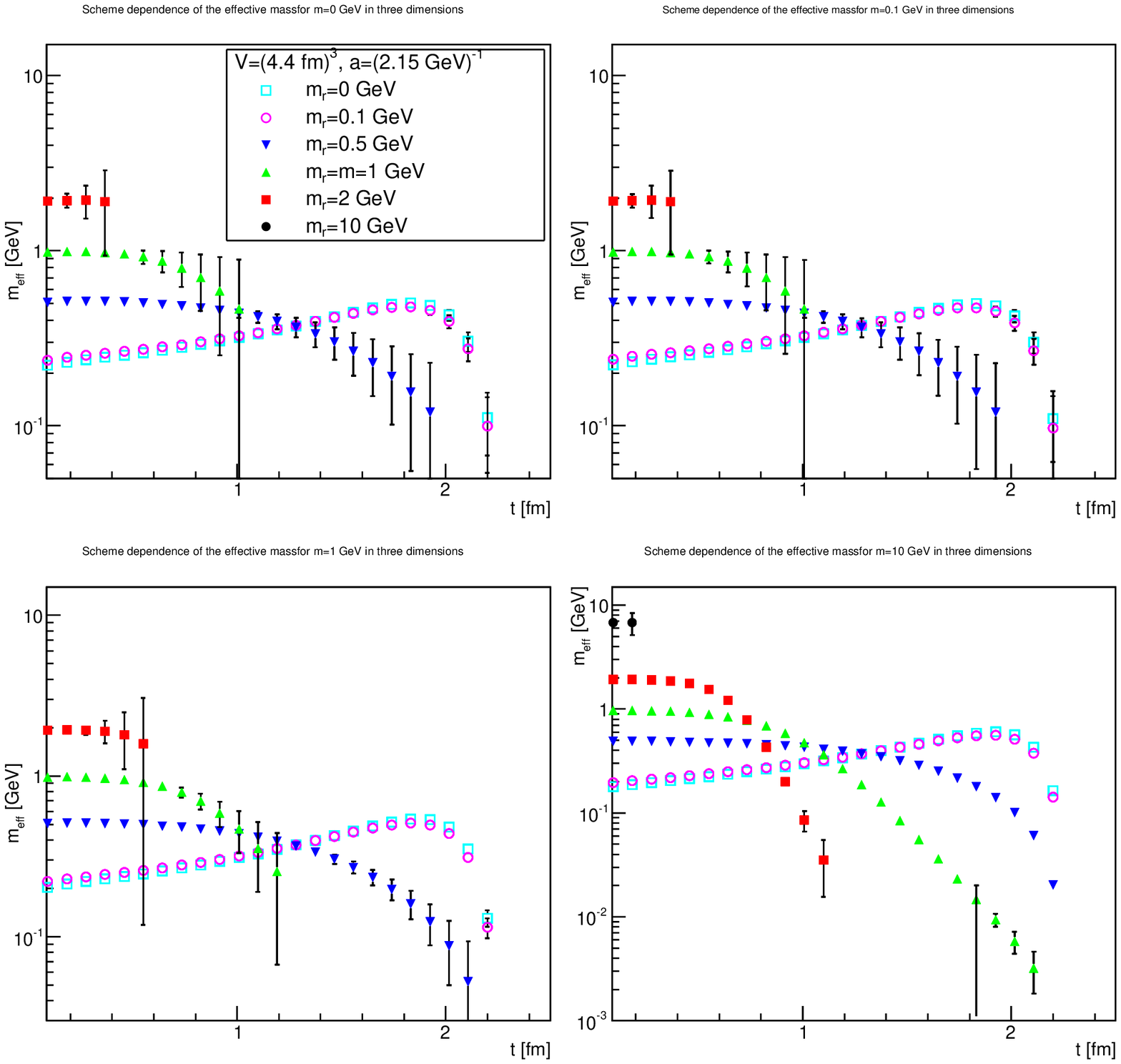}
\caption{\label{fig:sdm3d}The scheme-dependence of the effective mass in three dimensions for $m=0$ GeV (top-left panel), $m=0.1$ GeV (top-right panel), $m=1$ GeV (bottom-left panel), and $m=10$ GeV (bottom-right panel). Points with a relative error larger than 100\% have been omitted.}
\end{figure}

\begin{figure}[!ht]
\includegraphics[width=\linewidth]{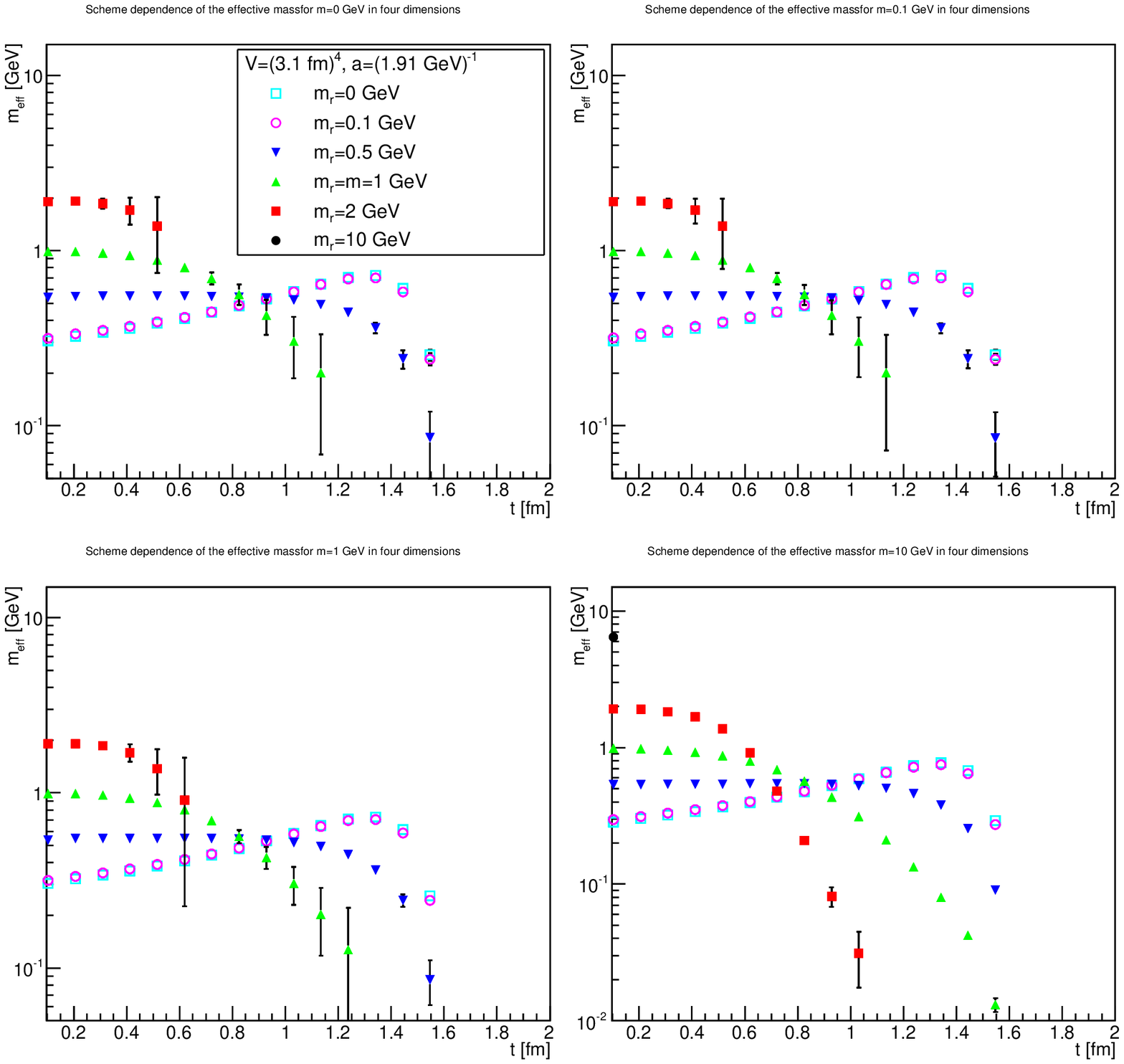}
\caption{\label{fig:sdm4d}The scheme-dependence of the effective mass in four dimensions for $m=0$ GeV (top-left panel), $m=0.1$ GeV (top-right panel), $m=1$ GeV (bottom-left panel), and $m=10$ GeV (bottom-right panel). Points with a relative error larger than 100\% have been omitted.}
\end{figure}

This becomes even more evident when considering the scheme dependence of the effective mass shown in figures \ref{fig:sdm2d}-\ref{fig:sdm4d}. It is quite visible that the effective mass strongly drifts as a function of the renormalized mass. There is a particular interesting observation to be made.

The effective mass is found to be of the order of the renormalized mass, as long as the renormalized mass is larger than about 200 MeV. Below this, the effective mass is not lowered further. This suggests again an intrinsic scale, independent of the bare or renormalized mass, of about this size. That the tree-level mass plays no substantial role in this is probably the most unexpected feature. It suggests that the long-distance behavior is only dominated by the interaction, not by the particle properties, as even a classically heavy particle is at most impeded in its movement by the mass scale introduced by the interaction.

This scale is not a real mass scale, as the unphysical bending of the effective mass shows. In fact, this bending increases the lower the renormalized mass and the lower the bare mass. At sufficiently large renormalized mass, the effective mass shows no sign of an unphysical bending, while it is very strong, even for large bare masses, at small renormalized mass. This unphysical bending is reduced when increasing the physical volume, as is seen in figures \ref{fig:m2d}-\ref{fig:m4d}, where the same unphysical bending is observed for small renormalized masses and small physical volumes.

The renormalized mass therefore seems to screen the unphysical properties of the correlator. The systematic uncertainty due to the extrapolation of the propagator to zero momentum cannot be the source of this problem, as this only decreases, but not increases, the correlator. The best explanation is the presence of a volume-dependent effective mass, which, however, only hampers long distance propagation. Therefore, the mass increases when going to long times. This is very different from the behavior of any finite-volume artifact for physical particles \cite{Luscher:1985dn,Luscher:1986pf}.

Speculating, the results are consistent with an effective mass, and correlator, which becomes more and more in line with a physical correlator with a single, simple mass pole in the infinite-volume limit. However, the particle is still gauge-dependent, and the mass scheme-dependent, and therefore neither is observable. Furthermore, there is a lower limit to this potential pole mass, below which no renormalization scheme can lower the mass.

It would be quite interesting to check for such a behavior also in the full case, especially in Higgs-like physics, along the lines of \cite{Maas:2013aia}. In the context of the Fr\"ohlich-Morchio-Strocchi mechanism \cite{Maas:2013aia,Frohlich:1980gj,Frohlich:1981yi,Maas:2012tj} any such lower mass bound for the pole mass could have rather interesting implications for the validity of perturbative spectrum calculations.

\section{Conclusions}\label{s:sum}

Summarizing, a systematic study of the quenched, fundamental scalar propagator has been performed in two, three, and four dimensions. The renormalization properties coincide with those of perturbation theory, as anticipated in an asymptotically free theory.

The analytical structure, however, is less trivial to interpret. It appears that the propagator behaves more and more like that of an ordinary particle with a single mass pole the larger the physical volume is, though at any finite volume it retains unphysical contributions. Furthermore, the pole mass, for sufficiently large renormalized mass, coincides with the renormalized mass, provided the lattice spacing is sufficiently small to resolve the corresponding scale. However, it appears not to be possible to lower the pole mass by the renormalization condition below a certain threshold, independent of the bare mass. This threshold is, rather independent of the dimensionality, about 200-250 MeV. Thus, the propagator seems to have an intrinsic non-perturbative mass scale. This may be due to a similar effect as the constituent (screening) mass of quarks \cite{Alkofer:2003jj}. A direct comparison is, due to the additive mass renormalization, not trivial.

Still, even if true, this does not make the pole mass of the fundamental scalar any more physical, as a change in renormalization scheme can shift it to larger masses. It is still a tantalizing question whether a similar intrinsic mass scale exists in the unquenched case, where this has not yet been investigated systematically. It would also be very interesting to see how the situation in the quenched adjoint case is \cite{Greensite:2006ns,Macher:2011ad,Capri:2012ah}, and whether any differences are present, a question which is currently under investigation \cite{Maas:unpublished,Maas:2011yx}.\\

\no{\bf Acknowledgments}\\

This work was supported by the DFG under grant numbers MA 3935/5-1, MA-3935/8-1 (Heisenberg program) and the FWF under grant number M1099-N16. Simulations were performed on the HPC clusters at the Universities of Jena and Graz. The author is grateful to the HPC teams for the very good performance of the clusters. The ROOT framework \cite{Brun:1997pa} has been used in this project.

\appendix

\section{Lattice setups}\label{a:ls}

The various lattice setups are listed in table \ref{tcgf}. The determination of the lattice spacings has been performed as in \cite{Maas:2014xma}.

\begin{longtable}{|c|c|c|c|c|c|c|c|}
\caption{\label{tcgf}Number and parameters of the configurations used, ordered by dimension, lattice spacing, and physical volume. In all cases $2(10N+100(d-1))$ thermalization sweeps and $2(N+10(d-1))$ decorrelation sweeps of mixed updates \cite{Cucchieri:2006tf} have been performed, and auto-correlation times of local observables have been monitored to be at or below one sweep. The number of configurations were selected such as to have a reasonable small statistical error for the renormalization constants determined in section \ref{s:ren}. The value $m_0$ denotes the value of the mass parameter in \pref{cov} to yield a tree-level mass of 1 GeV. The other tree-level masses are obtained by multiplying or dividing this number by 10, or setting it to zero for tree-level mass zero.}\\
\hline
$d$	& $N$	& $\beta$	& $a$ [fm] & $a^{-1}$ [GeV]	& L [fm]	&  $m_0$	& config.	\endfirsthead
\hline
\multicolumn{8}{|l|}{Table \ref{tcgf} continued}\\
\hline
$d$	& $N$	& $\beta$	& $a$ [fm] & $a^{-1}$ [GeV]	& L [fm]	&  $m_0$	& config.	\endhead
\hline
\multicolumn{8}{|r|}{Continued on next page}\\
\hline\endfoot
\endlastfoot
\hline
2	& 92	& 6.23	& 0.228		& 0.863	& 21	& 1.159		& 2146	\cr
\hline
2	& 80	& 6.40	& 0.225		& 0.875	& 18	& 1.143		& 3957	\cr
\hline
2	& 58	& 6.45	& 0.224		& 0.879	& 13	& 1.138		& 3386	\cr
\hline
2	& 18	& 6.55	& 0.222		& 0.886	& 4.0	& 1.129		& 3661	\cr
\hline
2	& 34	& 6.64	& 0.221		& 0.893	& 7.5	& 1.120		& 2970	\cr
\hline
2	& 68	& 6.64	& 0.221		& 0.893	& 15	& 1.120		& 3456	\cr
\hline
2	& 10	& 6.68	& 0.220		& 0.895	& 2.2	& 1.117		& 2192	\cr
\hline
2	& 50	& 6.68	& 0.220		& 0.895	& 11	& 1.117		& 3299	\cr
\hline
2	& 26	& 6.72	& 0.219		& 0.898	& 5.7	& 1.113		& 3410	\cr
\hline
2	& 42	& 6.73	& 0.219		& 0.900	& 9.2	& 1.112		& 3370	\cr
\hline
2	& 92	& 8.33	& 0.196		& 1.01	& 18	& 0.9933	& 1958	\cr
\hline
2	& 68	& 8.70	& 0.191		& 1.03	& 13	& 0.9708	& 3456	\cr
\hline
2	& 58	& 8.83	& 0.190		& 1.04	& 11	& 0.9632	& 3386	\cr
\hline
2	& 80	& 9.03	& 0.188		& 1.05	& 15	& 0.9519	& 3597	\cr
\hline
2	& 50	& 9.36	& 0.184		& 1.07	& 9.2	& 0.9341	& 3174	\cr
\hline
2	& 42	& 9.91	& 0.179		& 1.10	& 7.5	& 0.9066	& 3433	\cr
\hline
2	& 34	& 11.1	& 0.168		& 1.17	& 5.7	& 0.8543	& 2950	\cr
\hline
2	& 92	& 11.7	& 0.164		& 1.20	& 15	& 0.8312	& 4994	\cr
\hline
2	& 80	& 11.8	& 0.163		& 1.21	& 13	& 0.8275	& 3498	\cr
\hline
2	& 68	& 11.9	& 0.162		& 1.21	& 11	& 0.8239	& 3456	\cr
\hline
2	& 58	& 12.4	& 0.159		& 1.24	& 9.2	& 0.8065	& 3304	\cr
\hline
2	& 26	& 13.1	& 0.154		& 1.28	& 4.0	& 0.7838	& 3410	\cr
\hline
2	& 50	& 13.8	& 0.150		& 1.31	& 7.5	& 0.7629	& 3174	\cr
\hline
2	& 92	& 15.5	& 0.142		& 1.39	& 13	& 0.7185	& 2146	\cr
\hline
2	& 80	& 16.3	& 0.138		& 1.43	& 11	& 0.7001	& 2279	\cr
\hline
2	& 42	& 16.8	& 0.136		& 1.45	& 5.7	& 0.6893	& 3350	\cr
\hline
2	& 68	& 16.9	& 0.135		& 1.46	& 9.2	& 0.6872	& 3420	\cr
\hline
2	& 58	& 18.4	& 0.130		& 1.52	& 7.5	& 0.6578	& 3304	\cr
\hline
2	& 18	& 20.6	& 0.122		& 1.61	& 2.2	& 0.6208	& 3660	\cr
\hline
2	& 92	& 21.5	& 0.120		& 1.65	& 11	& 0.6074	& 1924	\cr
\hline
2	& 34	& 22.2	& 0.118		& 1.67	& 4.0	& 0.5974	& 2970	\cr
\hline
2	& 80	& 23.2	& 0.115		& 1.71	& 9.2	& 0.5841	& 3498	\cr
\hline
2	& 50	& 23.6	& 0.114		& 1.73	& 5.7	& 0.5791	& 3351	\cr
\hline
2	& 68	& 25.2	& 0.110		& 1.79	& 7.5	& 0.5600	& 3420	\cr
\hline
2	& 92	& 30.5	& 0.100		& 1.97	& 9.2	& 0.5082	& 4234	\cr
\hline
2	& 58	& 31.6	& 0.0983	& 2.00	& 5.7	& 0.4991	& 3300	\cr
\hline
2	& 42	& 33.6	& 0.0953	& 2.07	& 4.0	& 0.4838	& 3680	\cr
\hline
2	& 80	& 34.7	& 0.0938	& 2.10	& 7.5	& 0.4759	& 3498	\cr
\hline
2	& 26	& 42.4	& 0.0847	& 2.33	& 2.2	& 0.4300	& 2720	\cr
\hline
2	& 68	& 43.2	& 0.0839	& 2.35	& 5.7	& 0.4260	& 3505	\cr
\hline
2	& 92	& 45.7	& 0.0816	& 2.42	& 7.5	& 0.4140	& 3848	\cr
\hline
2	& 50	& 47.4	& 0.0801	& 2.46	& 4.0	& 0.4064	& 3215	\cr
\hline
2	& 80	& 59.7	& 0.0713	& 2.76	& 5.7	& 0.3618	& 3505	\cr
\hline
2	& 58	& 63.7	& 0.0690	& 2.86	& 4.0	& 0.3501	& 3276	\cr
\hline
2	& 34	& 72.3	& 0.0647	& 3.04	& 2.2	& 0.3285	& 3549	\cr
\hline
2	& 68	& 87.3	& 0.0589	& 3.35	& 4.0	& 0.2988	& 3472	\cr
\hline
2	& 42	& 110	& 0.0524	& 3.76	& 2.2	& 0.2660	& 3122	\cr
\hline
2	& 80	& 120	& 0.0502	& 3.93	& 4.0	& 0.02546	& 3631	\cr
\hline
2	& 50	& 155	& 0.0441	& 4.47	& 2.2	& 0.2239	& 3105	\cr
\hline
2	& 58	& 209	& 0.0380	& 5.19	& 2.2	& 0.1928	& 3304	\cr
\hline
2	& 68	& 287	& 0.0324	& 6.08	& 2.2	& 0.1644	& 2006	\cr
\hline
2	& 80	& 398	& 0.0275	& 7.16	& 2.2	& 0.1396	& 1760	\cr
\hline
\hline
3	& 60	& 3.30	& 0.234		& 0.841	& 14	& 1.189		& 2752	\cr
\hline
3	& 48	& 3.35	& 0.230		& 0.858	& 11	& 1.166		& 1800	\cr
\hline
3	& 8	& 3.40	& 0.225		& 0.874	& 1.8	& 1.144		& 3000	\cr
\hline
3	& 54	& 3.43	& 0.223		& 0.884	& 12	& 1.131		& 2295	\cr
\hline
3	& 14	& 3.44	& 0.222		& 0.887	& 3.1	& 1.127		& 3600	\cr
\hline
3	& 20	& 3.46	& 0.220		& 0.894	& 4.4	& 1.119		& 3160	\cr
\hline
3	& 26	& 3.47	& 0.220		& 0.897	& 5.7	& 1.115		& 2840	\cr
\hline
3	& 36	& 3.47	& 0.220		& 0.897	& 7.9	& 1.115		& 3300	\cr
\hline
3	& 42	& 3.47	& 0.220		& 0.897	& 9.2	& 1.115		& 1725	\cr
\hline
3	& 32	& 3.48	& 0.219		& 0.900	& 7.0	& 1.111		& 2996	\cr
\hline
3	& 54	& 3.68	& 0.204		& 0.966	& 11	& 1.035		& 2210	\cr
\hline
3	& 60	& 3.73	& 0.201		& 0.982	& 12	& 1.018		& 2752	\cr
\hline
3	& 36	& 3.82	& 0.195		& 1.01	& 7.0	& 0.9883	& 3462	\cr
\hline
3	& 48	& 3.86	& 0.192		& 1.03	& 9.2	& 0.9756	& 1800	\cr
\hline
3	& 42	& 3.92	& 0.189		& 1.04	& 7.9	& 0.9572	& 3450	\cr
\hline
3	& 60	& 4.01	& 0.183		& 1.07	& 11	& 0.9308	& 2752	\cr
\hline
3	& 32	& 4.10	& 0.178		& 1.10	& 5.7	& 0.9058	& 3006	\cr
\hline
3	& 54	& 4.25	& 0.171		& 1.15	& 9.2	& 0.8671	& 1976	\cr
\hline
3	& 26	& 4.28	& 0.169		& 1.16	& 4.4	& 0.8597	& 2840	\cr
\hline
3	& 42	& 4.33	& 0.167		& 1.18	& 7.0	& 0.8477	& 3277	\cr
\hline
3	& 48	& 4.38	& 0.165		& 1.20	& 7.9	& 0.8360	& 1800	\cr
\hline
3	& 54	& 4.83	& 0.147		& 1.34	& 7.9	& 0.7439	& 3744	\cr
\hline
3	& 48	& 4.84	& 0.146		& 1.35	& 7.0	& 0.7420	& 3600	\cr
\hline
3	& 36	& 4.52	& 0.159		& 1.24	& 5.7	& 0.8050	& 3300	\cr
\hline
3	& 20	& 4.60	& 0.155		& 1.27	& 3.1	& 0.7883	& 3160	\cr
\hline
3	& 60	& 4.64	& 0.154		& 1.28	& 9.2	& 0.7802	& 2496	\cr
\hline
3	& 32	& 5.09	& 0.138		& 1.43	& 4.4	& 0.6993	& 3070	\cr
\hline
3	& 42	& 5.15	& 0.136		& 1.45	& 5.7	& 0.6897	& 3400	\cr
\hline
3	& 60	& 5.29	& 0.132		& 1.50	& 7.9	& 0.6685	& 4602	\cr
\hline
3	& 54	& 5.36	& 0.130		& 1.52	& 7.0	& 0.6583	& 1800	\cr
\hline
3	& 14	& 5.39	& 0.129		& 1.53	& 1.8	& 0.6540	& 3600	\cr
\hline
3	& 36	& 5.64	& 0.122		& 1.61	& 4.4	& 0.6206	& 3300	\cr
\hline
3	& 26	& 5.76	& 0.119		& 1.65	& 3.1	& 0.6057	& 2768	\cr
\hline
3	& 48	& 5.78	& 0.119		& 1.66	& 5.7	& 0.6033	& 2485	\cr
\hline
3	& 54	& 6.41	& 0.106		& 1.87	& 5.7	& 0.5361	& 3893	\cr
\hline
3	& 42	& 6.45	& 0.105		& 1.88	& 4.4	& 0.5323	& 3450	\cr
\hline
3	& 32	& 6.91	& 0.0970	& 2.03	& 3.1	& 0.4925	& 3070	\cr
\hline
3	& 48	& 7.27	& 0.0917	& 2.15	& 4.4	& 0.4653	& 1800	\cr
\hline
3	& 20	& 7.39	& 0.0900	& 2.19	& 1.8	& 0.4569	& 3160	\cr
\hline
3	& 36	& 7.69	& 0.0861	& 2.29	& 3.1	& 0.4371	& 3387	\cr
\hline
3	& 54	& 8.08	& 0.0815	& 2.42	& 4.4	& 0.4139	& 1976	\cr
\hline
3	& 42	& 8.84	& 0.0739	& 2.67	& 3.1	& 0.3750	& 5359	\cr
\hline
3	& 60	& 8.89	& 0.0734	& 2.68	& 4.4	& 0.3727	& 2496	\cr
\hline
3	& 26	& 9.38	& 0.0692	& 2.84	& 1.8	& 0.3515	& 2840	\cr
\hline
3	& 48	& 10.0	& 0.0646	& 3.05	& 3.1	& 0.3280	& 3894	\cr
\hline
3	& 54	& 11.1	& 0.0577	& 3.41	& 3.1	& 0.2931	& 4973	\cr
\hline
3	& 32	& 11.3	& 0.0566	& 3.48	& 1.8	& 0.2875	& 3208	\cr
\hline
3	& 36	& 12.7	& 0.0500	& 3.94	& 1.8	& 0.2539	& 3410	\cr
\hline
3	& 42	& 14.6	& 0.0432	& 4.57	& 1.8	& 0.2191	& 3357	\cr
\hline
3	& 48	& 16.6	& 0.0377	& 5.22	& 1.8	& 0.1914	& 3769	\cr
\hline
3	& 54	& 18.6	& 0.0335	& 5.88	& 1.8	& 0.1700	& 1924	\cr
\hline
3	& 60	& 20.6	& 0.0301	& 6.54	& 1.8	& 0.01529	& 2400	\cr
\hline
\hline
4	& 14	& 2.179	& 0.221		& 0.889	& 3.1	& 1.124		& 2930	\cr
\hline
4	& 10	& 2.181	& 0.220		& 0.894	& 2.2	& 1.119		& 3000	\cr
\hline
4	& 26	& 2.183	& 0.219		& 0.898	& 5.7	& 1.114		& 3152	\cr
\hline
4	& 22	& 2.185	& 0.218		& 0.902	& 4.8	& 1.109		& 3186	\cr
\hline
4	& 6	& 2.188	& 0.217		& 0.908	& 1.3	& 1.101		& 2220	\cr
\hline
4	& 18	& 2.188	& 0.217		& 0.908	& 3.9	& 1.101		& 3284	\cr
\hline
4	& 30	& 2.188	& 0.217		& 0.908	& 6.5	& 1.101		& 3000	\cr
\hline
4	& 32	& 2.190	& 0.216		& 0.912	& 6.9	& 1.096		& 1646	\cr
\hline
4	& 30	& 2.241	& 0.190		& 1.03	& 5.7	& 0.9667	& 3639	\cr
\hline
4	& 26	& 2.252	& 0.185		& 1.06	& 4.8	& 0.9396	& 3406	\cr
\hline
4	& 22	& 2.268	& 0.177		& 1.11	& 3.9	& 0.9007	& 3096	\cr
\hline
4	& 18	& 2.279	& 0.172		& 1.14	& 3.1	& 0.8743	& 3277	\cr
\hline
4	& 30	& 2.305	& 0.160		& 1.23	& 4.8	& 0.8136	& 1650	\cr
\hline
4	& 14	& 2.311	& 0.158		& 1.25	& 2.2	& 0.7999	& 2910	\cr
\hline
4	& 26	& 2.328	& 0.150		& 1.31	& 3.9	& 0.7618	& 3256	\cr
\hline
4	& 22	& 2.349 & 0.141		& 1.40	& 3.1	& 0.7162	& 2996	\cr
\hline
4	& 10	& 2.376 & 0.130		& 1.52	& 1.3	& 0.6600	& 3000	\cr
\hline
4	& 30	& 2.376 & 0.130		& 1.52	& 3.9	& 0.6600	& 1652	\cr
\hline
4	& 18	& 2.395	& 0.123		& 1.61	& 2.2	& 0.6222	& 3105	\cr
\hline
4	& 26	& 2.403	& 0.120		& 1.65	& 3.1	& 0.6067	& 3253	\cr
\hline
4	& 30	& 2.448 & 0.103		& 1.91	& 3.1	& 0.5246	& 1747	\cr
\hline
4	& 22	& 2.457	& 0.100		& 1.96	& 2.2	& 0.5092	& 3052	\cr
\hline
4	& 14	& 2.480	& 0.0929	& 2.12	& 1.3	& 0.4714	& 2930	\cr
\hline
4	& 26	& 2.507	& 0.0847	& 2.33	& 2.2	& 0.4299	& 3239	\cr
\hline
4	& 30	& 2.548	& 0.0734	& 2.68	& 2.2	& 0.3726	& 1686	\cr
\hline
4	& 18	& 2.552	& 0.0724	& 2.72	& 1.3	& 0.3674	& 3280	\cr
\hline
4	& 22	& 2.609	& 0.0591	& 3.33	& 1.3	& 0.3001	& 3252	\cr
\hline
4	& 26	& 2.656	& 0.0501	& 3.93	& 1.3	& 0.2543	& 3706	\cr
\hline
4	& 30	& 2.698	& 0.0434	& 4.54	& 1.3	& 0.2204	& 1755	\cr
\hline
\end{longtable}

\bibliographystyle{bibstyle}
\bibliography{bib}


\end{document}